\documentclass[natbib,numberedappendix,twocolappendix]{emulateapj}
\usepackage{natbib}
\input epsf
\def\s2n{S^{\prime}/N}

\def\msun{M$_{\odot}$}

\graphicspath{{FIGURES/}}

\usepackage{amssymb,amsmath}
\def\bs{\boldsymbol}

\slugcomment{Submitted to ApJ, \today}
\shorttitle{The Origin of Massive Stars}

\shortauthors{Padoan et al.}

\begin{document}
\title{The Origin of Massive Stars: The Inertial--Inflow Model}

\author{Paolo Padoan,}
\affiliation{Institut de Ci\`{e}ncies del Cosmos, Universitat de Barcelona, IEEC-UB, Mart\'{i} i Franqu\`{e}s 1, E08028 Barcelona, Spain; ppadoan@icc.ub.edu}
\affiliation{ICREA, Pg. Llu\'{i}s Companys 23, 08010 Barcelona, Spain}
\author{Liubin Pan}
\affil{School of Physics and Astronomy, Sun Yat-sen University, 2 Daxue Road, Zhuhai, Guangdong, 519082, China; panlb5@mail.sysu.edu.cn}
\author{Mika Juvela,}
\affiliation{Department of Physics, PO Box 64, University of Helsinki, 00014, Helsinki, Finland}
\author{Troels Haugb{\o}lle,}
\affiliation{Centre for Star and Planet Formation, Niels Bohr Institute and Natural History Museum of Denmark, University of Copenhagen, {\O}ster Voldgade 5-7, DK-1350 Copenhagen K, Denmark}
\author{{\AA}ke Nordlund,}
\affiliation{Centre for Star and Planet Formation, Niels Bohr Institute and Natural History Museum of Denmark, University of Copenhagen, {\O}ster Voldgade 5-7, DK-1350 Copenhagen K, Denmark}

\begin{abstract}

We address the problem of the origin of massive stars, namely the origin, path
and timescale of the mass flows that create them. 
Based on extensive numerical simulations, we propose a scenario where massive stars are
assembled by large-scale, converging, inertial flows that naturally occur in supersonic turbulence. 
We refer to this scenario of massive-star formation as the {\it Inertial-Inflow Model}. 
This model stems directly from the idea that the mass distribution of stars is primarily 
the result of {\it turbulent fragmentation}. 
Under this hypothesis, the statistical properties of the turbulence determine the formation 
timescale and mass of prestellar cores, posing definite constraints on the formation mechanism of massive stars.
We  quantify such constraints by the analysis of a simulation of supernova-driven turbulence in a 250-pc region of 
the interstellar medium, describing the formation of hundreds of massive stars over a time of approximately 30 Myr.
Due to the large size of our statistical sample, we can say with
full confidence that {\it massive stars in general do not form from the collapse of massive cores, nor from 
competitive accretion}, as both models are incompatible with the numerical results. 
We also compute synthetic continuum observables in Herschel and ALMA bands.
We find that, depending on the distance of the observed regions, estimates of core mass based on commonly-used  
methods may exceed the actual core masses by up to two orders of magnitude, and that there is essentially no
correlation between estimated and real core masses.

\end{abstract}


\section{Introduction}

Supersonic turbulence maintains molecular clouds (MCs) in a chaotic state characterized by a complex system of crisscrossing shocks, leading to an intricate 
network of intersecting filaments and to a very broad, approximately log-normal, gas density distribution 
\citep[e.g.][]{Vazquez-Semadeni94,Padoan95,Nordlund+Padoan_Puebla98}. Intersecting filaments generate density peaks that 
can be gravitationally unstable and collapse into protostars. Because the turbulence naturally produces unstable density peaks with a broad range of masses,
the origin of stars of all masses can be understood as a direct effect of supersonic turbulence \citep{Padoan+97imf,Padoan+Nordlund02imf}. We refer to this general 
scenario as \emph{turbulent fragmentation}.
Analytical models of turbulent fragmentation have been developed with the common goal of converting a statistical description of supersonic turbulence 
into a statistical theory of star formation that inherits the universal nature of the turbulence. Both the stellar initial mass function (IMF) \citep{Padoan+Nordlund97imf,Padoan+Nordlund02imf,Hennebelle+Chabrier08,Hopkins12imf} and the star-formation rate (SFR)
\citep{Krumholz+McKee05sfr,Padoan+Nordlund11sfr,Hennebelle+Chabrier11sfr,Federrath+Klessen12,Burkhart18} have been modeled following 
this approach. In this work, the formation of massive stars is conceived in the context of our own
turbulent-fragmentation model of the IMF \citep{Padoan+Nordlund02imf,Padoan+Nordlund11imf}, where prestellar cores are assembled by the turbulence 
through the compression of regions of inertial-range scale that are not required to be gravitationally bound. This picture is at odds with other IMF models based 
on turbulent fragmentation where stars originate from gravitationally-bound overdensities induced by the turbulence
 \citep{Hennebelle+Chabrier08imf,Hennebelle+Chabrier09imf,Chabrier+Hennebelle2011,Hopkins12imf}.

From the viewpoint of our turbulent fragmentation model, high-mass stars have the same origin as low-mass stars, both being the consequence of a local pileup 
of gas by the random velocity field of a MC. While massive density peaks are all gravitationally unstable, low-mass peaks may not reach high enough density to
exceed their own critical Bonnor-Ebert mass, so only a fraction of them collapse into low-mass stars, while the rest are transient and eventually disperse. This 
selection by gravity results in the IMF turnover, an approximately log-normal distribution of stellar masses that reflects the density distribution of the turbulent 
gas \citep{Padoan+97imf}. Far from the turnover, massive stars follow a power-law mass distribution, presumably related to the scale-free nature of the turbulence. 

Despite their common origin, low and high-mass stars achieve their final mass on different timescales. Because of the velocity scaling of the turbulence, the turnover 
time increases with increasing scale. Larger stellar masses require converging flows from larger scales \citep{Padoan+Nordlund02imf}, so the time to accumulate the final 
stellar mass (of the order of the turnover time) increases with mass \citep{Padoan+Nordlund11imf}. For typical conditions in MCs, the turnover time of the converging 
flows is longer than the free-fall time of the prestellar cores, except possibly for the smallest-scale compressions responsible for the origin of the lowest-mass stars and 
brown dwarfs. Thus, we view the formation of a star as a three-step process: (1) the formation of a gravitationally unstable core exceeding the critical Bonnor-Ebert mass, 
(2) the collapse of the core into a low to intermediate-mass star, (3) the accretion of the remaining mass (through a circumstellar disk) driven by a large-scale 
converging flow, with the gradual buildup of the stellar mass over a longer time than the initial collapse time of the core. For low-mass stars, the third step may be 
relatively brief and contributes only a small fraction of their final mass. On the contrary, most of the mass of a massive star is assembled during the third step over 
many (core) free-fall times, as the final stellar mass is much larger than the critical Bonnor-Ebert mass of the core.

Because most of the final stellar mass is channeled towards the accreting star by the random velocity field from large scale, unaffected by the stellar 
gravity during most of its path towards the star, we refer to this process as \emph{inertial inflow}. Thus, we propose to name this scenario for the origin 
of massive stars the \emph{inertial-inflow model}, to distinguish it from the \emph{core-collapse model} that requires a much larger initial core mass, and 
from the \emph{competitive-accretion model} that only accounts for the mass accretion due to the gravity of the growing star, neglecting the preexisting 
inertial inflow at larger scale.\footnote{In simulations of turbulent clouds with sufficient numerical resolution the inertial flow feeding the formation of 
massive stars is naturally present, so one can erroneously interpret the growth of a massive star by its Bondi-Hoyle accretion rate as a confirmation 
of competitive accretion, while in reality the accreting mass is controlled by the inertial inflow from much larger-scale.} 
The main goal of this work is to use a numerical simulation to test and quantify this scenario that stems directly from our turbulent fragmentation model. 
Using 4-pc scale simulations that yield a full and realistic stellar IMF, we have already shown that the formation of a star requires a time that grows with the 
final stellar mass \citep{Padoan+14accr,Haugbolle+18imf}. Here, we use a 250-pc simulation to obtain a much larger sample of massive stars, besides a more 
realistic description of the formation and evolution of star-forming regions.  

This work is organized as follows. In the next section, we define the basic terminology adopted to address the multi-scale nature of our scenario. The 
numerical simulation is described in \S~\ref{sec_method}. We then present several simulation results, starting with the star-formation timescale in 
\S~\ref{sec_time} and the evolution of the accretion rates in \S~\ref{sec_history}. The analysis of the initial conditions for star formation is presented 
in \S~\ref{sec_ic}, where we focus on the mass of prestellar cores, and in \S~\ref{sec_ic_inflow}, where we study the inflow region around the cores. 
In \S~\ref{sec_model} we present the new scenario based on the numerical results of the preceding sections, and argue that all current 
models of massive star formation, as well as models of the stellar IMF, require fundamental revisions. We then address, in \S~\ref{sec_observations}, 
the observational properties of the prestellar cores, by generating synthetic sub-mm observations, and briefly discuss other works related to our scenario,
as well as some observational constraints, in \S~\ref{sec_discussion}. The main results and conclusions are summarized in \S~\ref{sec_conclusions}.

\section{Inertial Inflow, infall, and Accretion} \label{sec_def}

\begin{figure}[t]
\includegraphics[width=\columnwidth]{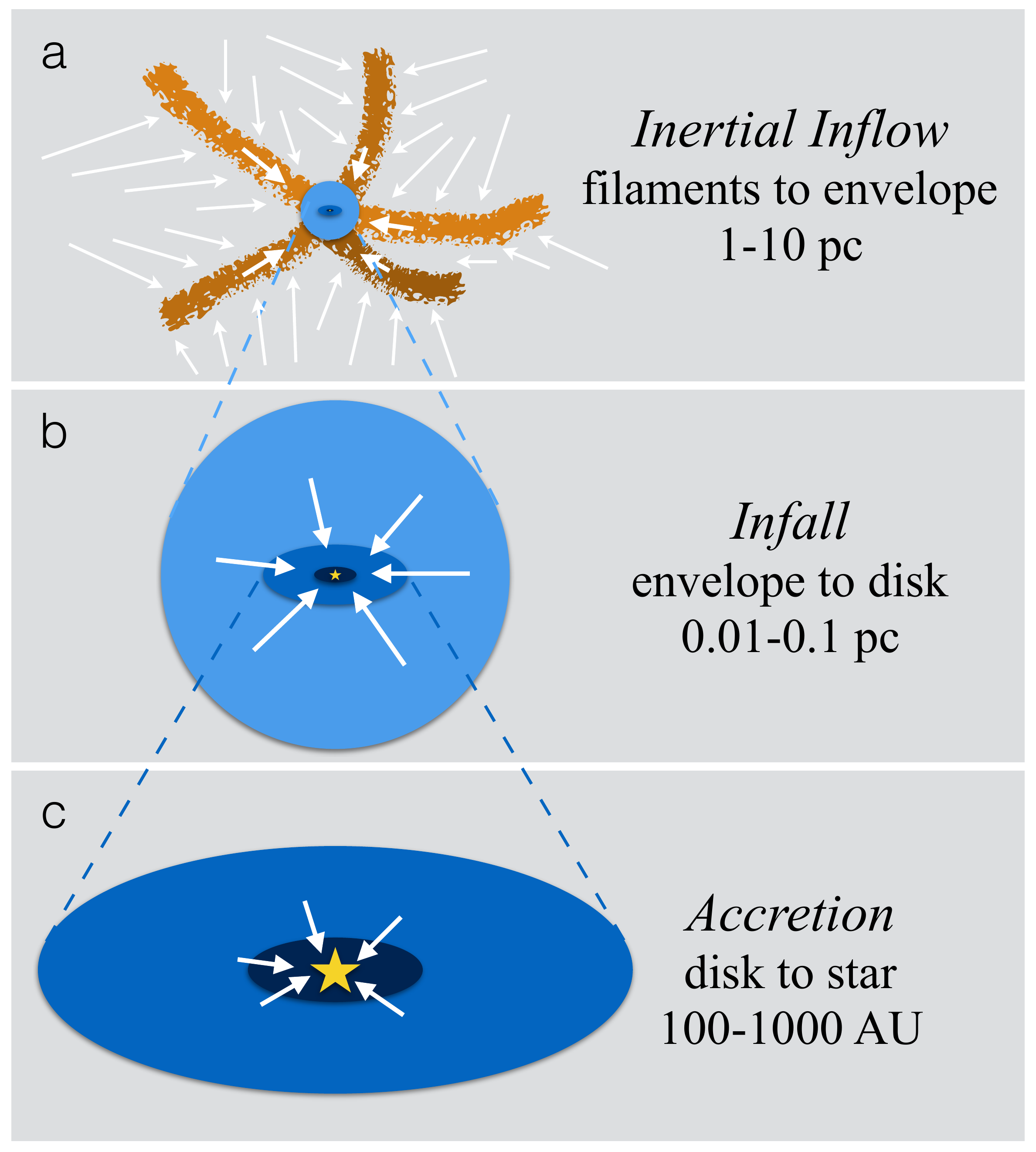}
\caption[]{Sketch of the different scales and corresponding terminology adopted in our inertial-inflow model. The infall and disk-accretion scales
inherit the filamentary structure of the larger scale, but are here depicted as smooth regions for simplicity.  
}
\label{sketch}
\end{figure}

In this work we study a scenario for massive-star formation where the origin and subsequent growth of a star are addressed self-consistently in the 
context of the large-scale ISM turbulence. Because of the multi-scale nature of our perspective, we refer to the mass flow of interstellar gas onto 
a growing star with the following terminology that emphasizes the different physical nature of this mass flow at three different scales: {\it inertial inflow},
{\it infall}, and {\it accretion}. This terminology is illustrated by the sketch in Figure~\ref{sketch}. 

We adopt the term {\it inertial inflow} to refer to a converging motion on a scale of few to several pc in the turbulent flow of a MC. Regions of converging 
motion arise naturally in supersonic turbulence, and we view them as inertial because the kinetic energy of the turbulence on that scale usually 
exceeds both the thermal energy (velocities are supersonic) and the gravitational energy, for characteristic virial parameter values and scaling relations 
of MCs. Because supersonic turbulence yields a filamentary morphology (due to intersections of postshock sheets), and dense cores are formed at the 
intersections of dense filaments, the converging motion occurs predominantly through filaments feeding the emerging prestellar core. After the collapse of 
the core, this inertial inflow from pc-scale filaments may continue, providing the mass reservoir for the growth of a massive star. The size of this mass reservoir
at the start of the prestellar-core collapse will be defined in \S~\ref{sec_ic_def} and we will refer to it as the {\it inflow radius}. 
As illustrated in panel a of Figure~\ref{sketch}, the inflow region is highly turbulent, so the velocity field is dominated by random motions, not by the inflow
component along the filaments.\footnote{In \S~\ref{sec_ic_inflow} --mid, left panel of Figure~\ref{inflow_profiles}-- we show that the radial component of the velocity is much 
smaller than the random component.} 

At smaller scale, self-gravity exceeds the kinetic energy of the turbulence and has a strong effect on the converging motion, thus we refer to this motion as {\it infall}. 
The size of the infall region at the start of the prestellar-core collapse will be defined in \S~\ref{sec_ic_def} and we will refer to it as the {\it infall radius}, which 
we will find to be typically larger than the prestellar-core radius. At later phases, when the gravitational potential of the star is dominant, the infall region is a 
dense envelope that feeds a circumstellar disk, and its size is of the order of the gravitational accretion radius of the star. For example, for a 10 M$_{\odot}$ star, the 
infall-dominated region may extend to $\sim 1$ pc in the case of subsonic inflow (the Bondi radius, $GM/c_{\rm s}^2$), or stay within $\sim 0.08$ pc for 
supersonic motion with a velocity of 1 km~s$^{-1}$ (the Hoyle-Lyttleton radius, $2GM/v_{\infty}^2$).

At even smaller scale, the gas finally accretes from the circumstellar disk onto the stellar surface. We reserve the term {\it accretion} for this process, on 
scales below the characteristic disk size of 100 to 1000~AU. Due to its spatial resolution, our simulation does not describe the accretion process, but 
addresses both the inflow and infall phases, within an approximation that neglects radiative feedback, as discussed below. Thus, when we measure 
the growth rate of a sink particle, we refer to it as infall rate, which, following the scenario of this work, is driven by the inertial inflow.

In the middle and bottom panels of Figure~\ref{sketch}, the infall and disk-accretion scales are depicted as smooth regions for simplicity, to stress that 
the role of inertial inflows is no longer dominant on those scales. However, the filamentary nature of the turbulent inflow region is certainly inherited 
by the smaller scales, as demonstrated by recent multi-scale zoom-in simulations covering a range of scales from 40~pc to 2~AU \citep{Kuffmeier+17,Kuffmeier+19}.   

\section{Numerical Approach} \label{sec_method} 

\begin{figure*}[t]
\centering
\includegraphics[width=\columnwidth]{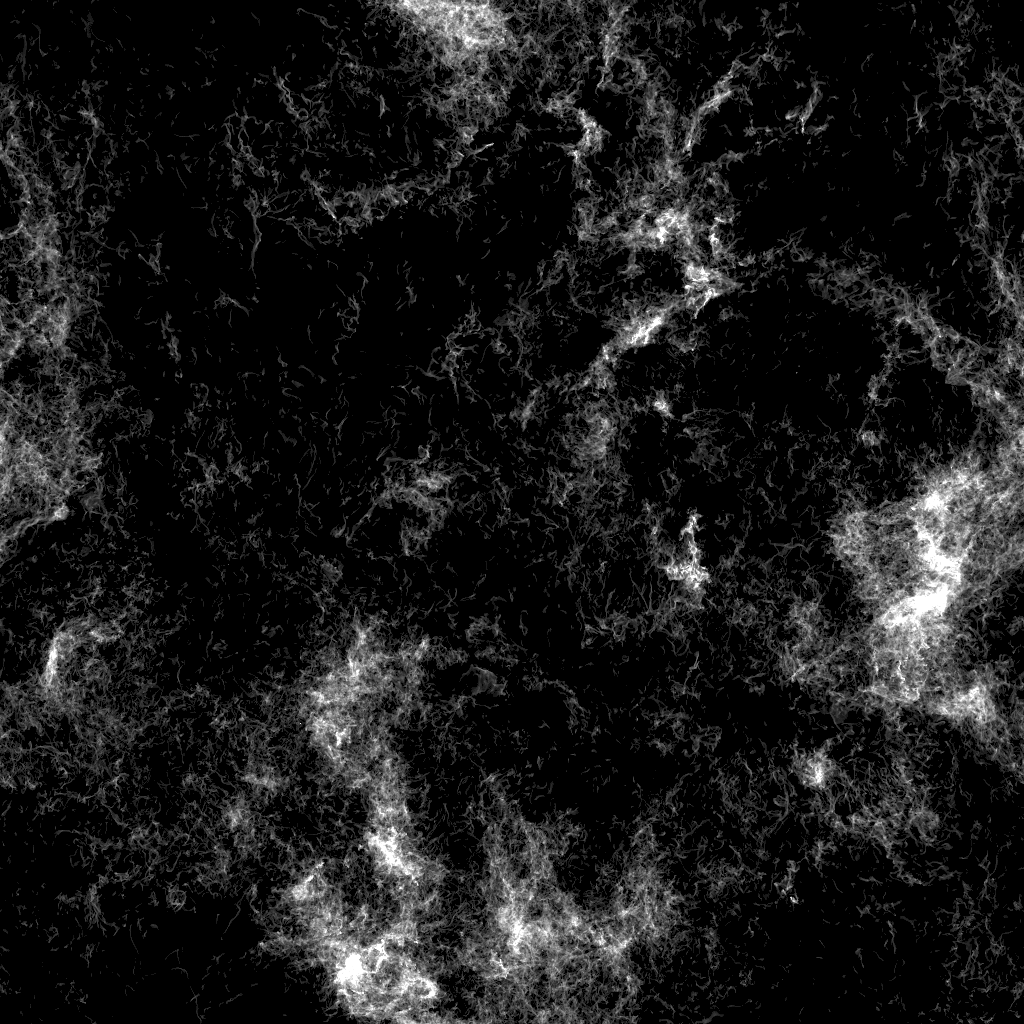}
\centering
\includegraphics[width=\columnwidth]{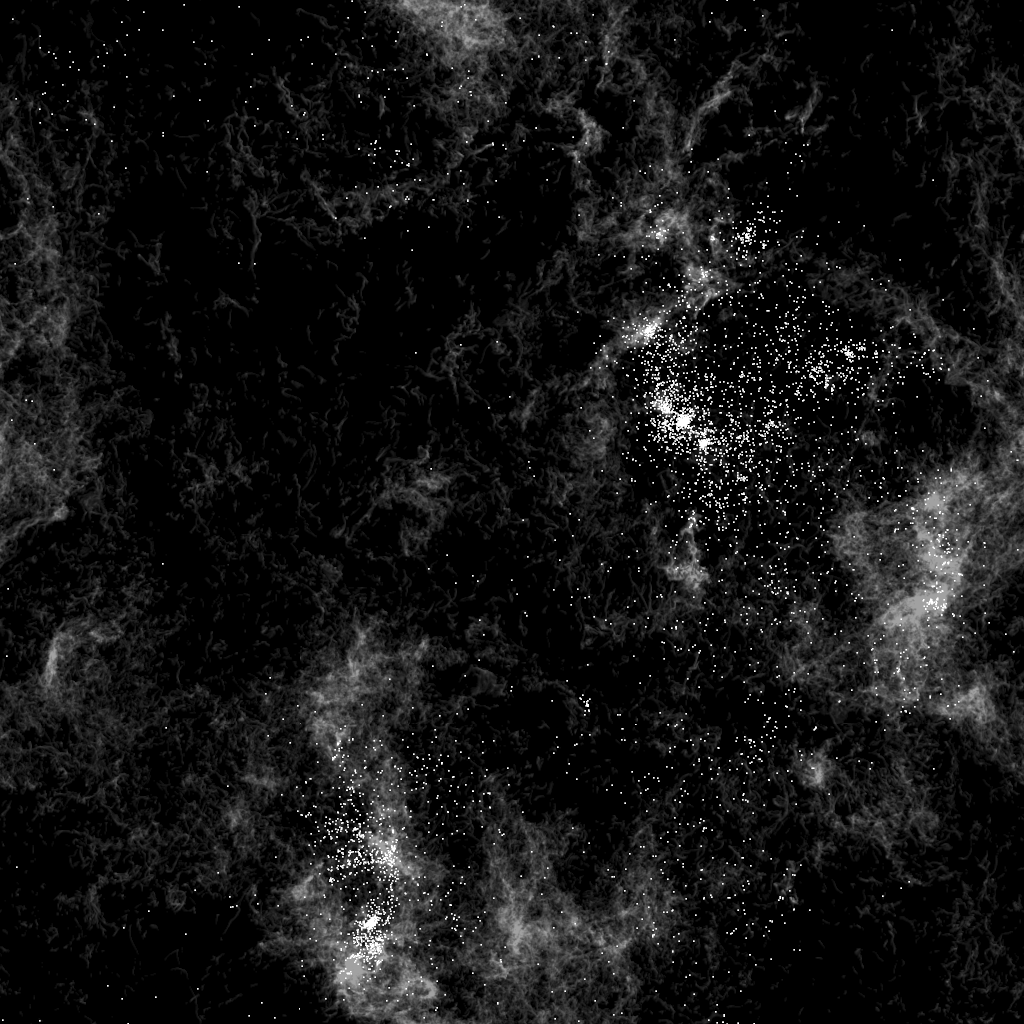}
\caption[]{Left panel: column density at the end of the simulation over the whole 250-pc volume. Right panel: same as the left panel, but
including also the positions of the approximately 3,000 stars more massive than $2.5\,M_{\odot}$ (the brightness of the column density
has been reduced relative to the left panel).    
}
\label{images}
\end{figure*}

This work is based on the same supernova (SN) driven magneto-hydrodynamic (MHD) simulation as in \citet{Padoan+17sfr}. Details of the 
numerical methods can be found there and in \citet{Padoan+16SN_I}. Here we only briefly summarize the numerical setup. The 3D MHD 
equations are solved with the Ramses adaptive-mesh-refinement (AMR) code \citep{Teyssier02,Fromang+06,Teyssier07} within a cubic 
region of size $L_{\rm box}=250$ pc, total mass $M_{\rm box}=1.9\times 10^6$ $M_{\odot}$, and periodic boundary conditions. The initial 
conditions are taken from a SN-driven simulation that was integrated for 45 Myr without self-gravity \citep{Padoan+16SN_I} with a mean 
density $n_{\rm H,0}=5$ cm$^{-3}$ and a mean magnetic field $B_0=4.6$ $\mu$G. The rms magnetic field generated by the turbulence has 
a value of 7.2 $\mu$G and an average of $|\bs B|$ of 6.0 $\mu$G, consistent with the value of $6.0 \pm 1.8$ $\mu$G 
derived from the `Millennium Arecibo 21-cm Absorption-Line Survey' by \citet{Heiles+Troland05}. 

The only driving force is from SN feedback. SNe are randomly distributed in space and time during the first period of the simulation without
self-gravity, while they are later determined by the position and age of the massive sink particles formed when self-gravity is included. In the 
initial phase without gravity, the minimum cell size is $dx=0.24$ pc, achieved with a $128^3$ root grid and three AMR levels, until $t=45$ Myr. 
It is then decreased to $dx=0.03$ pc, using a root-grid of $512^3$ cells and four AMR levels, during an additional period of 10.5 Myr without 
self-gravity. Finally, at $t=55.5$ Myr, gravity is introduced and the minimum cell size is further reduced to $dx=0.0076$ pc (1568 AU) by adding two more 
AMR levels. With this final setup we can follow the star-formation process (see details below), and the simulation is continued for that purpose 
for an additional period of approximately 30~Myr, The simulation also includes 250 million passively advected tracer particles, each representing 
a fluid element with a characteristic mass of approximately 0.008 $M_{\odot}$. The tracer particles record all the hydrodynamic variables and are 
tagged once they accrete onto a sink particle. 

To follow the collapse of prestellar cores, sink particles are created in cells where the gas density is larger than $10^6$ cm$^{-3}$, if the following 
conditions are met at the cell location: i) The gravitational potential has a local minimum value, ii) the three-dimensional velocity divergence is negative, 
and iii) no other previously created sink particle is present within an exclusion radius, $r_{\rm excl}$ ($r_{\rm excl}=16 dx=0.12$~pc in this simulation). 
We have verified that these conditions, similar to those in \citet{Federrath+10sinks}, avoid the creation of spurious sink particles in regions where the 
gas is not collapsing \citep{Haugbolle+18imf}. Sink particles gradually accrete the gravitationally-bound surrounding gas within an accretion radius 
$r_{\rm accr} =4 \mathrm{dx}=0.03$~pc, with an efficiency $\epsilon_{\rm out}=0.5$, meaning that only half of the infalling gas contributes to the growth of
the sink-particle mass. 

The resolution of the simulation is high enough to interpret individual sink particles as individual stars.
When a sink particle of mass larger than 7.5 M$_{\odot}$ has an age equal to the corresponding stellar lifetime for that mass \citep{Schaller+92},
a sphere of $10^{51}$ erg of thermal energy is injected at the location of the sink particle to simulate the SN explosion, as described in detail in
\citet{Padoan+16SN_I}. We refer to this driving method as \emph{real SNe}, as it provides a SN feedback that is fully consistent with the SFR,
the stellar IMF, and the ages and positions of the individual stars whose formation is resolved in the simulation. 

The simulation has so far been run for approximately 30 Myr with self-gravity, star formation and \emph{real SNe}, generating $\sim 3,000$ stars
with mass $> 2.5\,M_{\odot}$ and $\sim 800$ stars with mass $> 8\,M_{\odot}$. The left panel of Figure~\ref{images} shows the column density of 
the whole computational volume at the end of the simulation. The gas distribution is highly filamentary on all scales and densities, with large voids 
created by the explosions of multiple SNe. The stars with mass $> 2.5\,M_{\odot}$ are shown on the right panel of the same figure, where the grayscale 
intensity range has been compressed. Young stars are found inside the densest filaments, while older ones have already left their parent clouds. 
Most of the stars in the simulation are formed in clusters or associations, some of which have cleared their surrounding gas thanks to SN explosions 
of their most massive members. We have identified seven clusters with mass $>10^4\,M_{\odot}$, whose structural and dynamical properties will be 
the focus of future works. 

For the purpose of this work, we select a subsample of stars by retaining only sink particles formed before the last 1~Myr of the simulation and with a 
negligible final accretion rate averaged over the last 1~Myr of the simulation (we require that the time to double the final stellar mass at that average
rate is longer than 1~Gyr), so the final stellar masses are well defined. This selection yields a sample of 1,503 stars with mass $> 2.5\,M_{\odot}$, of 
which $\sim 447$ stars have mass $> 8\,M_{\odot}$. 

The simulation snapshots are saved every 30~kyr, so we have a total of approximately 1,000 snapshots (nearly 200 TB of data).  
The star formation is distributed over many different clouds with realistic values of the SFR, and the global SFR corresponds to a mean gas depletion 
time in the computational volume of almost 1 Gyr, also realistic for a 250-pc scale \citep{Padoan+17sfr}.

\subsection{Caveats and Limitations} \label{sec_caveats}

In the 70s and 80s, the main problem of massive star formation was to understand how accretion could overcome the very high radiation 
pressure of the star \citep[e.g.][]{Kahn74,Yorke+Kruegel77,Wolfire+Cassinelli87}. It was later understood that if the accretion proceeds through 
optically thick blobs and fingers and an optically thick disk, and much of the radiation escapes through optically thin channels created by the outflow, 
radiation pressure does not impede the growth of a massive star \citep[e.g.][]{Krumholz+05,Keto07,Krumholz+09,Kuiper+11,Klassen+16}. 
Radiative
bubbles around massive protostars cannot prevent the accretion of the infalling gas onto the star-disk system, because such bubbles are 
Rayleigh-Taylor unstable at early times \citep{Rosen+16} and the instability is expected to occur even in the magnetized case, though
with a longer growth time-scale \citep{Yaghoobi+Shadmehri18}. Although the precise role of the various radiative feedback mechanisms remains difficult 
to quantify, here we focus on the {\it origin} of massive stars, that is the initial conditions responsible for their creation, and on the {\it source and timescale} 
of the accretion process, neglecting radiative feedback. Thus, the final stellar mass we derive is not computed precisely. 

To account for the overall effect of jets and outflows, we assume an efficiency factor  $\epsilon_{\rm out}=0.5$, 
meaning that only half of the infalling mass is accreted onto the sink particle, following our previous works \citep{Padoan+14acc,Haugbolle+18imf}. 
While this is a reasonable approximation for low-mass stars \citep{Matzner+McKee2000}, the efficiency may decrease with increasing final mass 
and with decreasing surface density of the prestellar core, at least in the context of models were all the mass reservoir is initially contained in a 
dense core \citep[e.g.][]{Tanaka+17}. Nevertheless, recent simulations including radiation forces, photoionization feedback and protostellar outflows
show that a value close to $\epsilon_{\rm out}=0.5$ is not unreasonable even for very massive stars \citep{Kuiper+Hosokawa18}.

The role of radiation feedback mechanisms in the case of a longer formation time and highly filamentary morphology, 
as in our simulation, should be addressed systematically in future studies, accounting for the effect of the accretion rate on the stellar 
structure \citep[see][]{Jensen+Haugboelle18}. 
For example, the accretion rates in our simulation may maintain our stars bloated until they reach the main sequence as an intermediate-mass star 
and ionization feedback would not play a role in that initial phase, as also confirmed observationally by the high luminosity of young, massive 
protostars \citep{Ginsburg+17}. There is also tentative observational evidence that stellar radiation cannot strongly affect the mass inflow when 
this occurs through dense filaments \citep{Watkins+19}. Radiative feedback mechanisms may also assist the formation of massive stars, by 
suppressing fragmentation in the neighborhood of a massive star, which increases the mass reservoir available for its growth, while preventing 
the formation of lower-mass stars \citep{Krumholz+07rad}. 

On the other hand, the limited spatial resolution of the simulation may lead us to overestimate the final stellar mass, as the fragmentation in the 
neighborhood of a massive star is not fully resolved. The final stellar masses must be corrected for this resolution effect, which we do by 
multiplying them by a mass-correction factor, $f_{\rm m}<1$, which is derived in the following. This mass correction is not applied to the sink 
masses in the simulation, but only applied a posteriori as we interpret the results, so it does not affect mass conservation in the simulation. 
However, when we estimate stellar lifetimes to decide if and when a sink particle should explode as a SN, we do account for this mass correction, 
to avoid overestimating the SN feedback.  

\begin{figure}[t]
\includegraphics[width=\columnwidth]{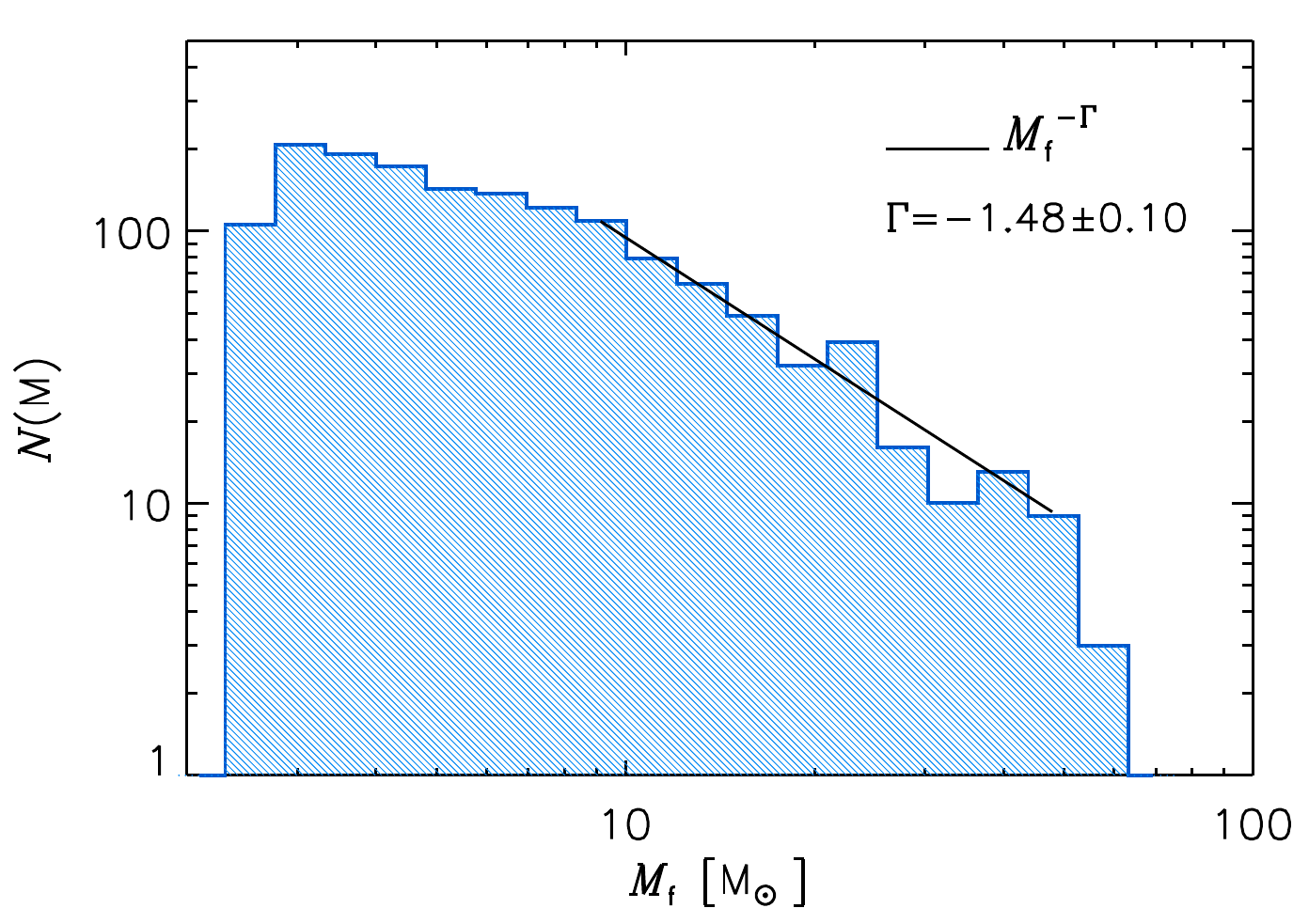}
\caption[]{Mass distribution of sink particles formed in the simulation over a period of approximately 30 Myr. Only stars with final masses 
$>2.5$ M$_{\odot}$ and negligible final accretion rates are shown, and the sink masses have been multiplied by $f_{\rm m}=0.53$ (see text for details).
The IMF is consistent with a power law with slope $\Gamma\approx -1.5$ between 10 and 50~M$_{\odot}$.  
}
\label{imf}
\end{figure}

The lack of fragmentation caused by the limited spatial resolution is illustrated by the incompleteness of the numerical IMF below a few solar
masses, so we can use the numerical IMF to derive an estimate for the mass-correction factor, $f_{\rm m}$. Because not all of the missing 
low-mass and intermediate-mass stars should originate from the same mass reservoir as the high-mass stars, we expect this IMF-based correction 
to be too large, hence the final stellar masses somewhat underestimated, irrespective of the precise outcome of the two radiative
feedback mechanisms mentioned above. 

The simulation was designed to yield a complete IMF for stellar masses above approximately 8 \msun, as one of the main goals is to achieve a realistic 
SN feedback by resolving the formation of all the individual stars that end their life as SNe (see \S~\ref{sec_method}). Figure \ref{imf} shows that the IMF 
from the simulation (with masses already multiplied by $f_{\rm m}$) is consistent with a single power law above approximately 8 \msun, with a slope 
$\Gamma=1.49\pm0.09$, slightly steeper than Salpeter's value \citep{Salpeter55}, but consistent with the result of a study of many stellar 
clusters in M31, covering a similar range of stellar masses as in our simulation \citep{Weisz+15}. Below approximately 8 \msun, the IMF is significantly 
shallower, and essentially flat below 2~\msun\ (not shown in Figure \ref{imf}). Thus, in this work, we only consider stars with final masses $>2.5$~M$_{\odot}$. 
The IMF in Figure \ref{imf} exhibits a cutoff above 50-60~M$_{\odot}$, which is probably real as it corresponds to the maximum stellar mass for this simulation 
as proposed in \S~\ref{sec_mmax}.

In previous isothermal simulations representing regions of a few pc with higher spatial resolution than here \citep{Padoan+14acc,Haugbolle+18imf}, 
we obtained complete stellar IMFs consistent with the observations, meaning Chabrier's IMF \citep{Chabrier05} below 2 \msun\ and Salpeter's IMF 
at larger masses. In \citet{Haugbolle+18imf}, a convergence test was carried out in the range of cell-size resolution from 800 AU to 50 AU, 
showing strong evidence of numerical convergence of the IMF. Furthermore, previous work has shown that simulations of rather low spatial resolution 
can achieve numerical convergence of the star-formation rate (SFR) even with a very incomplete IMF \citep{Padoan+Nordlund11sfr,Padoan+12sfr,Haugbolle+18imf}. 
Thus, it can be expected that, with a higher spatial resolution, 
1) the numerically-converged IMF would be the same Chabrier+Salpeter IMF as in the observations and 2) the SFR in the simulation is already numerically 
converged. These two assumptions imply that the IMF incompleteness is the result of overestimating the final mass of our sink particles, $M_{\rm s,f}$, because 
some of the mass assigned to them should instead have resulted into a few lower-mass stars in the same neighborhood that were not resolved. The correct 
final stellar mass, $M_{\rm f}$, should then be given by $M_{\rm f}=f_{\rm m} M_{\rm s,f}$, with $f_{\rm m}<1$ assumed to be constant (independent of mass).
We assume $f_{\rm m}$ to be independent of mass because the slope of the mass distribution of sink particles for $M_{\rm s,f} > 16$~M$_{\odot}$ is already 
consistent with the observed IMF.  

Based on the above assumptions, we estimate the mass-correction coefficient, $f_{\rm m}$, as follows. We normalize both the observational IMF, $N_{\rm obs}$,
and the sink IMF, $N_{\rm sink}$, to a total probability of unity, $\int{N_{\rm obs}\,dm}=\int{N_{\rm sink}\,dm}=1$, in line with our assumption that the SFR in the simulation 
is correct (numerically converged). Of course, with such a normalization, the sink IMF has a too large ratio of high to low-mass sinks, relative to the correct 
(observed) stellar IMF. Thus, we derive $f_{\rm m}$ by imposing the condition that, for large masses where the sink IMF is a power law, between the masses $m_1$ and 
$m_2$, the total mass of sink particles, multiplied by $f_{\rm m}$, is equal to the total stellar mass derived from the observational IMF in the corresponding interval between 
the masses $m_1f_{\rm m}$ and $m_2f_{\rm m}$:
\begin{equation}
f_{\rm m}\,\,\int_{m_1}^{m_2}{N_{\rm sink}(m)\,dm}=\int_{m_1f_{\rm m}}^{m_2f_{\rm m}}{N_{\rm obs}(m)\,dm}.
\label{fm}
\end{equation}
We have adopted $m_1=20$~M$_{\odot}$ and $m_2=100$~M$_{\odot}$, as these values define the range of sink masses where $N_{\rm sink}(m)$ is approximately 
a power law. We solve the implicit Equation~(\ref{fm}) by iteration, and obtain a mass-correction factor $f_{\rm m}=0.53$. Thus, the final stellar masses 
are assumed to be equal to the sink masses multiplied by $f_{\rm m}$, $M_{\rm f}=f_{\rm m} M_{\rm s,f}$. Figure \ref{imf} shows the current IMF, after 
evolving the simulation for approximately 30 Myr with self-gravity. The mass-completeness limit of approximately 8 \msun\ (16 \msun\ for the sink masses) 
and the general IMF shape was already clear at the beginning of the star-formation process, so the mass-correction factor could be estimated early on in 
the simulation and was applied to compute the lifetime of all the sink particles. In \citet{Haugbolle+18imf}, a resolution study showed that at a cell-size 
resolution of 800 AU the IMF is complete to $\sim 3\,M_{\odot}$, which, scaling to the current simulation, implies a completeness limit of approximately
8~M$_{\odot}$, in good accordance with the above, more quantitative analysis.

Finally, it should be stressed that the simulation was not tailored to represent any specific star-formation region in the Galaxy, nor particularly extreme conditions 
such as those found near the Galactic center or in other very dense regions of massive star formation. With a total mass of $1.9\times 10^6$ M$_{\odot}$, the
mean column density of the simulation is 30~M$_{\odot}$pc$^{-2}$, so our computational volume may be viewed as a generic dense section of a spiral arm. 
For example, the total column density in the Perseus arm of the Milky Way is 23 M$_{\odot}$pc$^{-2}$ \citep{Heyer+Terebey98}. In fact, we have shown in
previous works \citep{Padoan+16SN_I,Pan+16,Padoan+16SN_III} that the lower-resolution version of this simulation generated MCs with properties consistent
with those of real MCs from the $^{12}$CO FCRAO Outer Galaxy Survey \citep{Heyer+98,Heyer+01}. Because a significant fraction of massive stars may be formed 
under more extreme conditions than those found in our simulation, the star-formation time and the maximum stellar mass should be rescaled accordingly when 
more extreme regions are considered, which is discussed in \S~\ref{sec_scaling} and \S~\ref{sec_mmax}. However, the qualitative conclusions 
of this work have a general validity.

\section{The Star-Formation Timescale} \label{sec_time}

The formation timescale of massive stars may differ significantly between alternative models. For example, in the turbulent-core model 
\citep{McKee+Tan02,McKee+Tan03}, the formation timescale is very rapid, of the order of the free-fall time of the massive prestellar core. 
On the contrary, the formation time could last much longer in the case of competitive accretion \citep{Bonnell+2001a,Bonnell+2001b}, 
as well as in the formation scenario implied by our turbulent fragmentation model \citep{Padoan+Nordlund11imf}.
Furthermore, the time evolution of the accretion rate of a massive star is also an important test of the theoretical models, particularly for 
those predicting a long formation timescale. For example, in the competitive accretion model, the accretion is strongly dependent on the 
evolution of the stellar mass, while in our scenario the infall that controls the accretion rate is determined by converging flows on scales 
too large to be affected by the stellar gravity, thus insensitive to the increase of the stellar mass over time.

To study the formation timescale in our simulation, we use the sample of 1503 stars with mass $>2.5~M_{\odot}$ and negligible final accretion rate 
as described in \S~\ref{sec_method}, and define the final stellar mass, $M_{\rm f}$, as the final sink mass, $M_{\rm s,f}$, multiplied by the 
mass-correction factor, $f_{\rm m}$, described in \S~\ref{sec_caveats}, $M_{\rm f}=f_{\rm m}M_{\rm s,f}$. We define the formation time, $t_{95}$, 
as the time interval between the sink-particle creation (approximately the time when the prestellar core starts to collapse) and the moment when 
the sink particle reaches 95\% of its final mass. The results of this study are not very sensitive to this precise percentage\footnote{The slope 
of the relation between formation time and final mass is only slightly increased as smaller percentages of the final mass are used in the definition 
of the formation time, varying from 0.47 to 0.55 when we adopt from 95\% to 50\% of the final mass.}.

In \citet{Padoan+14acc}, using a large-dynamic-range simulation of a 4-pc volume, with periodic boundaries, isothermal equation of state, and random driving, we 
obtained nearly 1300 sink particles over a time of 3.2 Myr, with a mass function closely following a Chabrier IMF at small masses and a Salpeter
IMF at masses larger than 1-2 M$_{\odot}$. We used that simulation to argue that the large-scale mass flow from the turbulent inertial flows feeding the protostars
(through an accretion disk in nature) could explain the observed luminosity distribution of protostars, and that the later Bondi-Hoyle phase could also account 
for the observed accretion rates of pre-main-sequence stars. We also showed that, on average, the time to gather 95\% of the final
stellar mass, $t_{95}$, increased with increasing final stellar mass, $M_{\rm f}$, according to
$t_{95} = 0.45\, {\rm Myr} \times (M_{\rm f}/1 {\rm M}_{\odot} )^{0.56}$, so it took on average nearly 2~Myr to form a 10~M$_{\odot}$ star
(see Figure 13 in \citet{Padoan+14acc}). However, we did not see an accelerated accretion rate as the stars gain mass, so our results
were at odds with the predictions of the competitive accretion scenario \citep{Bonnell+2001a,Bonnell+2001b,Bonnell+Bate2006}. 
These results have been confirmed by the highest-resolution simulation in our recent 
IMF study \citep{Haugbolle+18imf}, with physical and numerical parameters similar to those in \citet{Padoan+14acc}. The power-law fit 
to the relation between formation times and final masses in this more recent work is $t_{95} = 0.51\, {\rm Myr} \times (M_{\rm f}/1 {\rm M}_{\odot} )^{0.58}$,
essentially indistinguishable from the previous one.   

\begin{figure}[t]
\includegraphics[width=\columnwidth]{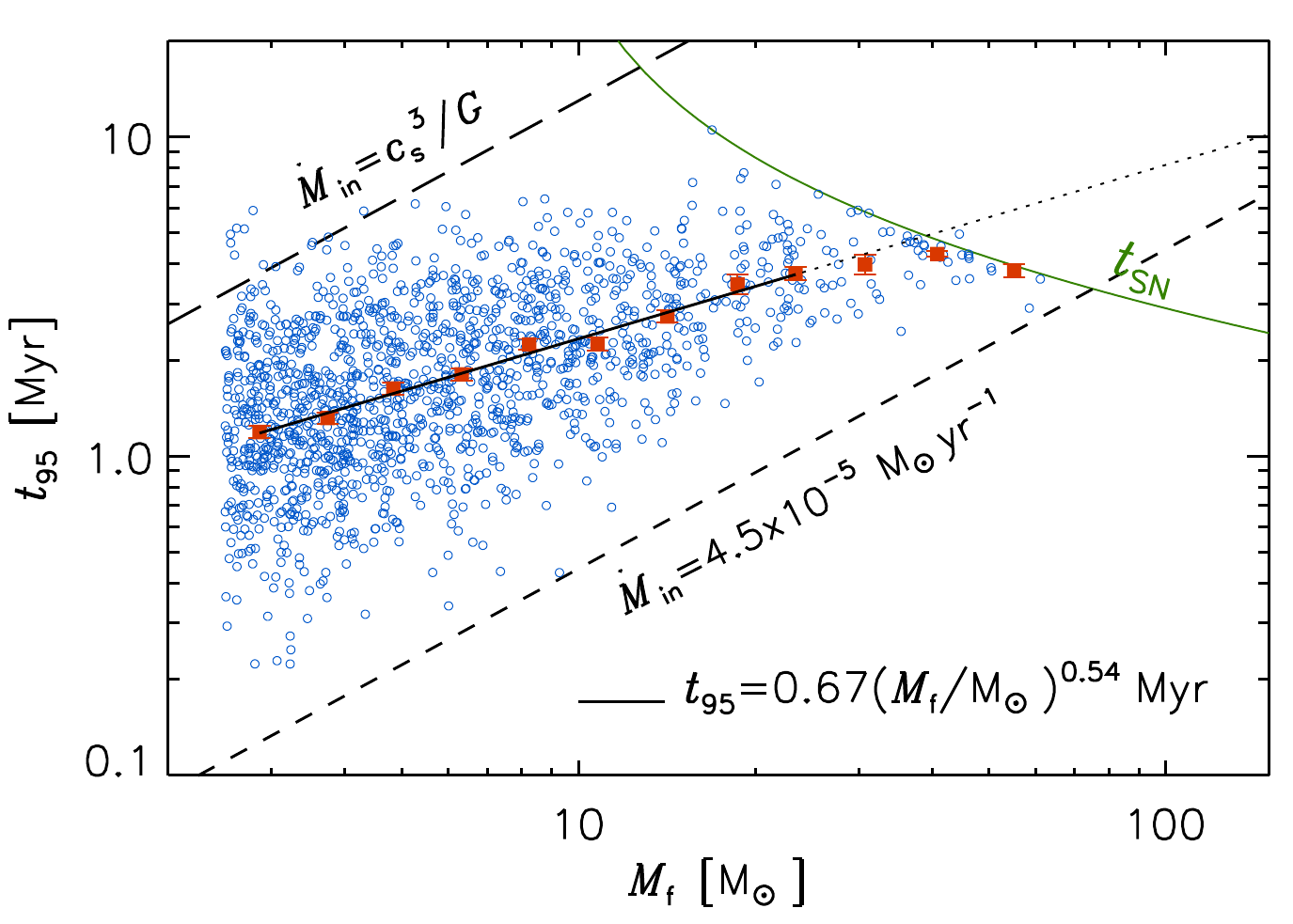}
\caption[]{Timescale to reach 95\% of the final mass versus final mass for the 1,503 stars in the simulation with mass $\ge 2.5$ M$_{\odot}$ that have 
stopped accreting at the time $t=30$~Myr. The curved solid line shows and analytic fit to the relation between stellar lifetime, $t_{\rm SN}$, and mass from 
\citet{Schaller+92}. The short-dashed line is an approximate lower envelope of the scatter plot, while the long-dashed line is the infall
rate from the collapse of a critical isothermal sphere \citep{Shu+87}, assuming $T=10$~K.
The square symbols with error bars show the median values of $t_{95}$ in logarithmic intervals of $M_{\rm f}$. The straight solid
line is a power-law fit to those median values up to 23 M$_{\odot}$, giving $t_{95}=0.67\, (M_{\rm f}/{\rm M}_{\odot})^{0.54}$ Myr. The power-law fit is not
extended to larger masses where the growth time of stars becomes limited by $t_{\rm SN}$ (the extrapolation of the power law to larger masses
is shown by the dotted line). 
}
\label{t95_v1}
\end{figure}
%

While those studies barely reached a maximum stellar mass of approximately 10 M$_{\odot}$, due to the limited volume and total mass, the current 
work yields a large number of stars more massive than 10 M$_{\odot}$, even after applying the mass-correction factor, $f_{\rm m}$, described in \S~\ref{sec_caveats}. 
Thus, we can verify if the relation between formation time and final mass extends to very massive stars as well. The relation from our SN-driven simulation is shown
in Figure~\ref{t95_v1}, where we have plotted only the 1,503 sinks with $M_{\rm f}>2.5$ M$_{\odot}$ and negligible accretion rate at the time $t=30$~Myr. As discussed 
in \S~\ref{sec_caveats}, the IMF of our sink particles is incomplete, essentially flat, at lower masses, which may cause biases in the relation between $t_{95}$ and 
$M_{\rm f}$, so stars with $M_{\rm f}< 2.5$ M$_{\odot}$ are not included in this study.
The power-law fitting of the median values of $t_{95}$ in logarithmic intervals of $M_{\rm f}$ gives the relation
\begin{equation}
t_{95} = 0.67 \, {\rm Myr} \times (M_{\rm f}/1 {\rm M}_{\odot} )^{0.54},
\label{t95}
\end{equation}
consistent with the relations we previously derived with lower-mass stars \citep{Padoan+14acc,Haugbolle+18imf}, discussed above.

Despite the large scatter in the plot, its lower envelope is well defined (short-dashed line in Figure \ref{t95_v1}). It is even better defined in our previous
4-pc runs, as those plots extend over approximately three orders of magnitude in $M_{\rm f}$ (Figure 13 in \citet{Padoan+14acc} and Figure 11
in \citet{Haugbolle+18imf}). The lower envelope corresponds to a linear dependence of $t_{95}$ on $M_{\rm f}$. Because the ratio $M_{\rm f}/t_{95}$ 
gives the average accretion rate over the formation time of a star, the lower envelope shows that the maximum average accretion rate is independent 
of the final stellar mass. The average infall rate, $\dot M_{\rm in}=\epsilon_{\rm out}^{-1}M_{\rm f}/t_{95}$, is twice larger, because we have assumed 
that half of the infalling mass is lost through jets and outflows, $\epsilon_{\rm out}=0.5$. The short-dashed line in Figure \ref{t95_v1} corresponds to a constant 
average infall rate of $4.5\times 10^{-5}$ M$_{\odot}\,$yr$^{-1}$, or a twice lower accretion rate. 

The actual infall rate in the simulation is approximately twice larger than $\dot M_{\rm in}$. Recall that a mass-correction factor, $f_{\rm m}$, was applied, 
so that the final sink mass is $M_{\rm s,f}=M_{\rm f}/f_{\rm m}$, with $f_{\rm m}=0.53$, as explained in \S~\ref{sec_caveats}. Thus, the maximum infall 
rate in the simulation is larger than the rate based on the growth of the stellar mass given above. Based on our interpretation of the IMF incompleteness in 
\S~\ref{sec_caveats}, with higher spatial resolution the simulation would yield a few more lower-mass stars around each massive star, and this extra infall 
rate corresponds to the fraction that would be accreted by such stars. Thus, the maximum infall rate in the simulation is approximately 
$0.9\times 10^{-4}$ M$_{\odot}\,$yr$^{-1}$. Furthermore, at a pc scale, the maximum {\it inflow rate} is typically 10 times larger than the infall rate,
as shown in \S~\ref{sec_ic_inflow}, so the the maximum inflow rate in the simulation is of order $10^{-3}$ M$_{\odot}\,$yr$^{-1}$.
 
The long-dashed line in Figure~\ref{t95_v1} shows the infall rate from the collapse of a critical isothermal sphere, $\dot{M}_{\rm iso}=0.975\,c_{\rm s}^3/G$ \citep{Shu+87},
assuming $T=10$~K. Virtually all our stars have $\dot M_{\rm in}>\dot{M}_{\rm iso}$, because the infall rate is driven by inertial inflows from larger scale that have mass-flow 
rates significantly larger than $c_{\rm s}^3/G$. If these large inflow rates were present during the initial buildup phase of the prestellar cores as well, a prestellar 
core could accumulate a mass in excess of its critical mass (as shown in \S~\ref{core_properties}).    

The largest stellar masses in the simulation are limited by the lifetime of massive stars. The curved, solid line in Figure \ref{t95_v1} shows the stellar 
lifetime, $t_{\rm SN}$, as a function of the stellar mass from \citet{Schaller+92}, which was adopted in the simulation to determine the SN time of the 
sink particles (using the mass $f_{\rm m}M_{\rm s}$, where $M_{\rm s}$ is the sink mass). The plot shows that stars in the approximate mass range 
20-60 M$_{\odot}$ may have their growth time limited by their lifetime, and no star can grow much above 60 M$_{\odot}$, at the maximum accretion rate
values of this run. In star-forming regions with larger mean density than assumed here and/or larger velocity dispersion and sound speed, accretion rates 
may be larger, resulting in shorter growth timescales and larger maximum stellar masses. This is further discussed in \S~\ref{sec_scaling} and \ref{sec_mmax}, 
where we interpret the $t_{95}-M_{\rm f}$ plot based on the velocity scaling of supersonic turbulence.

\section{The Time Evolution of the Infall Rate} \label{sec_history}

\begin{figure}[b]
\includegraphics[width=\columnwidth]{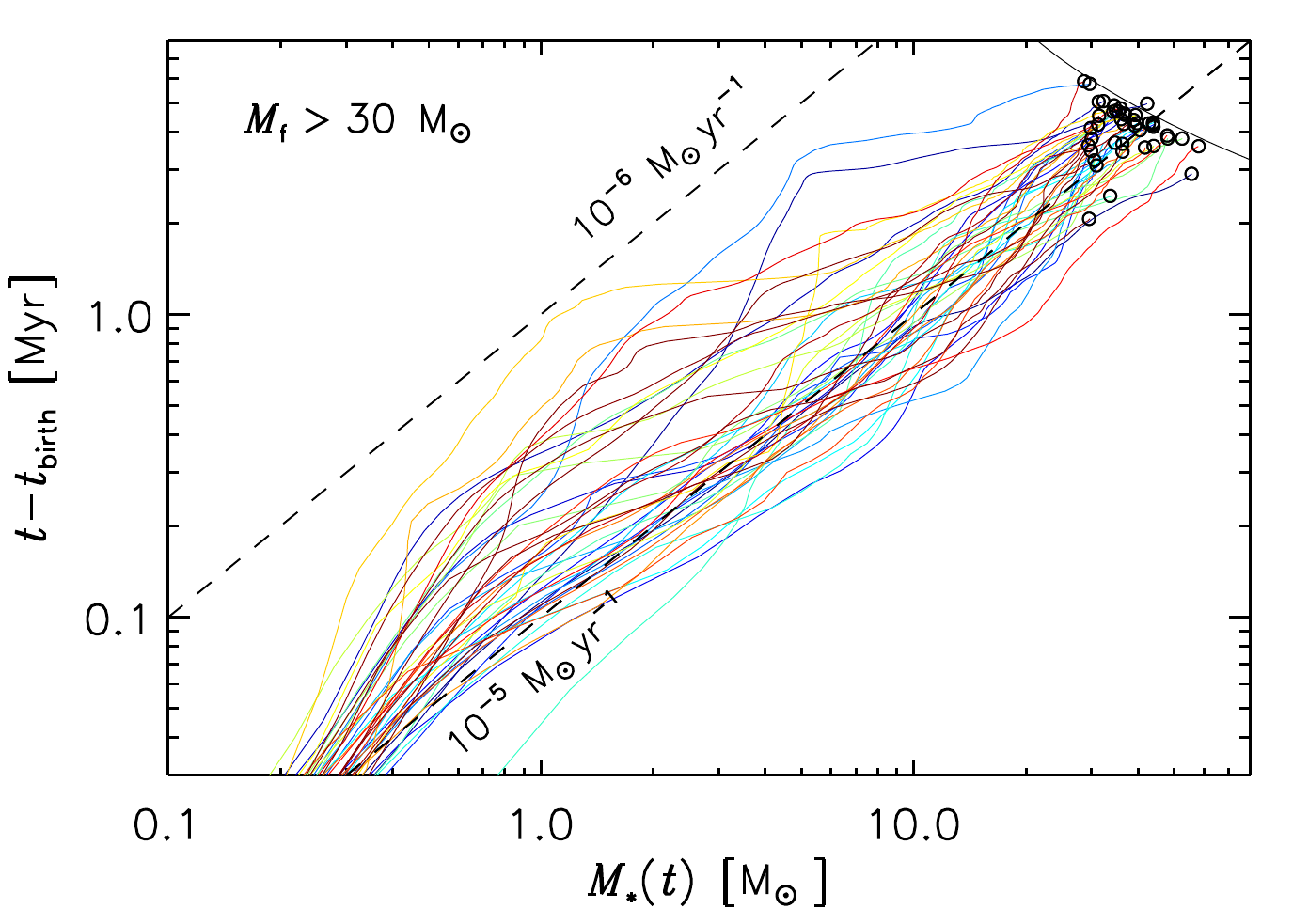}
\caption[]{Evolution of the stellar mass versus time for the 43 stars in the simulation with $M_{\rm f} > 30$ M$_{\odot}$, plotted by setting the 
initial time equal to the birth time, $t_{\rm birth}$, for each star. The empty circles mark the mass at $t=t_{95}$, not the final stellar mass. 
The dashed lines correspond to constant values of the average accretion rate (the actual infall rate in the simulation is approximately four times larger, 
because only a fraction $\epsilon_{\rm out}=0.5$ of the infalling gas is accreted onto the sink particles, and because the stellar mass is taken to be
a fraction $f_{\rm m}=0.53$ of the sink mass). The curved, solid line in the top-right corner of the plot shows $t_{\rm SN}$, as in Figure \ref{t95_v1}.
}
\label{t95_tracks_massive}
\end{figure}

\begin{figure}[b]
\includegraphics[width=\columnwidth]{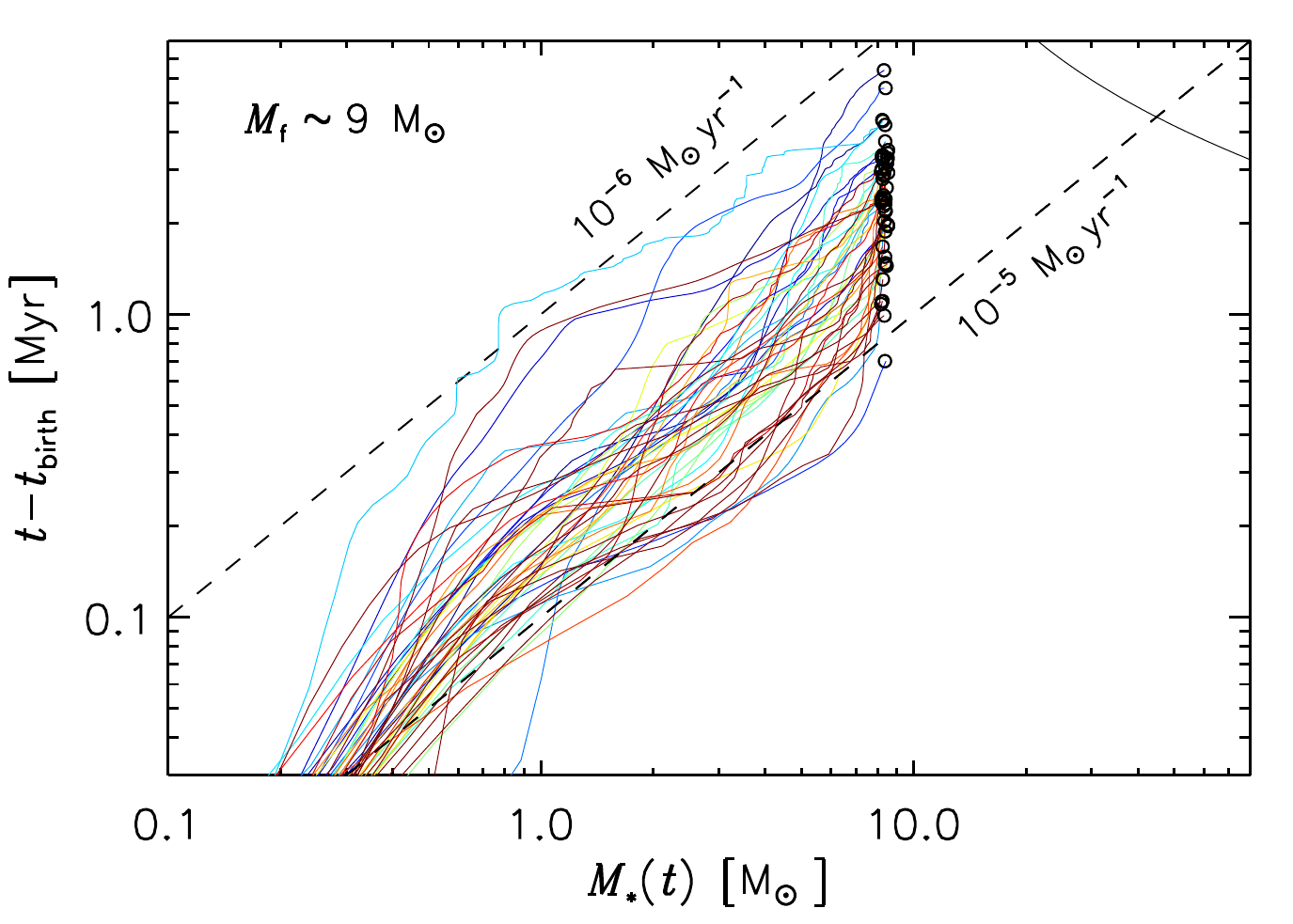}
\caption[]{The same as Figure \ref{t95_tracks_massive}, but for the 49 stars with $8.6 < M_{\rm f} < 9$ M$_{\odot}$. As for the most massive stars, the average 
accretion rates are relatively constant after the first solar mass has been accreted, and are mostly between $10^{-6}$ and $2\times 10^{-5}$ M$_{\odot}$yr$^{-1}$.  
}
\label{t95_tracks}
\end{figure}

To illustrate how the final stellar mass is assembled over time, we plot individual stellar tracks showing the stellar mass versus time, where the time is shifted by the birth time of each star, $t-t_{\rm birth}$. The tracks are shown for a subset of the most massive stars in Figure~\ref{t95_tracks_massive} and for stars with $8.6 < M_{\rm f} < 9$ M$_{\odot}$
in Figure~\ref{t95_tracks}. We only use the mass values recorded at each simulation snapshot, so the mass increments are averaged over intervals of $30$~kyr. 
The plots show that the average infall rate, $M_*(t)/(t-t_{\rm birth})$, along a stellar track is not a systematic function of time or mass, at least for $M_*> 1$~M$_{\odot}$. 
Although many of the tracks show relatively large oscillations (large variations of the infall rate), they are approximately parallel to the dashed lines 
corresponding to constant infall rates. Thus, on average, stars destined to become massive grow with an approximately constant mean infall rate, irrespective 
of their current mass. As discussed in \S~\ref{sec_competitive} this is definitive evidence against the competitive accretion scenario, despite the long star-formation time, 
and is consistent with the idea that massive stars are assembled by inertial inflows.

\begin{figure*}[t]
\centering
\includegraphics[width=\textwidth]{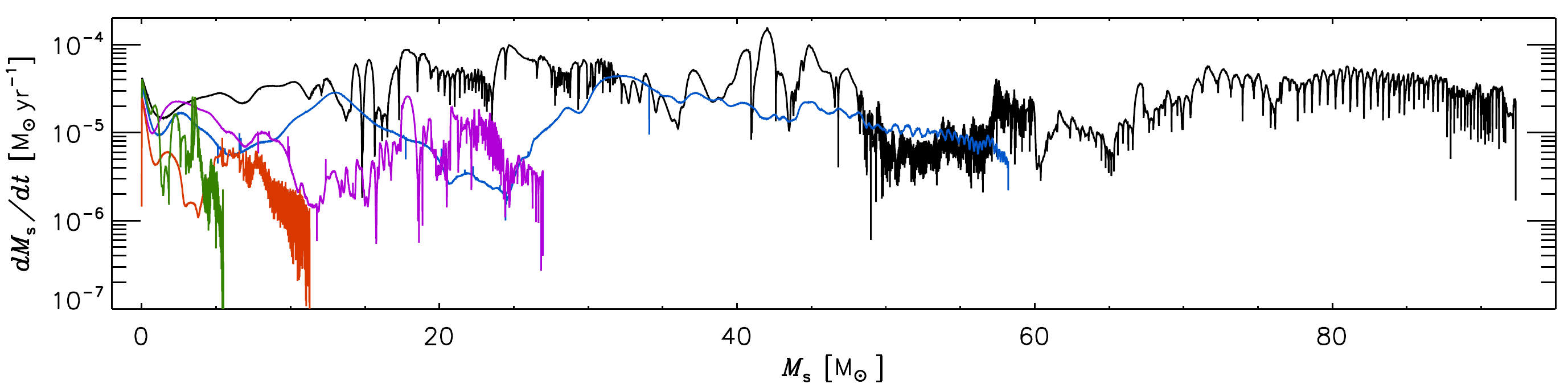}
\centering
\includegraphics[width=\textwidth]{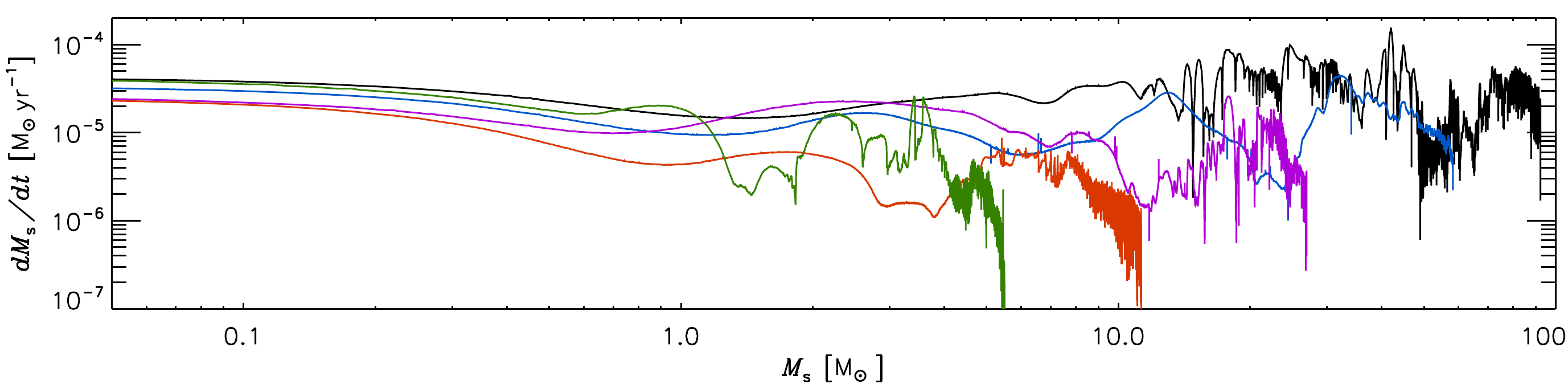}
\caption[]{Evolution of the sink-particle accretion rate versus sink-particle mass for five typical sinks with final masses 
$M_{\rm s,f}=5.5$, 11.3, 27.0, 58.2, and 92.3 M$_{\odot}$. The corresponding infall rate is twice larger, because only a fraction 
$\epsilon_{\rm out}$ of the infall rate is accreted onto the sink particles. The accretion rate is averaged over a timescale of order 
300~yr. The bottom panel is the same as the top panel, except for the logarithmic mass scale to show the initial evolution dominated
by the collapse of the prestellar cores.      
}
\label{mdot_mass}
\end{figure*}
\begin{figure*}[t]
\centering
\includegraphics[width=\textwidth]{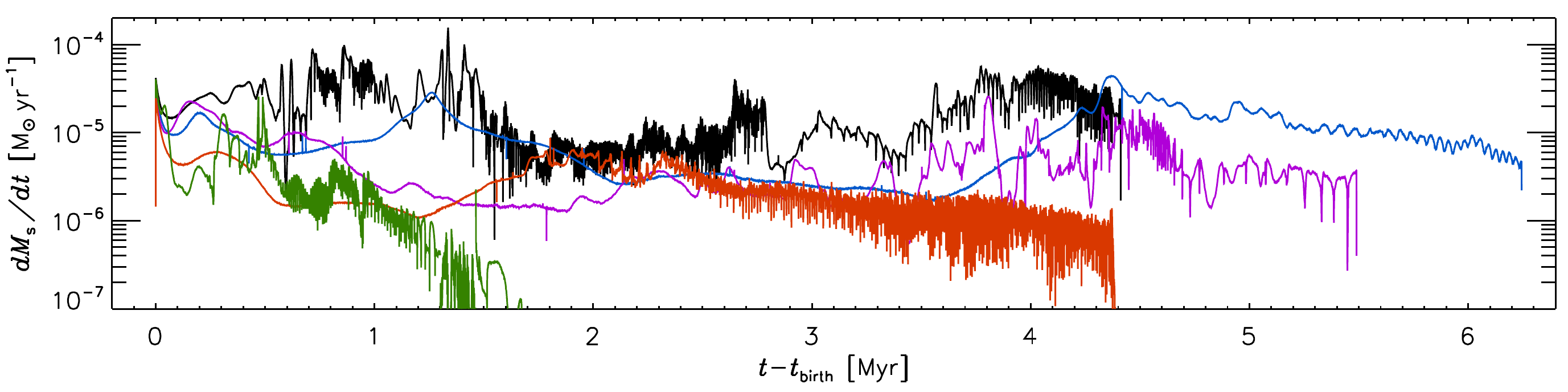}
\centering
\includegraphics[width=\textwidth]{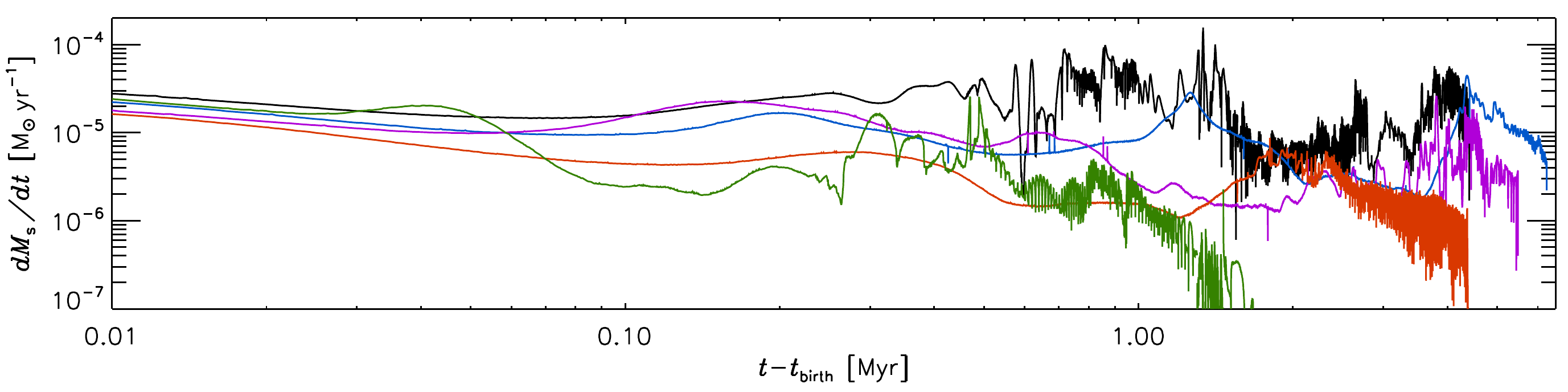}
\caption[]{Same as Figure~\ref{mdot_mass}, but plotted as a function of time instead of sink mass.  
}
\label{mdot_time}
\end{figure*}

As shown by Figure~\ref{t95_tracks}, the average infall rate varies from star to star, with a total scatter of approximately one order of magnitude even for stars with 
nearly the same final mass. Thus, the final stellar mass depends both on the average infall rate and on the duration of the infall process (controlled by the duration of 
the corresponding inertial inflow). The most massive stars are formed in regions where a large infall rate can be maintained for a long time, fed by coherent 
inertial motions over a scale of several pc.  

We have also computed the infall rate of the sink particles with a much higher time frequency, of order $300$~yr, the average time-step size of the simulation at the 
root-grid resolution of $0.49$~pc. Figures~\ref{mdot_mass} and \ref{mdot_time} show the infall rate evolution for five typical sink particles with final sink masses
$M_{\rm s,f}=5.5$, 11.3, 27.0, 58.2, and 92.3 M$_{\odot}$. As a sink grows in mass, its infall rate experiences oscillations, often of one or two orders of magnitude,
but no systematic dependence on mass, as pointed out above. The plots with linear mass and time axis (top panels of Figures~\ref{mdot_mass} and \ref{mdot_time}), 
show that some of the strongest oscillations are approximately periodic. They are associated with the orbital motion of the sink particles in bound multiple systems, 
as already discussed in \citet{Padoan+14acc} and \citet{Jensen+Haugboelle18}. We do not pursue a study of such oscillations here, because the dynamics of binaries 
and multiple systems (or the accretion-disk instabilities that modulate the actual accretion rate from the disk to the star) cannot be properly addressed at the 
spatial resolution of this simulation.

The bottom panels of Figures~\ref{mdot_mass} and \ref{mdot_time} show the same plots as the top panels, but with logarithmic mass and time axes. The initial evolution
is characterized by infall rates of order 2-4$\times 10^{-5}$~M$_{\odot}\,$yr$^{-1}$ for all final masses. This is clearly due to the collapse of the prestellar core that lasts less than 
100~kyr. As commented in the previous section, these early infall rates are well in excess of $c_{\rm s}^3/G$ because the prestellar cores are fed by inflow rates larger than that.
After the collapse, the infall is controlled by the larger-scale inertial inflow, as shown by the stochastic nature of its evolution. 
Interestingly, the bottom panel of Figure~\ref{mdot_mass} shows that the collapsing prestellar core has a mass of order 1~M$_{\odot}$ (this is true for most 
sink particles, not only for the five shown here), irrespective of the final mass of the sink. This result is consistent with the characteristic prestellar-core virial mass 
derived below in \S~\ref{core_properties}.

\section{The Initial Conditions for Massive Star Formation: Prestellar Core and Infall Region} \label{sec_ic}

A major goal of this work is to characterize the initial conditions that lead to the formation of a massive star. Most computational studies or analytical models
of massive star formation are based on ad hoc initial conditions, typically an isolated and very dense core that may collapse into a single object or a stellar cluster.
However, it is unlikely that an isolated core is a realistic representation of the initial conditions, because star-forming cores are typically found at the intersection of dense 
filaments, both in the simulations and in real MCs. This filamentary morphology reflects the dynamical coupling between small and large scales in the ISM turbulence,
and the ongoing mass accretion driven by the compressive part of the turbulent flow. Furthermore, because of the stochastic nature of the turbulence in star-forming regions, 
it is possible that a variety of conditions result into massive stars. 

Thanks to the large volume of our SN-driven simulation and the large number of massive stars it generates, we are able to explore a vast parameter space of initial and 
boundary conditions, besides ensuring that such conditions are consistent with the larger-scale environment and that their statistical distributions are realistic. Furthermore, thanks to
the large number of tracer particles embedded in the simulation, we can accurately trace the full path of gas elements that contribute to the final mass of a sink particle
representing a massive star. A detailed study of the Lagrangian time evolution of such gas elements will be attempted in future works. Here, we focus on the initial conditions
for massive star formation at a single specific time, defined numerically as the time when the sink particle representing the star is created. In practice, we study the initial 
conditions at the first available snapshot after the birth time of the sink particle, a delay between 0 and 30~kyr (the time separation between snapshots), 15~kyr on average.

The numerical implementation of sink particles (see \S~\ref{sec_method}) guarantees that a sink particle is created only when a dense core has emerged and has just 
started to collapse. Thus, despite being defined numerically by a threshold density, the time of creation can be identified as the approximate time of the beginning of the 
gravitational collapse of the prestellar core. Because the core-collapse time ($\sim 10^5$~yr) is much shorter than the star-formation time ($\sim 10^6$~yr), and 
because the sink particle is usually created at the very start of the collapse, the uncertainty in the definition of this birth time (including the time interval between snapshots) 
is small enough for the purpose of defining an initial time of star formation. 

The beginning of the gravitational collapse of the core and the creation of the sink particle representing the protostar mark the transition of the core from {\it prestellar} to
{\it protostellar}. Thus, the core mass we derive is the largest mass the prestellar core achieves prior to the formation of the protostar. The earlier build up and evolution 
of the prestellar core is also of interest to understand the origin of massive stars and for statistical comparisons with observational surveys of prestellar 
cores and will be addressed in a future study.

\subsection{Prestellar Core Definition} \label{sec_ic_def}

We identify prestellar cores as density enhancements centered around the positions of the sink particles in the first simulation snapshot after the formation 
of the sink particle. As in \S~\ref{sec_time}, we consider only the 1,503 sinks with final mass $M_{\rm s,f}\ge 5$~M$_{\odot}$ and negligible accretion rate
at the time $t=30$~Myr. Because we have saved approximately 1,000 snapshots at fixed intervals of 30~kyr after self-gravity and star formation were included 
(see \S~\ref{sec_method}), we have typically one or two prestellar cores per snapshot (though the number tends to increase with time initially, when the 
SFR is still increasing). With the core center marked by the birth position of the sink particle, we then need a criterion to determine the size of cores, which is non-trivial 
because cores are usually found within dense filaments where the core edges are not clearly defined. Given the large number of cores in our sample, the criterion 
should be relatively straightforward to compute based on average core properties, to avoid a detailed inspection of every single core. To that purpose, 
we compute radial profiles of the gas density, rms velocity, virial parameter and other quantities (see \S~\ref{sec_ic_inflow}), centered at the birth positions of the sink-particles. 
In the definition of the virial parameter, $\alpha_{\rm vir}$, we include both the turbulent kinetic energy and the thermal energy, $\alpha_{\rm vir}=2 (E_{\rm k} + E_{\rm th}) / E_{\rm g}$,
and compute $E_{\rm g}$ as the gravitational energy of a sphere, $E_{\rm g}=3/5 \, a\, G M^2/R$, where the coefficient $a$ is chosen based on the estimated average slope 
of the core density profile, according to \citet[][Appendix A]{Bertoldi+McKee92}.

\begin{figure}[t]
\includegraphics[width=\columnwidth]{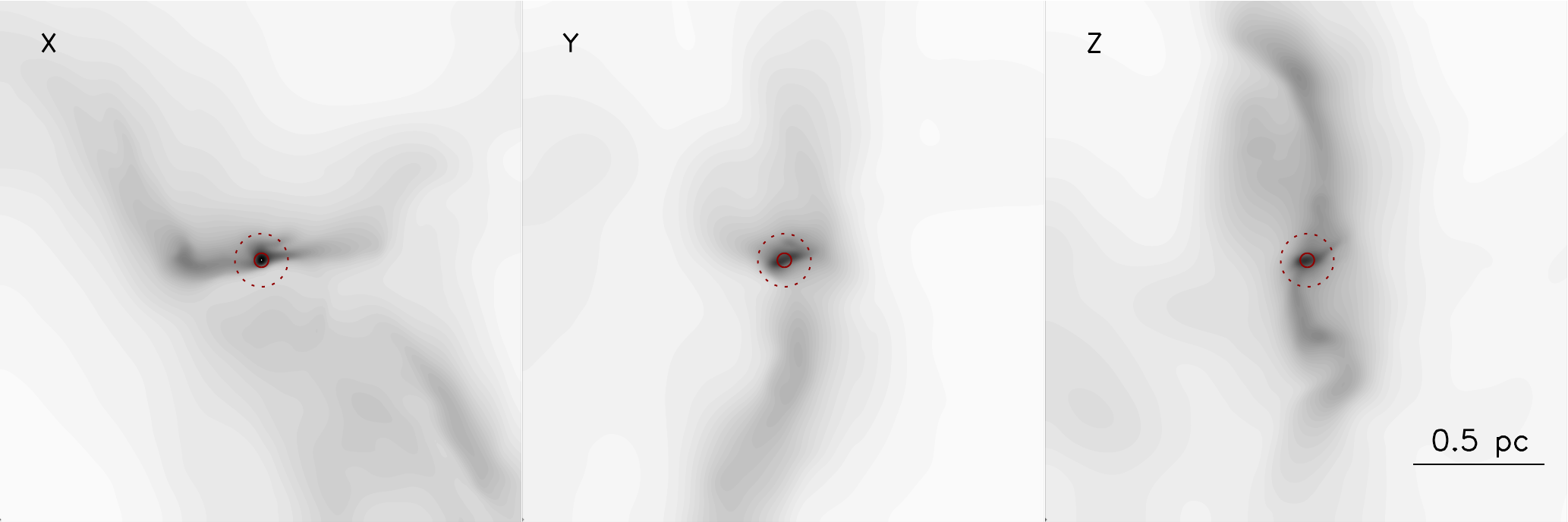}
\includegraphics[width=\columnwidth]{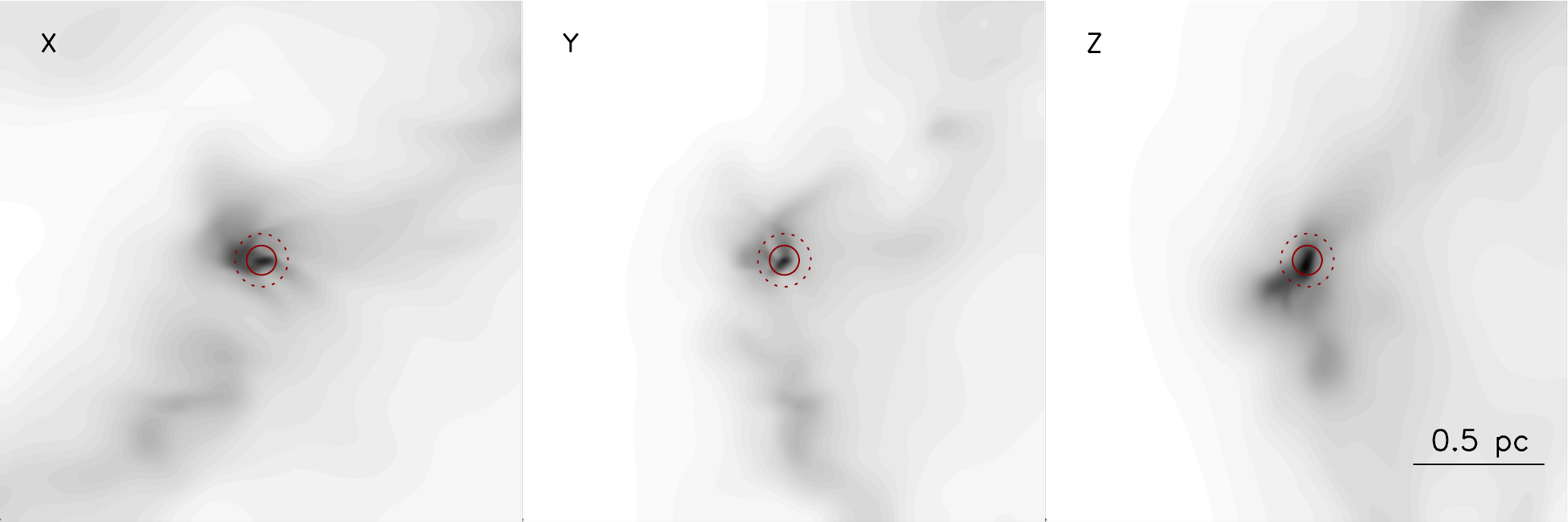}
\includegraphics[width=\columnwidth]{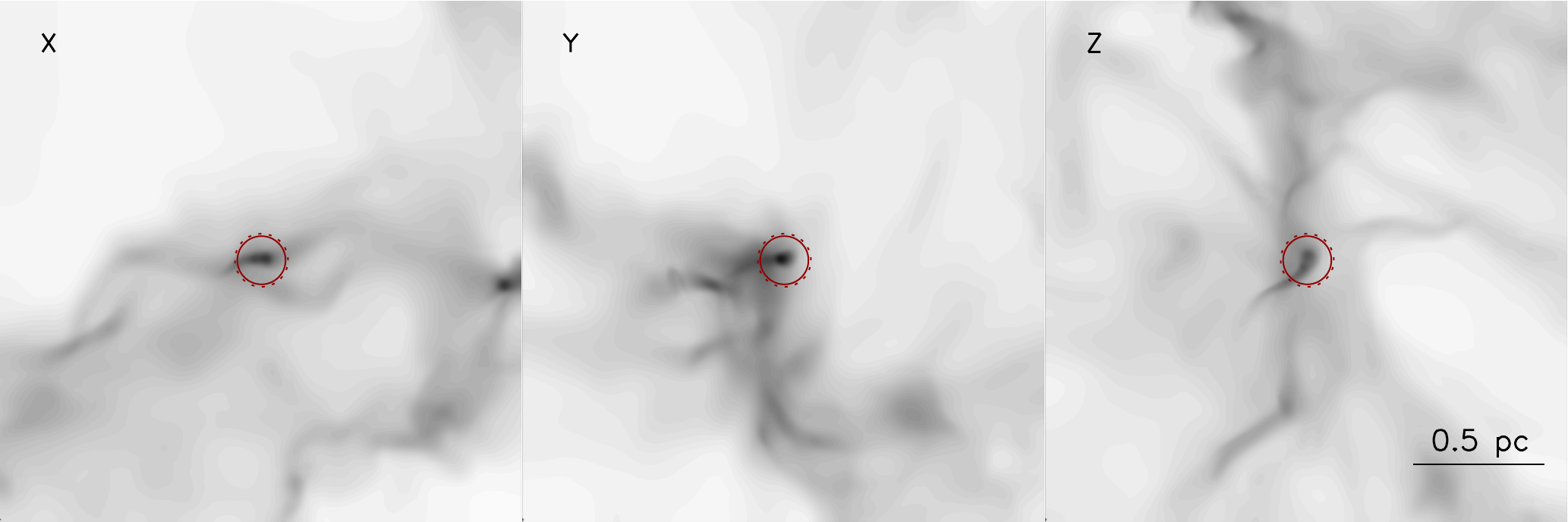}
\caption[]{Square root of column density within (2 pc)$^3$ volumes centered around recently-created sink particles that will become massive stars.
Each row shows a relatively isolated sink particle, with the three columns corresponding to the three orthogonal lines of sight. The dotted-line 
circle has a radius of 0.1 pc; the solid-line circle a radius equal to $R_{\rm c,vir}$.
}
\label{sink_simple}
\end{figure}
\begin{figure}[t]
\includegraphics[width=\columnwidth]{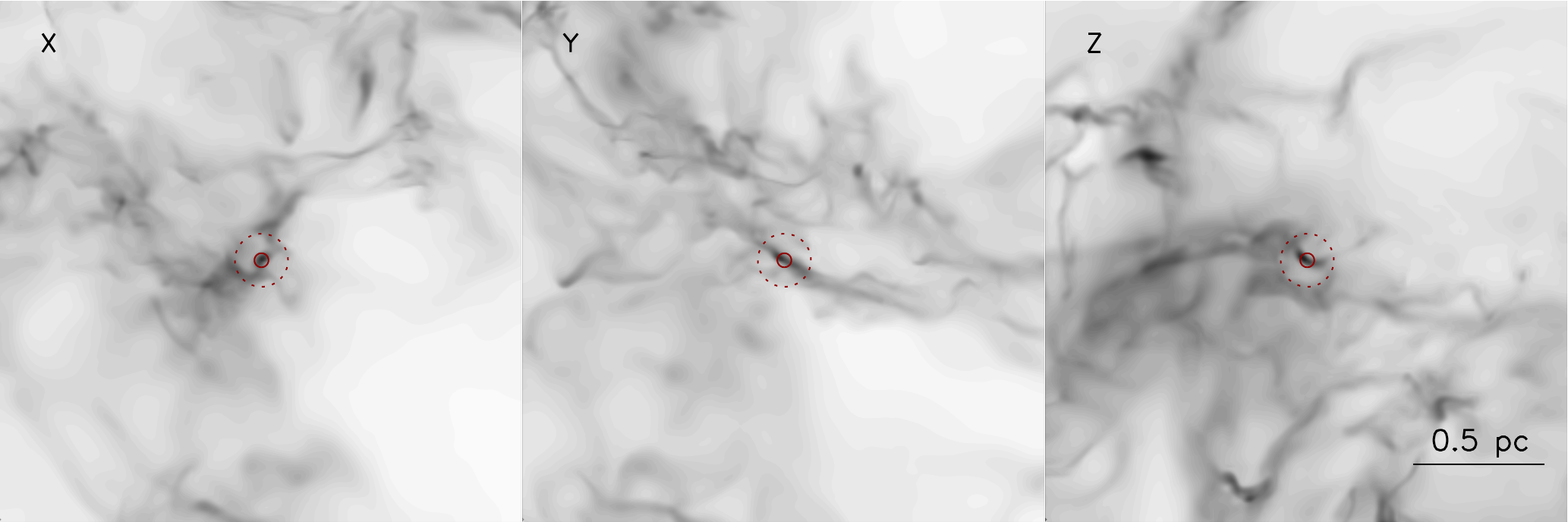}
\includegraphics[width=\columnwidth]{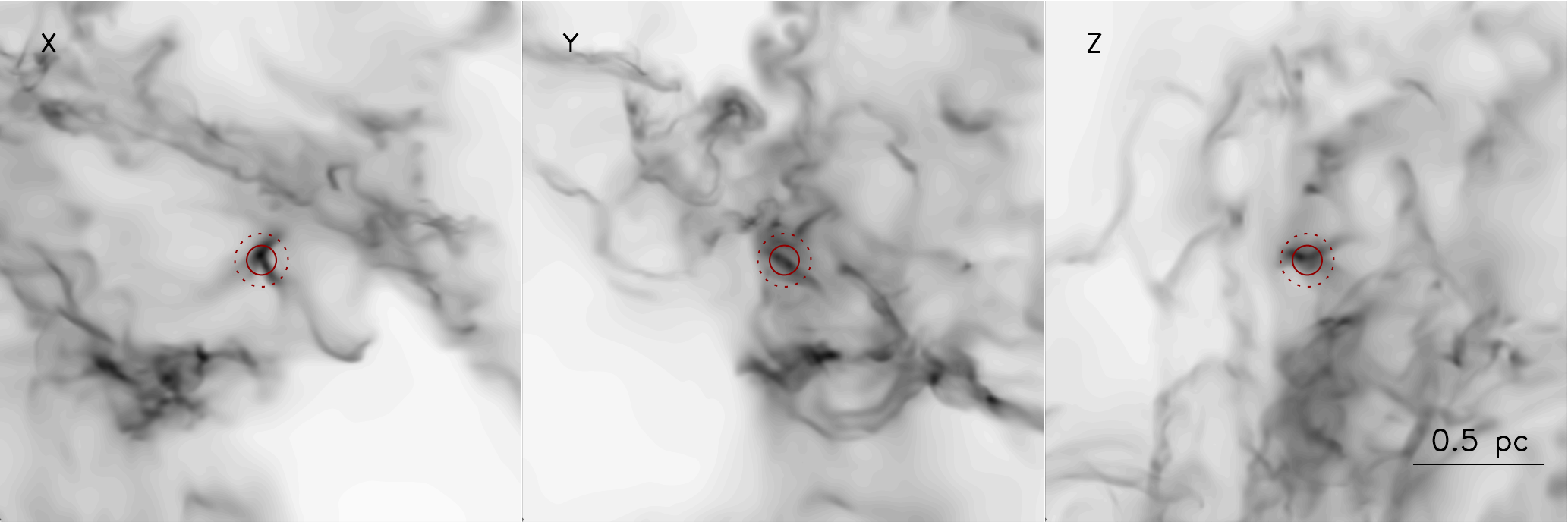}
\includegraphics[width=\columnwidth]{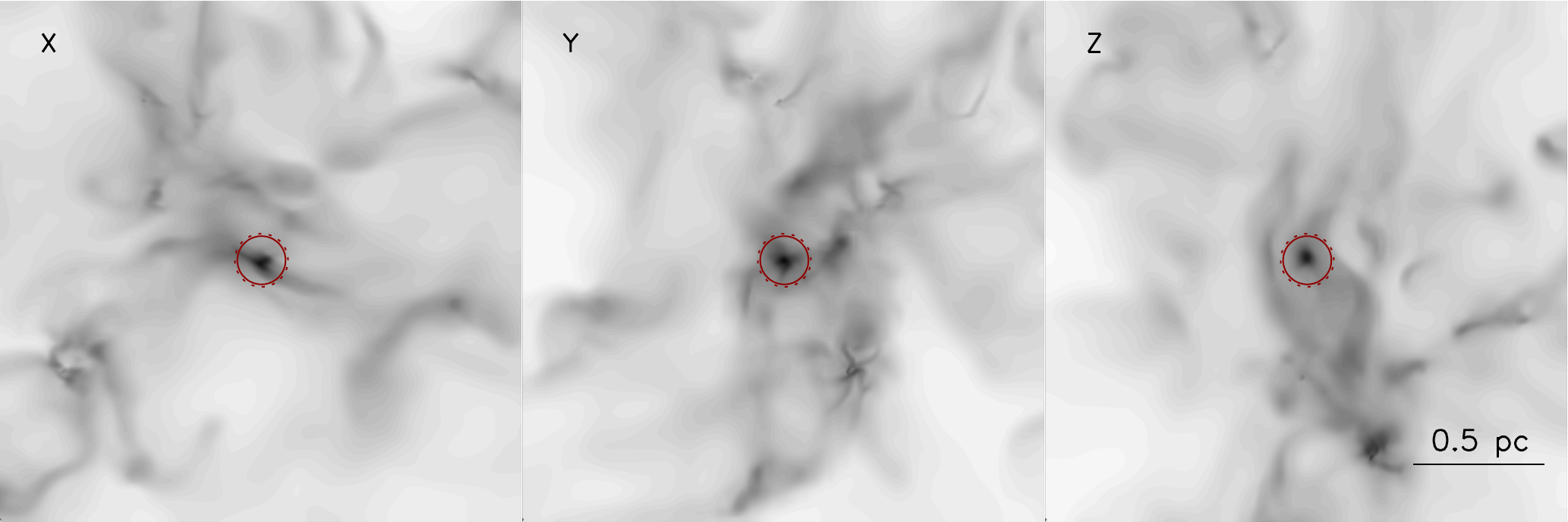}
\caption[]{The same as Figure \ref{sink_simple}, but for sinks created in morphologically more complex regions, usually sites of future stellar clusters.  
}
\label{sink_complex}
\end{figure}

Observations show that prestellar cores are quiescent \citep{Barranco+Goodman98,Goodman+98,Andre+07}, meaning that their internal rms velocity is 
subsonic, while supersonic line widths are found near their edges \citep{Goodman+98,Pineda+10}. This is consistent with the picture of turbulent fragmentation, 
where cores are formed by shocks in the turbulent flow \citep{Myers83,Padoan+2001cores,Chen+19}, and the kinetic energy in the postshock gas, at the 
intersection of filaments, is mostly dissipated. A detailed inspection of our cores shows that this picture is qualitatively confirmed, so we could in principle use 
the drop in turbulent velocity dispersion with decreasing radius to define the core size. However, the transition is often not very sharp, partly due to the shell 
averaging. Because magnetic pressure is expected to be dominant in the cores \citep{Padoan+Nordlund99MHD}, 
we could possibly use the radial dependence of the ratio of turbulent to magnetic pressure as well. However, the shell-averaged radial profiles of that ratio often fails to show a 
sharp transition around unity at the core boundaries, partly because the magnetic field is amplified not only by the compression that generates a core, but also by the turbulence 
outside the core, so the ratio of turbulent to magnetic pressure is often nearly constant with radius. Furthermore, the transition to velocity coherence cannot be the only criterion 
to define the core boundaries, because we must also verify that the core is gravitationally bound. On the other hand, we find that $\alpha_{\rm vir}$
increases monotonically with increasing radius in nearly all cores, and $\alpha_{\rm vir} = 1$ at a radius that corresponds approximately to the core boundaries in most of the 
cores where the pressure ratio shows a clear transition, or where we identify the core boundaries by a detailed inspection. Thus, we adopt the radius where $\alpha_{\rm vir}=1$ 
as a practical definition of the core radius, which also guarantees that the core is gravitationally bound. Because of rare cases where $\alpha_{\rm vir}$ is not a monotonic function 
of the radius, we actually choose the largest radius where the virial parameter is unity. We call this radius the core virial radius, $R_{\rm c,vir}$, and refer to the core mass within this radius as the core virial mass, $M_{\rm c,vir}$. 

The maximum size of prestellar cores extracted from sub-millimeter surveys is $\sim 0.1$~pc 
\citep{Motte+98,Motte+01,Johnstone+06,Konyves+15,Tige+17,Rayner+17,Bresnahan+18,Russeil+19}. In some studies, a maximum core size is imposed as one of the 
selection criteria to avoid including clumps in the core sample, for example a maximum size of 0.3~pc is adopted in \citet{Tige+17} and \cite{Russeil+19}. 
The turbulent-core model by \citet{McKee+Tan03} predicts a similar core size of $\sim 0.1$~pc for progenitors of massive stars and characteristic column densities 
of star-forming regions (see their equation (20)). Thus, we consider the core mass within a radius of 0.1~pc as well, and refer to it as $M_{\rm c,0.1}$. As shown below, 
we usually find $R_{\rm c,vir}<0.1$~pc, which is not surprising because the characteristic thickness of dense filaments where prestellar cores are found is also 
$\lesssim 0.1$~pc, in agreement with the observations \citep{Arzoumanian+11,Andre+16,Roy+19,Arzoumanian+19}.

\begin{figure}[t]
\includegraphics[width=\columnwidth]{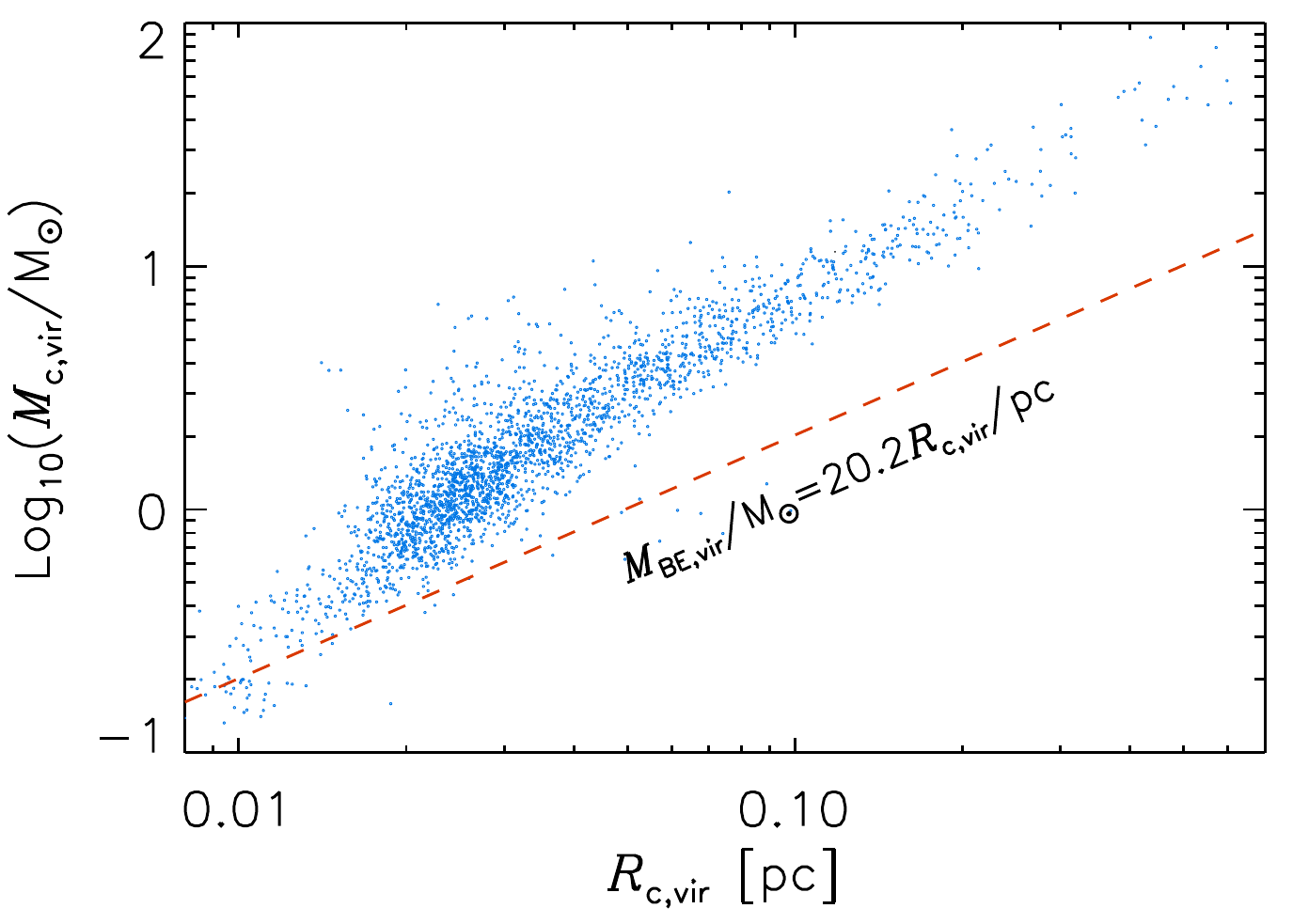}
\caption[]{Core mass versus core radius for cores selected near the moment of sink particle creation and defined by the largest radius where  
$\alpha_{\rm vir} \le 1$. The critical Bonnor-Ebert mass corresponding to the core virial radius and a temperature of 10 K is shown by the dashed line. 
}
\label{mcores_rcores}
\end{figure}
\begin{figure}[t]
\includegraphics[width=\columnwidth]{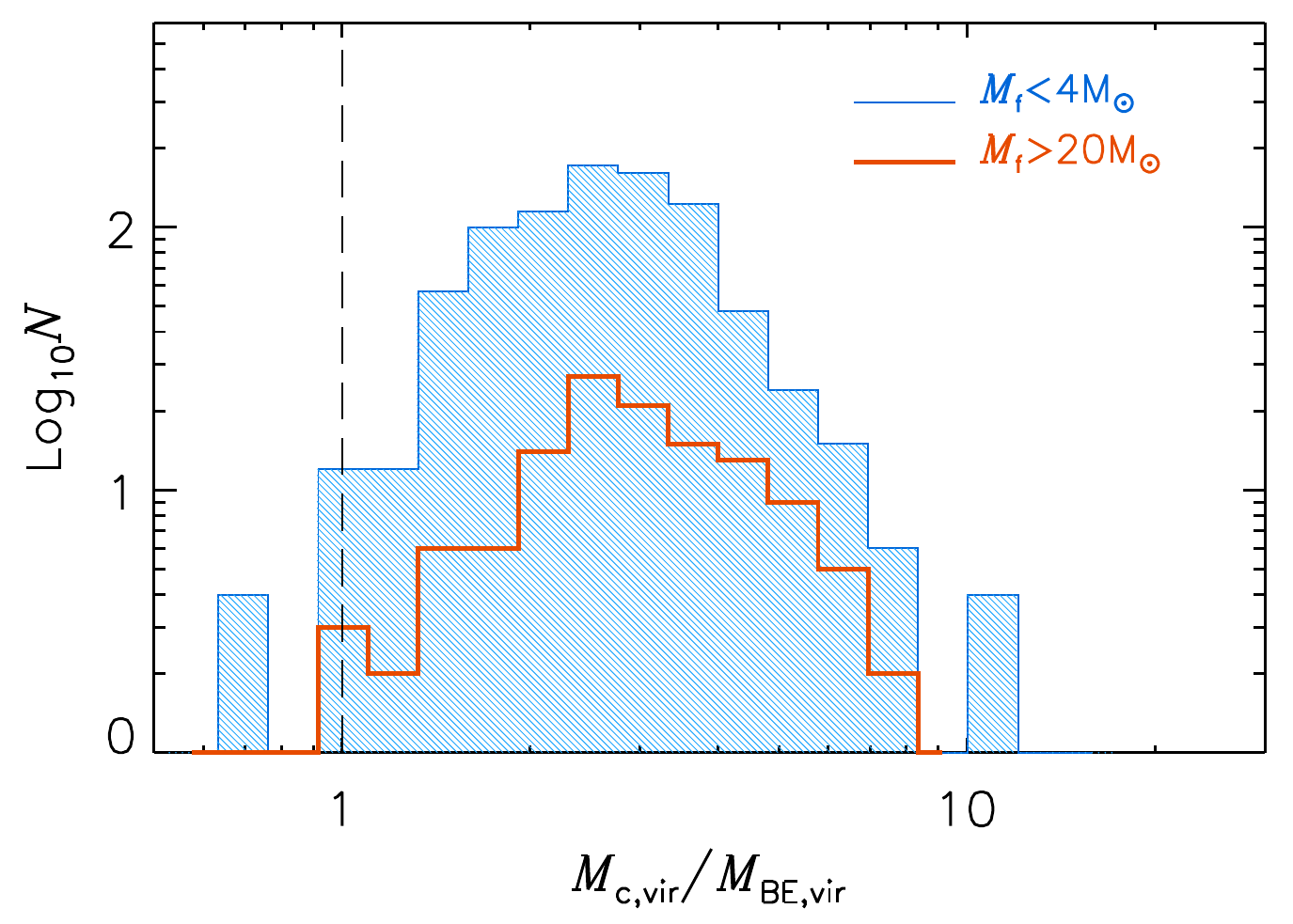}
\caption[]{Probability distribution of the ratio between virial core mass and critical Bonnor-Ebert mass corresponding to the core virial radius and
a temperature of 10 K. The unshaded thick-line histogram is for cores that will yield a final stellar mass $M_{\rm f} > 20$ M$_{\odot}$, while the 
shaded histogram for those resulting in stars with final mass $M_{\rm f} < 4$ M$_{\odot}$.   
}
\label{BE}
\end{figure}
\begin{figure}[t]
\includegraphics[width=\columnwidth]{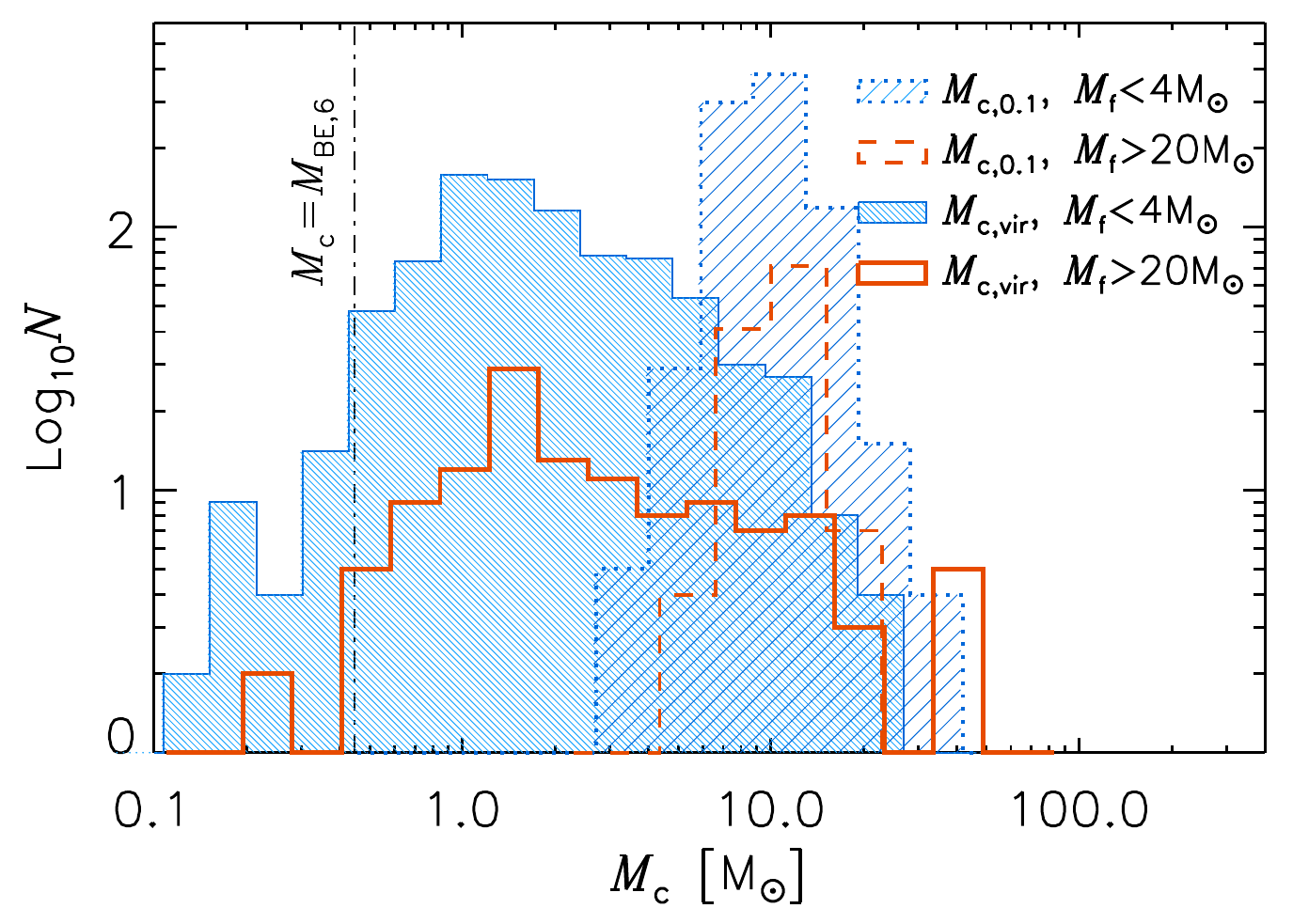}
\caption[]{Mass distribution of prestellar cores for cores that will yield a final stellar mass $M_{\rm f} > 20$ M$_{\odot}$ (unshaded, solid-line and
dashed-line histograms), and for cores resulting in stars with final mass $M_{\rm f} < 4$ M$_{\odot}$ (shaded solid-line and dotted-line histograms).
The solid-line histograms are for core masses within the virial radius, $M_{\rm c,vir}$, while the dashed and dotted-line histograms for masses within
a fixed 0.1 pc radius, $M_{\rm c,0.1}$. The vertical dotted-dashed line is the critical Bonnor-Ebert mass corresponding to the threshold density for 
sink formation of $10^{6}$~cm$^{-3}$. 
}
\label{mcores_hist}
\end{figure}
\begin{figure*}[t]
\includegraphics[width=\textwidth]{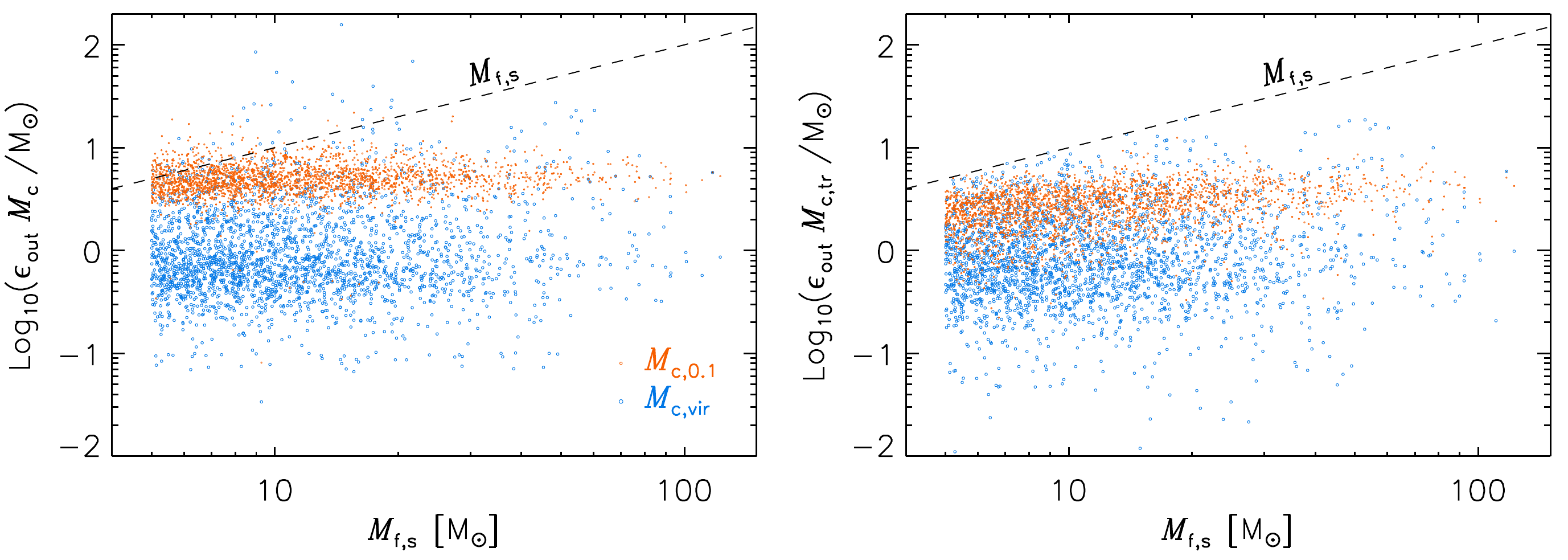}
\caption[]{{\it Left Panel:} Core mass multiplied by the core star-formation efficiency, $\epsilon_{\rm out}=0.5$, versus final sink mass, $M_{\rm f,s}$. 
Notice that the final stellar mass, $M_{\rm f}$, shown in other figures is $M_{\rm f}= f_{\rm m} M_{\rm f,s}$, with $f_{\rm m}=0.53$ to account for the
incompleteness of the IMF (see \S~\ref{sec_caveats}). Here we show $M_{\rm f,s}$ because we are comparing with the available mass reservoir in 
the prestellar cores. {\it Right Panel:} Same as left panel, but for $M_{\rm c,tr}$ instead of $M_{\rm c}$, where $M_{\rm c,tr}$ is the total mass 
of the tracer particles in the core that are eventually accreted onto the sink particle.  
}
\label{mcores}
\end{figure*}

Finally, we also measure the mass of the gravitationally--bound spherical region around the birth position of the sink particle, defined by the largest radius, $R_{\rm b}$,
where $\alpha_{\rm vir}=2$. We refer to $R_{\rm b}$ as the {\it infall radius}, because it marks the transition from the inertial inflow region to the infall region where self-gravity 
becomes dominant. In the core-collapse model of \citet{McKee+Tan02,McKee+Tan03} and in the IMF models of \citet{Hennebelle+Chabrier08imf,Hennebelle+Chabrier09imf} and \citet{Hopkins12imf}, the progenitors of massive stars are cores, or over-dense regions, where self-gravity overcomes the total pressure (primarily turbulent pressure in the
case of massive stars), so the virial parameter must be $\lesssim 2$. For example, \citet{McKee+Tan03} estimate a value $\alpha_{\rm vir}=1.34$ (see their Appendix A.1).  
Thus, our infall radius, $R_{\rm b}$, can be taken as an upper limit to the size of prestellar cores in those models. 

The inflow region outside $R_{\rm b}$ is studied in the following sections, were we will define the {\it inflow radius}, $R_{95}$, as the radius of the sphere that, at the
birth time of the sink, contains 95\% of the total mass (tracer particles) that will accrete onto the star (see \S~\ref{sec_ic_inflow} and \S~\ref{sec_scaling}).

Figures~\ref{sink_simple} and \ref{sink_complex} show examples of prestellar cores through images of the projected density of (2~pc)$^3$ volumes centered 
on the sink-particle positions. The core is usually a well-defined density enhancement even in projection, typically at the intersection of dense filaments. The 
virial radius, shown by the solid circle, is almost always smaller than 0.1~pc (dotted circle), although a few cores with more quiescent envelopes and $R_{\rm c,vir}\sim 0.1$~pc 
are also found, as shown by the bottom rows of panels in Figures~\ref{sink_simple} and \ref{sink_complex}.

\subsection{Distributions of Prestellar-Core Masses}  \label{core_properties}

Figure~\ref{mcores_rcores} shows the relation between the virial mass and the virial radius of all the prestellar cores. Although it is a small contribution, the mass 
of the newly-created sink particle has been added to the core mass, as our purpose is to define the final core mass {\it before} the protostar is created. 
The dashed line marks the mass-radius relation for the critical Bonnor-Ebert mass. As we have selected cores at the very beginning of their gravitational collapse, 
the great majority of them are above the critical Bonnor-Ebert line. Only a very small number of cores are found below the critical line, mostly because of an incorrect 
determination of the core radius\footnote{In a few cases, the radial profile of $\alpha_{\rm vir}$ oscillates around unity over a range of radii, so picking the largest 
radius where $\alpha_{\rm vir} = 1$ may overestimate the true core size.}. The purpose of this plot is primarily to verify that our sink-particle model does not result
in false positives, meaning sink particles created in a transient density enhancement above the given threshold. However, the fact that these sink particles
are known to achieve stellar masses already indicates that they are not numerical artifacts, because a mass reservoir of gravitationally bound gas must have been available 
for their growth. 

Despite the uncertainty in defining the size of prestellar cores, we should expect their masses to be in excess of the critical one, based on the infall rates estimated 
in \S~\ref{sec_time} and \ref{sec_history}. As we commented there, the inertial inflows assemble the prestellar cores at a mass-flow rate in excess of $c_{\rm s}^3/G$, 
so, by the time the cores are collapsing, their mass exceeds the critical one. 

Similar plots are often used to interpret the results of observational surveys of prestellar cores, in the absence of line observations 
\citep{Johnstone+06,Konyves+15,Bresnahan+18}: cores above the critical Bonnor-Ebert line are selected as prestellar, while cores below the 
dashed line are discarded. While some of the excluded cores may be of a transient nature (never massive enough to become unstable), some may be 
true prestellar cores caught in their growth process. Because the formation process of a core is most likely longer than its initial collapse, we should generally 
expect to find a large number of true prestellar cores below the critical Bonnor-Ebert line, depending on the sensitivity and angular resolution of a survey. 
We will address this issue in a future study by following the formation of the prestellar cores in our simulation.

\begin{figure}[b]
\includegraphics[width=\columnwidth]{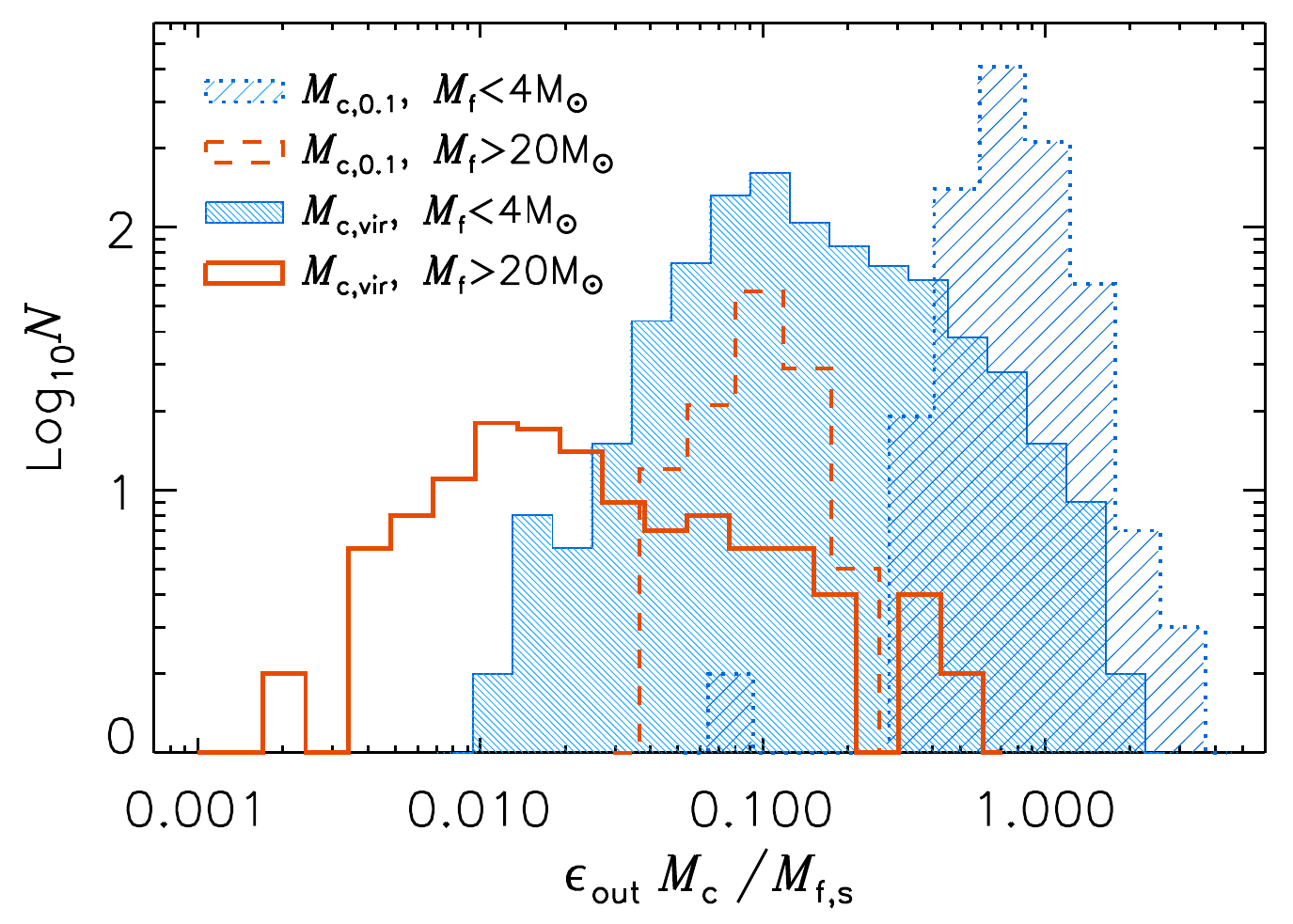}
\caption[]{Probability distributions of the ratio of core mass (multiplied by the star formation efficiency) and final sink particle mass.
The histogram plotting symbols are the same as in Figure \ref{mcores_hist}. 
}
\label{mvir_mstar}
\end{figure}
\begin{figure*}[t]
\includegraphics[width=\textwidth]{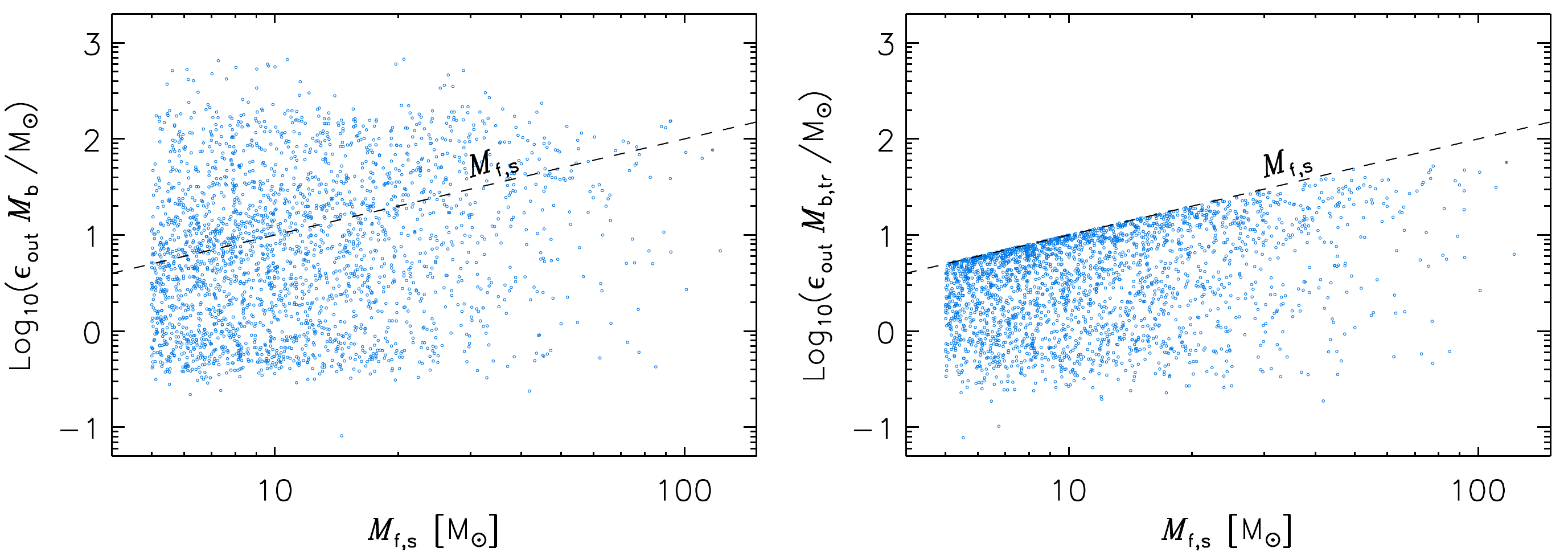}
\caption[]{Same as Figure~\ref{mcores}, but for the mass within the infall radius, $M_{\rm b}$, instead of the core mass.
}
\label{mcores_b}
\end{figure*}

The probability distribution of the ratio between the virial core mass and the critical Bonnor-Ebert mass, $M_{\rm BE,vir}$, corresponding to the core virial 
radius and a temperature of 10 K is shown in Figure~\ref{BE}, while Figure~\ref{mcores_hist} shows the core mass distribution. In both figures, 
we have separated the cores resulting in massive stars with $M_{\rm f} > 20$ M$_{\odot}$ from those yielding lower-mass stars with 
$M_{\rm f} < 4$ M$_{\odot}$. All cores, independent of the final stellar mass, have a ratio $M_{\rm c,vir}/M_{\rm BE,vir}$ in the approximate range of 1 to 10, 
with the peak of the distributions 
at a value of $\sim 2.5$. Notice that the Jeans mass is $M_{\rm J}=2.47 \,M_{\rm BE}$ \citep{McKee+Ostriker07}, so the core masses are on average 
of the order of the Jeans mass, as often found in the interferometric studies of massive clumps mentioned in \S~\ref{sec_obs}.
The mass distributions peak at $\sim 1$~M$_{\odot}$ for $M_{\rm f} < 4$ M$_{\odot}$ and $\sim 2$~M$_{\odot}$ for 
$M_{\rm f} > 20$ M$_{\odot}$, with the largest mass $\sim 40$~M$_{\odot}$ . This is a remarkable result, showing that \emph{massive stars are 
not the result of massive prestellar cores}. Even within a radius of 0.1~pc, the integrated mass is typically $\sim 10$~M$_{\odot}$, 
irrespective of the final stellar mass, as shown by the dashed-line and dotted-line histograms in Figure~\ref{mcores_hist}. Thus, within this
characteristic size of 0.1~pc, the precursors of massive stars are not particularly conspicuous. The peak of the core mass distribution is two orders 
of magnitude (one order of magnitude for $M_{\rm c,0.1}$) less massive than would be required to form a very massive star solely from the mass 
reservoir of the core.  
          
This result is shown again in the left panel of Figure~\ref{mcores}, through a comparison of the core mass with the final sink mass. Here, the core mass is multiplied 
by the value of the core star-formation efficiency, $\epsilon_{\rm out}=0.5$, adopted in the sink-particle model, and the final sink mass, $M_{\rm f,s}$, 
is used instead of the final stellar mass, $M_{\rm f}= f_{\rm m} M_{\rm f,s}$, because we are comparing the available prestellar-core mass reservoir 
with the total mass reservoir necessary to form the sink. The figure illustrates that the mass reservoir of most prestellar cores is much smaller than 
necessary to account for the final mass of massive stars. This is further quantified by the mass distribution of the ratio of core mass and final
sink-particle mass shown in Figure~\ref{mvir_mstar}. For massive stars with $M_{\rm f} > 20$ M$_{\odot}$ (unshaded, solid-line histogram in 
Figure~\ref{mvir_mstar}), the probability distribution peaks at a value of $\sim 0.01$, meaning that the most likely case is that only approximately 
1\% of the final mass of a massive star is contained in the prestellar core defined by the virial radius.    

The actual fraction of the final stellar mass contained in the prestellar core is even lower than estimated above, because only a fraction of the core mass 
is eventually accreted onto the star. This can be computed as the total mass of the tracer particles inside the core that are eventually accreted onto the star,
$M_{\rm c,tr}$, which is shown in the right panel of Figure~\ref{mcores}. Of course $\epsilon_{\rm out} M_{\rm c,tr} \le M_{\rm f,s}$, as shown in the figure. 

One may suspect that the stellar mass is at least contained within the gravitationally-bound region around the prestellar core, as for example required
by the IMF models of \citet{Hennebelle+Chabrier08imf,Hennebelle+Chabrier09imf} and \citet{Hopkins12imf}, but that is not the case. Figure~\ref{mcores_b} 
shows the mass within the bound region (left panel), $M_{\rm b}$, and the corresponding part of that mass that is eventually accreted onto the sink (right panel),
$M_{\rm b,tr}$, versus the final sink mass. Although $M_{\rm b}>M_{\rm c,vir}$, the right panel shows that even the bound region contains, on average, 
only a small fraction of the final sink mass, particularly in the case of the most massive stars. Evidently, a major fraction of the final stellar mass still resides
outside of the infall region when the prestellar core starts to collapse, showing the importance of inertial compressive motions from the more extended inflow region.

\begin{figure*}[t]
\centering
\includegraphics[width=\textwidth]{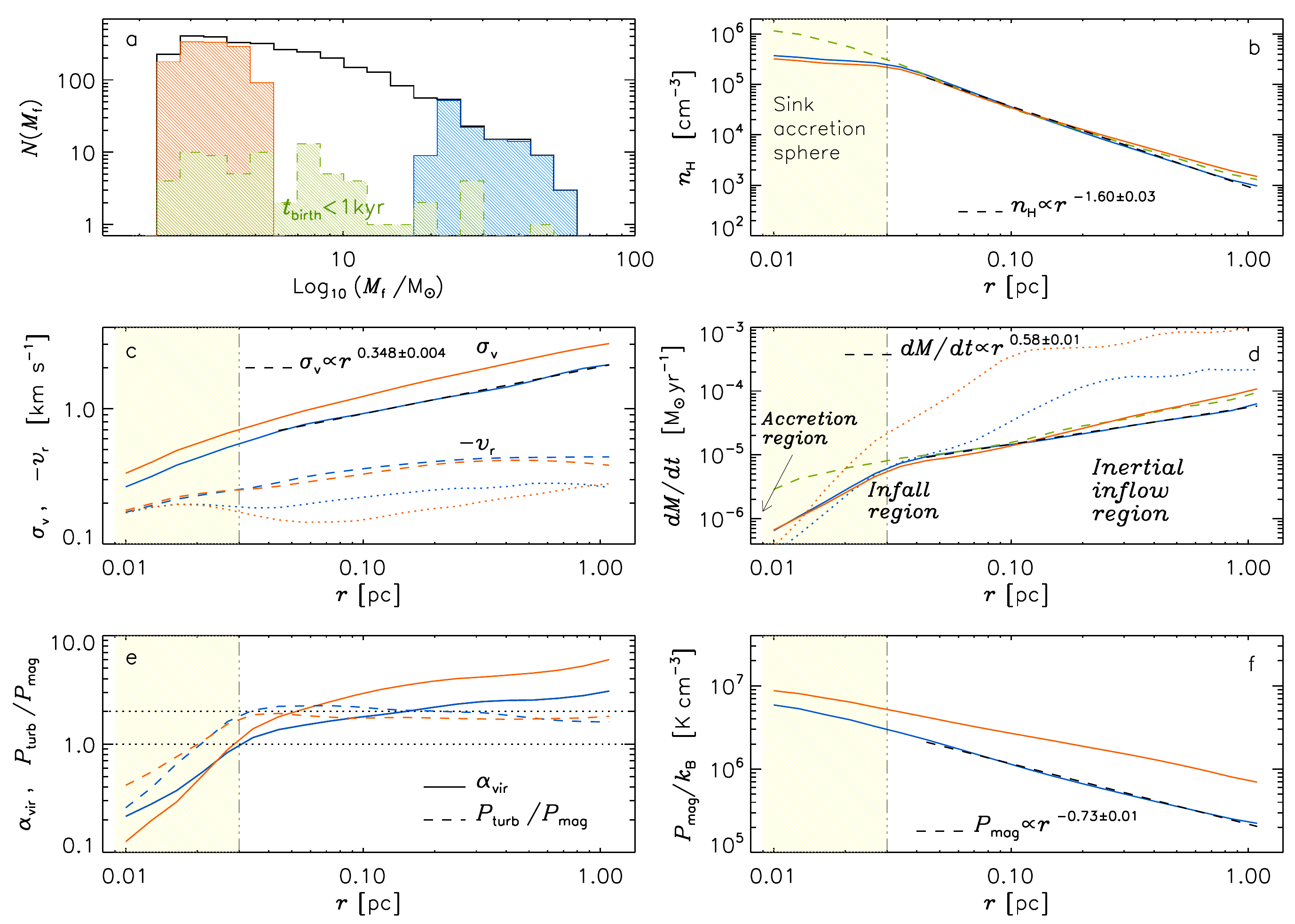}
\centering
\caption[]{Physical properties of the inflow regions around sink particles with different final stellar masses, either $M_{\rm f} < 5$~M$_{\odot}$ (red
plots) or $M_{\rm f} > 20$~M$_{\odot}$ (blue plots), or with very recent birth times (relative to the first simulation snapshot after the sink birth time), 
$t_{\rm birth}<1$~kyr (green plots). a) Mass distributions of the three samples (colored shaded histograms) relative to the total sample 
(black histogram). b) Radial profiles of gas density; the black dashed line is a power-law fit to the profile for massive stars (solid blue line) for $r>0.044$~pc, 
giving a slope of $-1.60\pm0.01$. c) Radial profiles of rms velocity (solid lines), radial velocity (dashed lines), and mass-weighted radial velocity (dotted lines). 
The black dashed line is a power-law fit to the profile of the rms velocity for massive stars (solid blue line) for $r>0.044$~pc, 
giving a slope of $0.348\pm0.004$.d) Radial profiles of mas-flow rate. The dotted lines show the profiles of the stars with the largest accretion rate at $r=0.2$~pc. 
The black dashed line is a power-law fit to the inflow-rate profile for massive stars (solid blue line) for $r>0.044$~pc, giving a slope of $0.58\pm0.01$.e) Radial profiles 
of virial parameter (solid lines) and ratio of turbulent to magnetic pressures (dashed lines). f) Radial profiles of magnetic pressure. The black dashed line is a 
power-law fit to the magnetic-pressure profile for massive stars (solid blue line) for $r>0.044$~pc, giving a slope of $-0.73\pm0.01$. 
}
\label{inflow_profiles}
\end{figure*}

\section{The Initial Conditions for Massive Star Formation: The Inflow Region} \label{sec_ic_inflow}

In the previous section, we have estimated the mass of prestellar cores at the beginning of their collapse, with the core size determined either by the core virial 
parameter or by a fixed radius of 0.1~pc. Here, we study the initial conditions further away from the protostar, on the scale of the inflow region, where the converging 
flows feeding the growing star have a kinetic energy in excess of the gravitational energy. For this purpose, we extract sub-volumes of $2\times2\times2$~pc$^3$
from the birth snapshot of each sink particle, centered around the birth position of the corresponding sink particle. This box size is appropriate to study the inflow region, even if in some
cases it includes only its inner portion. We define this region as the spherical volume around the sink particle containing 95\% of all the tracer particles that will be 
accreted onto the sink. As mentioned earlier, we refer to the radius of this stellar-mass reservoir, $R_{95}$, as the {\it inflow radius}, and find that its values cover 
a wide range, 0.1~pc~$\lesssim R_{95}\lesssim 33.5$~pc (see \S~\ref{sec_scaling}), with an average of 2.1~pc. 

In each sub-volume, we compute radius profiles of various  quantities by averaging over shells of different radii (20 logarithmically-spaced values) centered 
on the sink particle. Even with the smoothing effect of the shell or spherical averaging, individual profiles may exhibit complex radial variations and variations 
between sinks that have a purely stochastic origin, due to the turbulent nature of the inflow regions. Such stochastic fluctuations may hide the general trends 
from physical processes, so we further average the profiles of different sinks together. We perform this stacking procedure for two different groups of sink particles, 
based on the final stellar mass, either $M_{\rm f} < 5$~M$_{\odot}$ or $M_{\rm f} > 20$~M$_{\odot}$. The ranges of final stellar masses of the two samples,
in relation to the global stellar mass distribution, are shown in the top-left panel of Fig.~\ref{inflow_profiles}. The other panels of that figure follow 
the same color convention: blue plots for the profiles of the progenitors of very massive stars, red plots for those of the lower mass stars. The 
two samples have average values $\langle M_{\rm f} \rangle =3.5$~M$_{\odot}$ and $\langle M_{\rm f}\rangle = 29.1$~M$_{\odot}$, a mass 
ratio of one order of magnitude that should be sufficient to uncover any existing dependence on final stellar mass in the initial conditions.

The top-right panel of Fig.~\ref{inflow_profiles} shows that the shell-averaged mean density profiles for the two samples are nearly identical. 
A power-law fit to the profile corresponding to the most massive stars and for radii $r>0.044$~pc (dashed black line) gives a slope of $-1.60\pm0.03$.
Because our sink-accretion model transfers mass from the gas to the sink 
particle within an accretion sphere with radius of 0.03~pc, the profiles are somewhat artificial at $r\le0.03$~pc (meaning they obey not only the fluid 
equations, but also the sink sub-grid model), so that region is shaded in yellow as a reminder of that. On the other hand, we can still define
an average initial density profile that is almost insensitive to the sink-accretion model by selecting only the cores whose sink particles were 
born extremely close in time to the first subsequent snapshot. The snapshot time separation is 30~kyr, so the average time difference between 
the sink birth time and the snapshot time (when the profile is computed) is 15~kyr. As a compromise between adopting a time lag as short as 
possible and a number of cores as large as possible, we adopt 1.0~kyr for the time lag, resulting in a sample of 71 cores, whose corresponding final 
stellar masses are shown by the green histogram in the top-left panel of Fig.~\ref{inflow_profiles}. The average density profile for these pristine cores is plotted 
as a dashed green line in the top-right panel. It transitions smoothly from a $\propto r^{-2}$ power law to a lower slope as it crosses the accretion radius, 
reaching a central density of $10^6$~cm$^{-3}$, which is our threshold density to create a sink particle. Notice that a power-law slope of the shell-averaged density 
profile does not imply a spherical mass distribution that may be compared with predictions for isothermal spheres. In fact, outside of the virial radius, typically $\sim 0.03$~pc
(as reported below in relation to the left-bottom panel), the density field is highly fragmented and filamentary. The shell-averaged density of a single filament 
of constant density centered on the star scales as $r^{-2}$, so power-law slopes near $r^{-2}$ in the inflow region are more likely to be the result of a filamentary 
mass distribution than indicating any similarity to isothermal spheres.  

The velocity profiles are shown in the middle-left panel, where solid lines are for the shell-averaged rms velocity, $\sigma_{\rm v}$, 
dashed lines for the shell-averaged radial velocity, $v_{\rm r}$, and dotted lines for the mass-weighted shell-averaged radial velocity. The mean radial 
velocity is always negative, indicating inflow motion on the average for any radius up to at least 1~pc. The radial velocity grows monotonically towards
larger radii, reaching a maximum of $\approx 0.5$~km\,s$^{-1}$ at $r\approx 0.6$~pc. The inflow motion is transonic, or mildly supersonic, with the mass-weighted 
radial velocity always lower than the global shell-averaged radial velocity. This is an indication that the stellar mass is assembled through dense filaments: the 
inflowing lower-density gas is collected into such filaments through shocks, hence part of its pre-schock radial velocity component is lost as the gas is funneled 
towards the star through the filaments. Such dissipation of the radial component could only be avoided if all filaments were perfectly aligned in the radial direction,
which is a very unlikely arrangement. The rms velocity, instead, is highly supersonic, showing that the inflowing region is very turbulent. However, the velocity increase 
with radius is a bit shallower than in the global velocity-size relation (the power-law fit indicated by the black dashed line has a slope of $0.348\pm0.004$), 
perhaps because the negative mean radial velocity around the cores causes the transport of the 
larger velocity fluctuations at larger radii toward the center, or perhaps because of a slight amplification of the turbulence by compression, as the compression is stronger 
at smaller radii. The relatively shallow velocity scaling may be a fundamental property of inflowing regions feeding a central star through dense filaments, and deserves 
further investigation in future works. The velocity dispersion also appears to be slightly lower around the prestellar cores that form more massive stars.

These velocity profiles may be the key feature to distinguish our scenario from those where the large-scale mass reservoir is assumed to be collapsing. In the collapse
scenarios, the infall motion must be dominant over the random motion, in contradiction with our result that $v_{\rm r} < \sigma_{\rm v}$. Future observational studies
should try to separate the radial and random component of the velocity field in regions of massive-star formation, to discriminate between our prediction that 
$v_{\rm r} < \sigma_{\rm v}$ and the idea of global collapse. Another important feature that differentiates our velocity profiles from the collapse case is the monotonic increase of
$v_{\rm r}$ with distance up to nearly 1 pc. If gravity were to control such a flow, one would expect the gas to accelerate towards smaller radii, contrary to our results.
However, testing the radial dependence of $v_{\rm r}$ is probably beyond the capability of current observational methods. Because our definition of $v_{\rm r}$ involves
shell averages in 3D, it is not possible to compute the same quantity from observational data. A forward method should be used, where synthetic observations are 
performed with data from simulations dominated by either turbulence or global collapse and compared with observations of real star-forming clumps. 

In the middle-right panel, we plot the average profile of the mass-flow rate, $-4\pi\rho(r)r^2v_{\rm r}(r)$. The mass-flow rate decreases monotonically towards smaller 
radii (in the case corresponding to the most massive stars, the power-law fit has a slope of $0.58\pm0.01$). 
It varies from $\approx 10^{-4}$~M$_{\odot}$\,yr$^{-1}$ at 1~pc to $\approx 10^{-5}$~M$_{\odot}$\,yr$^{-1}$ at 0.1~pc. It may further decrease towards
even smaller radii within the sink accretion radius, as indicated by the dashed green line for the youngest cores in our sample, but those small values 
would only characterize the very initial stage of the collapse (the green, dashed-line profile is an average for cores that have started to collapse less than
1~kyr ago). The infall rate within the (gravitationally-bound) infall region must later control the actual accretion rate onto the star (through a disk), so the drop
in the infall rate within the inner 0.03~pc is just a feature of the sink-particle sub-grid model.\footnote{
The sink-particle accretion model could be tuned differently to make the infall rate inside the sink accretion sphere constant with radius, but a realistic 
physical picture would nevertheless require a description of the circumstellar disk.} The profiles for individual cores may deviate significantly from the 
average ones. To illustrate this, we also plot, as dotted lines, the profiles of the stars with the largest mass-flow rate measured at $r=0.2$~pc, showing that
the largest values can be approximately 10 times larger than the mean. 

The much larger value of the {\it inflow rate} in the inflow region, relative to that of the {\it infall rate} in the infall region should be viewed as another 
fundamental property of the feeding regions of massive stars. As shown in \S~\ref{sec_history}, the accretion rate of the sinks does not grow systematically
with time, so the mean radial dependence of the mass-flow rate implies that a significant fraction of the inflowing mass is not destined to accrete onto the central star. 
Because the inflow region is highly turbulent on a parsec scale, despite the mean radial motion, several intersecting shocks must be present, causing secondary 
convergence points around the main one feeding the central massive star. In other words, it is unlikely that a massive star is formed in isolation, and a significant 
fraction of the inflow rate ends up feeding secondary stars in the same region. Thus the inflow rate must be larger than the infall rate. This radial dependence of the 
mass-flow rate should be kept in mind when interpreting observations of mass-flow rate at different scales in regions of massive-star formation, as further discussed
in \S~\ref{sec_rates}. 

The bottom-left panel of Fig.~\ref{inflow_profiles} shows the average profile of the virial parameter, $\alpha_{\rm vir}$, and of
the ratio of turbulent to magnetic pressures, $P_{\rm turb}/P_{\rm mag}$. While the pressures are shell-averaged values, the virial parameter at a given 
radius, $r$, is computed as $2 (E_{\rm k}+E_{\rm th})/E_{\rm g}$, where $E_{\rm k}$, $E_{\rm th}$ and $E_{\rm g}$ are the kinetic, thermal and 
gravitational energies of the sphere of radius $r$. The virial parameter increases with increasing radius, starting from 
$\alpha_{\rm vir}=1$ at $r\approx 0.03$~pc for both core samples. The value of $\alpha_{\rm vir}$ is even lower at $r<0.03$~pc, but there the profiles 
may be affected by the sink-accretion model and by numerical dissipation of the velocity, so they cannot be fully trusted. In the inflow region, the virial 
parameter is a bit lower for the progenitors of the most massive stars than for those of lower-mass stars. As a result, the average infall radius, that is
the radius within which the gas is gravitationally bound, is a bit larger for the progenitors of the most massive stars, $\approx 0.15$~pc, than for
those of the lower-mass stars, $\approx 0.06$~pc. The pressure ratio in the inflow region of both groups of cores is $P_{\rm turb}/P_{\rm mag}\approx 2$,
showing that the turbulence in the inflow region is able to amplify the magnetic energy almost to equipartition with the kinetic energy. This is not representative 
of the average nature of the turbulence in the MCs of our simulation. We have shown in \citet{Padoan+16SN_I} that the turbulence in our MCs with mass 
$\gtrsim 10^{3}$~M$_{\odot}$ is always super-Alfv\'{e}nic also with respect to the rms magnetic-field strength. Thus, the near equipartition of turbulent and
magnetic energy is another distinguishing property of the inflow regions of stars of intermediate to large final masses. 

Finally, the bottom-right panel of Fig.~\ref{inflow_profiles} shows the average radial profiles of the magnetic pressure. Unlike the gas, the magnetic field is not accreted onto 
the sink particles, so the magnetic pressure keeps increasing with decreasing radius inside the accretion sphere, nearly unaffected by the sub-grid model for the sink formation 
and accretion. The magnetic pressure is a bit larger for the progenitor of the less massive stars, most likely a result of the slightly stronger turbulence there. 
The magnetic-pressure profiles are quite shallow in comparison with the density profiles. In the case corresponding to the most massive stars (blue solid line), 
the power-law fit (black dashed line) gives a slope of $-0.73\pm0.01$. Such shallow profiles suggest that the inflow motion must be directed predominantly
along magnetic field lines. Because we have already inferred that the inflow motion is organized in dense filaments (see the above discussion about the velocity profiles), 
the magnetic field within such filaments must be approximately aligned with the filaments, in agreement with recent results from ALMA polarization studies \citep[e.g.][]{DallOlio+19}.

\section{The Inertial-Inflow Scenario of Massive Star Formation}\label{sec_model}

Before describing our new scenario for the origin of massive stars, we show that the results presented in the previous sections rule out 
both the \emph{core-collapse model} and the \emph{competitive-accretion model}.

\subsection{Core Collapse and Competitive Accretion} \label{sec_competitive}

The main assumption of the core-collapse model \citep{McKee+Tan02,McKee+Tan03} is that a massive star originates from the collapse of a dense, massive core 
containing most of the final stellar mass. With the standard assumption that the core star-formation efficiency  $\lesssim 0.5$, the core mass at the beginning of its
gravitational collapse is at least more than twice the final stellar mass. Because thermal pressure alone cannot support a prestellar core of $\sim 100$~M$_{\odot}$,
the model assumes that the large critical mass is due to turbulent or magnetic support (hence the original ``turbulent-core" name of the model). This is also the
main assumption in the IMF models of \citet{Hennebelle+Chabrier08imf,Hennebelle+Chabrier09imf} and \citet{Hopkins12imf}, where the mass of a star comes from a
gravitationally-bound overdensity induced by the turbulence, while in our IMF model \citep{Padoan+Nordlund02imf,Padoan+Nordlund11imf} the mass reservoir
of a star comes from an inertial-range scale where the gas is not required to be gravitationally bound.

In \S~\ref{sec_ic} we have derived the mass distribution of prestellar cores defined as the progenitors of our sink particles with final stellar masses 
$M_{\rm f} > 2.5$~M$_{\odot}$. We have selected such cores at the beginning of their gravitational collapse, that is at the very transition between the prestellar 
and protostellar phases. Because we do not search for cores over the full volume, at a fixed time, independent of final stellar mass, nor from a 
sub-mm synthetic map, our mass distribution cannot be compared directly to those derived from observations of star-forming regions. However, 
because the mass we estimate is the largest one that can be assigned to the prestellar phase, it can be used to constrain theoretical models of massive-star
formation.  

\begin{figure}[t]
\includegraphics[width=\columnwidth]{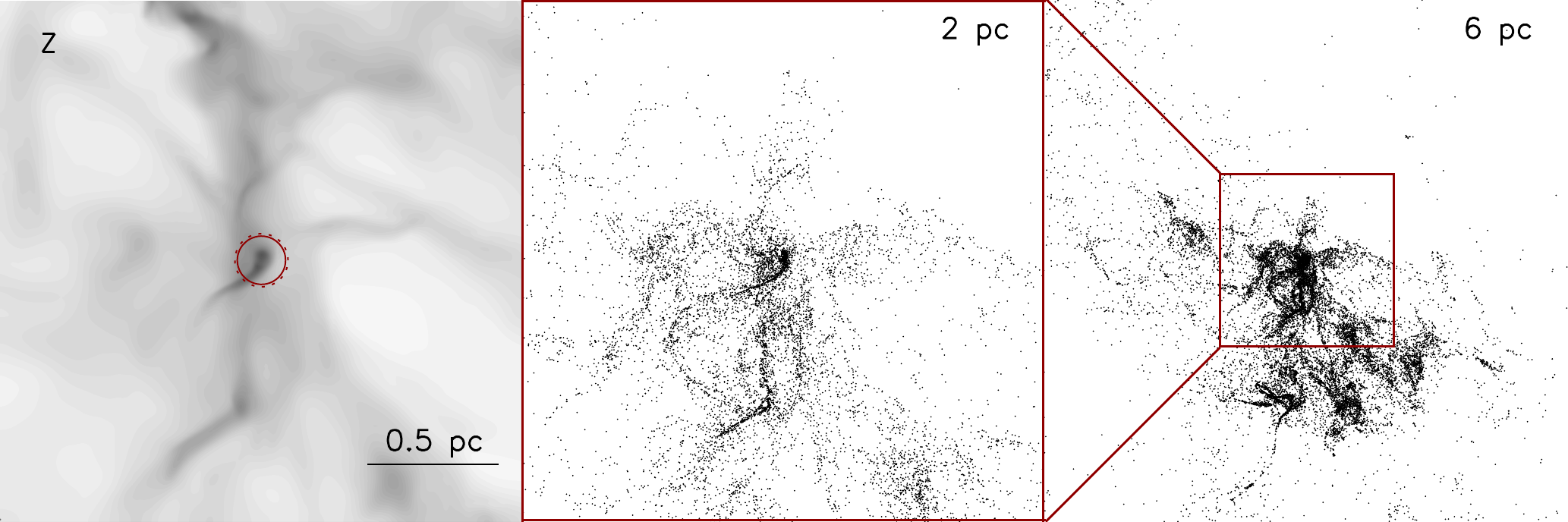}
\includegraphics[width=\columnwidth]{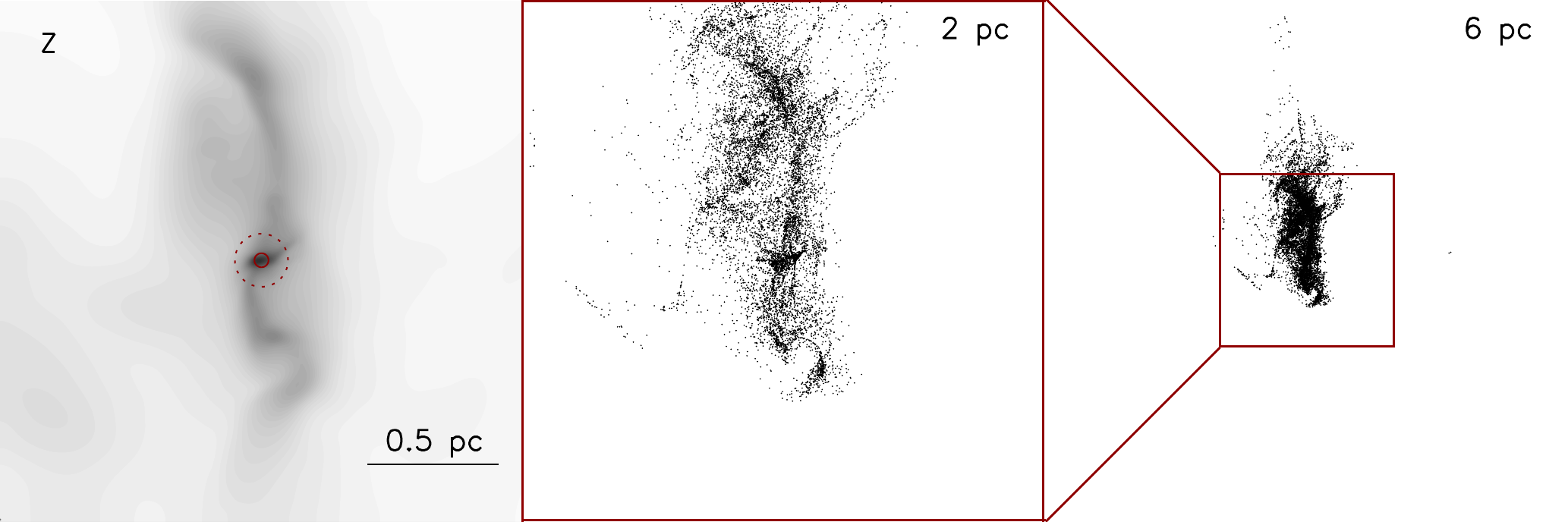}
\caption[]{Column density and tracer particles for two of the regions shown in Figure \ref{sink_simple}. The left panels show the square root of the
column density (as in Figure \ref{sink_simple}), the middle panels the distribution of tracer particles that will be accreted onto the central sink particle 
from the same (2 pc)$^3$ volume as the left panels, and the right panels the tracer particle distribution over a (6 pc)$^3$ volume. We have a selected 
these two sink particles to contrast one case (the most common one for massive stars) where the tracers that will accrete onto the final sink particle 
are distributed over a region larger than 6 pc (upper panel), with another case (less common) where all tracers are contained within a 3 pc region 
at the moment of creation of the sink particle (lower panels).  
}
\label{tracer_simple}
\end{figure}
\begin{figure}[t]
\includegraphics[width=\columnwidth]{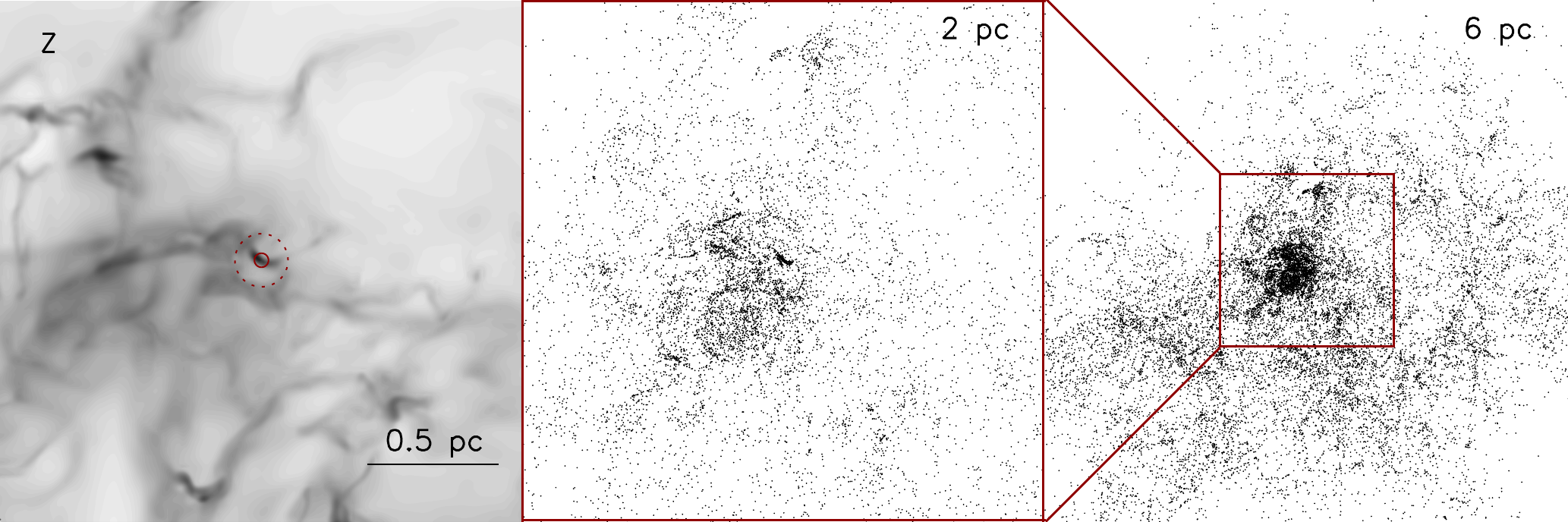}
\includegraphics[width=\columnwidth]{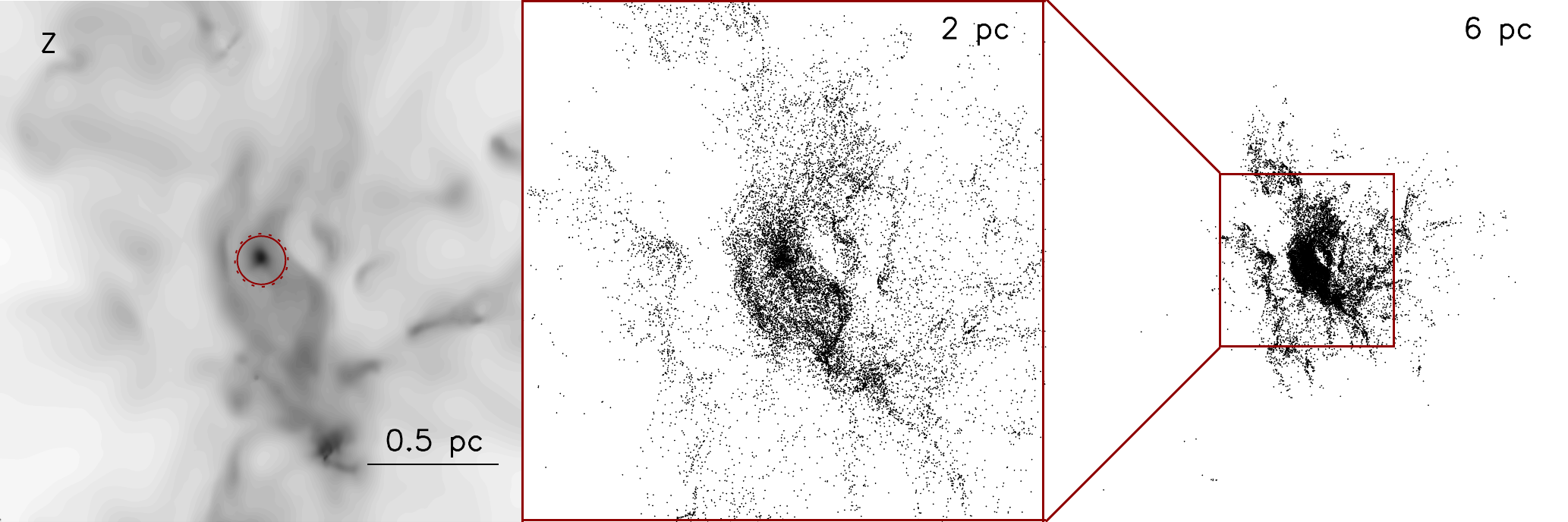}
\caption[]{The same as in Figure \ref{tracer_simple}, but for two of the regions shown in Figure \ref{sink_complex}. 
}
\label{tracer_complex}
\end{figure}

We have found that most cores forming stars with $M_{\rm f} > 20$~M$_{\odot}$ have virial masses between 0.4 and 40~M$_{\odot}$ and virial radii mostly 
between 0.01 and 0.5~pc, and their mass distribution peaks at $\sim 2$~M$_{\odot}$. We also found typical prestellar masses of $\sim 10$~M$_{\odot}$ 
within a fixed 0.1~pc radius, irrespective of the final stellar mass. Because the above models require turbulent-pressure support, we have also considered
the radius, $R_{\rm b}$, of the largest gravitationally-bound spherical volume around the prestellar cores, which can be considered as a strict upper limit
to the prestellar mass to be applied to those models. Even within this turbulent region, the total prestellar mass is still a small fraction of the final stellar mass.   
Thus, we conclude that the massive cores required by core-collapse models do not form in supersonic turbulence under characteristic MC conditions. Nevertheless, 
massive stars do form from lower-mass cores, because the mass reservoir that supplies the growth of the core is not exhausted, nor dispersed after the core collapses.  

As discussed in \S~\ref{sec_scaling} and in \S~\ref{sec_mmax}, the timescales and masses derived from our simulation can be rescaled to more extreme environments.
Our simulation was not tailored to describe extreme massive-star formation environments, and the infall rates we derive are rather low, so the final stellar mass, if 
stellar radiation were included, would certainly be somewhat lower than derived here. However, we argue that even rescaling to higher mean density, column density,
or turbulent velocity dispersion, would not change the qualitative picture given by our simulation. Our simulation shows that turbulent converging flows form 
gravitationally-unstable cores that collapse when their mass is only a few times their critical Bonnor-Ebert mass. In a more extreme environment, the density
of such cores would be larger than in our simulation, making the critical mass even smaller, and the collapse time shorter. Thus, the sequence of events would be
the same, namely the initial collapse of a small core followed by the growth of the star fed a mass flow through dense filaments. Because $R_{95} \gg R_{\rm b}$ 
and $M_{\rm f}\gg \epsilon_{\rm out} M_{\rm b}$, the mass reservoir feeding the star is much larger than predicted based on turbulent-pressure support. In other words,
massive stars are born with much lower masses and can potentially grow to much larger final masses than predicted by core-collapse models. We find no correlation 
between the prestellar mass (either $M_{\rm c,vir}$ or $M_{\rm b}$) and the final stellar mass, so the stellar mass cannot be constrained by the mass of the prestellar 
core, irrespective of specific environment or core definition.     

\begin{figure}[t]
\includegraphics[width=\columnwidth]{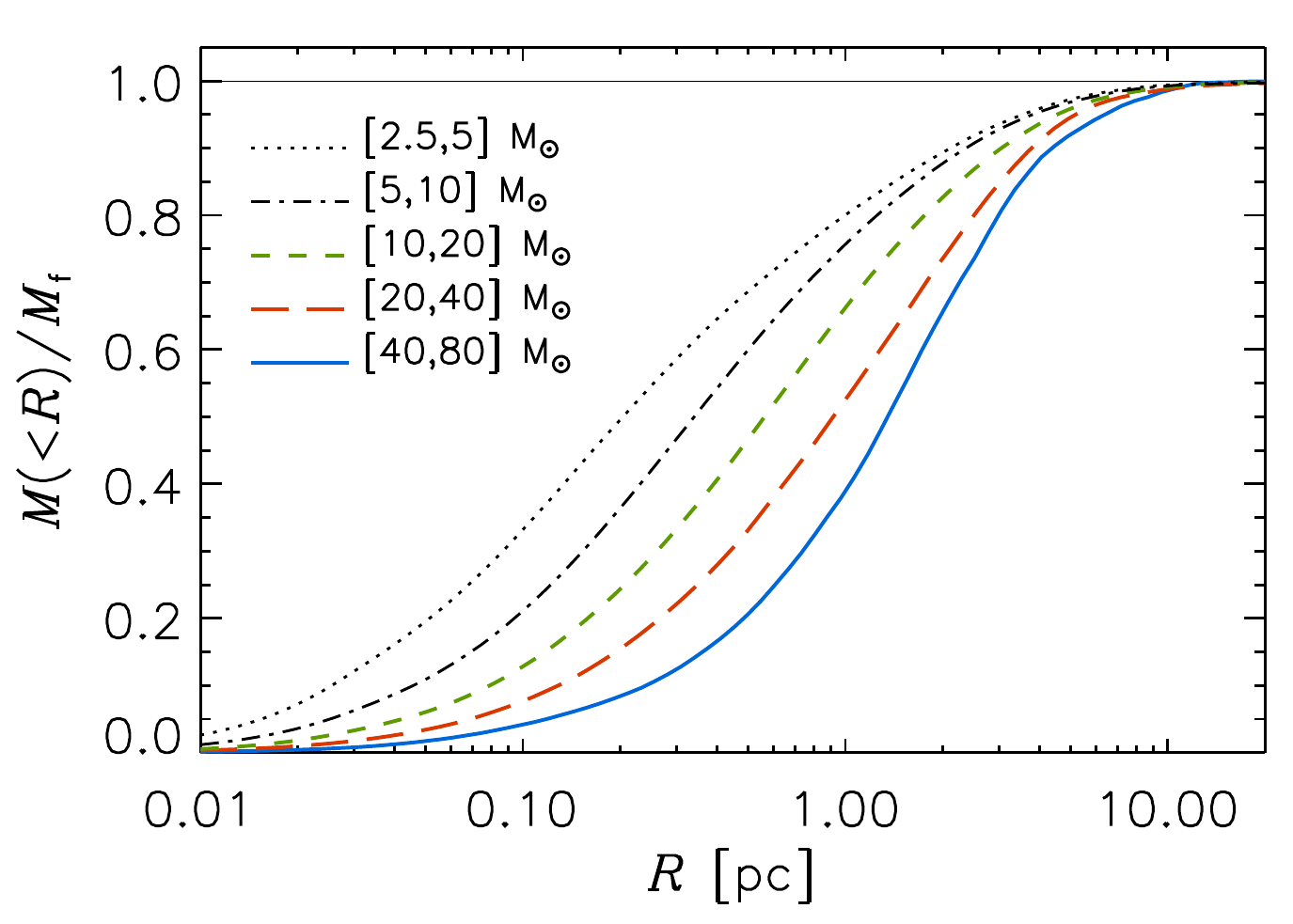}
\caption[]{Cumulative tracer-particle mass profiles, at the moment of sink-particle creation, for all the tracer particles that will accrete onto the sink.
The profiles are first computed for the tracers of each sink particle by summing up the mass of all tracers within a sphere of radius $R$, they are 
then normalized to the final stellar mass, $M_{\rm f}$, and are finally stacked to compute a single average profile. Five average profiles are shown,
for five different intervals of $M_{\rm f}$. For the most massive stars, the initial mass reservoir when the collapse of the prestellar core starts 
is 80-90\% of the final stellar mass within a region as large as 6-8 pc (in diameter) on average. The spatial extent of the mass reservoir decreases 
with decreasing $M_{\rm f}$. 
}
\label{tracer_cum}
\end{figure}

In order to characterize the extension of such mass reservoirs, we identify all the tracer particles accreted by each star, and consider their 
locations at the same time when the prestellar core is identified (the first simulation snapshot after the sink particle is created). Figures~\ref{tracer_simple} and 
\ref{tracer_complex} show the tracer particle positions in four representative cases. It is evident that most of the future stellar mass is still
distributed over a volume of a few pc when the prestellar core starts to collapse. To estimate the characteristic size of this volume, we compute
the cumulative tracer-particle mass profile for each star, and then average the profiles, each normalized to its own final stellar mass, within five different mass ranges.
These average profiles are plotted in Figure~\ref{tracer_cum}. In the case of the most massive stars (solid line), less than half of the total stellar mass is found,
on average, within a radius of 1~pc, and to include 90\% of the final stellar mass, we must consider a sphere with a radius of approximately 5~pc. 

 Although the competitive-accretion model \citep{Zinnecker82,Bonnell+2001a,Bonnell+2001b} predicts that the initial core mass is much smaller than the final stellar mass, 
 our results are in contradiction with that model as well. 
 This is easily understood based on the criticism of competitive accretion presented by \citet{Krumholz+2005}, where it is demonstrated that competitive
accretion can explain massive star formation only if the star-forming region feeding the accreting star has a very low virial parameter, $\alpha_{\rm vir}\ll 1$. If this 
condition is not satisfied, the Bondi-Hoyle accretion rate is too small and the low-mass stellar ``seed" cannot significantly increase its mass. In our simulation, the virial parameter
averaged within spheres centered on the sink particle increases with increasing radius, and the prestellar core has been defined by the radius where $\alpha_{\rm vir}=1$.
At larger radii, where most of the future stellar mass is contained, the virial parameter is $\gg 1$, so competitive accretion must be negligible. In other words, because the 
large-scale region where the future stellar mass is located is not gravitationally bound, and since the initial stellar mass is only a small fraction of the total mass in the
region, the stellar gravity has a negligible effect on the mass inflow towards the star (except at very short distances from the star). 

The accretion-rate history discussed in \S~\ref{sec_history} serves as further evidence against competitive accretion. Competitive accretion predicts that the
accretion rate increases with the stellar mass, with ${\dot M}\propto M^{2/3}$ in the case of gas-dominated potentials, or ${\dot M}\propto M^2$, in the case of 
stellar-dominated potentials \citep[e.g.][]{Bonnell+2001a,Bonnell+2001b}. As shown by \citet{Bonnell+2001b}, massive stars acquire most of their mass during the 
stellar-dominated phase, with ${\dot M}\propto M^2$. Despite being a fundamental prediction of the model, such a dependence of the accretion rate on stellar 
mass has never been derived from star-formation simulations, not even when the simulations are purported to generate a stellar mass spectrum because of
competitive accretion \citep[e.g.][]{Bonnell+2004,Bonnell+Bate2006}. At best, scatter plots of accretion rate versus sink mass at a fixed time have been 
shown \citep{Maschberger+2014,Ballesteros-Paredes+2015,Kuznetsova+2018}, but these do not prove that the accretion rate of a single star grows over 
time as the stellar mass increases (the accretion rates in \citep{Ballesteros-Paredes+2015} deviate by orders of magnitude from the Bondi-Hoyle prediction).
In our simulation, the accretion rate during the formation of a massive star has strong time variations of a random nature, but no systematic increase with time. 
This demonstrates that the infall rate (we assume the accretion rate is proportional to the infall rate times $\epsilon_{\rm out}$) is not controlled by the stellar 
gravity as in the competitive-accretion scenario, but by the large-scale mass inflow, which is just a consequence of shocks in the supersonic turbulence.

\subsection{Inertial-Inflow Scenario}

As mentioned in the introduction, our turbulent-fragmentation model of the stellar IMF implies that the formation of any star can be viewed as a sequence 
of three main steps: (1) the formation of a gravitationally unstable core exceeding the critical Bonnor-Ebert mass, (2) the collapse of the core into a low or 
intermediate-mass star, (3) the accretion of the remaining mass driven by a large-scale converging flow, with the gradual buildup of the stellar mass over a 
number of free-fall times. For a massive star, most of the growth occurs during the third step, as the final stellar mass is much larger than the critical 
Bonnor-Ebert mass of the prestellar core.\footnote{
This third step becomes gradually less important for stars of decreasing final stellar mass for masses
in the neighborhood of the IMF peak, as we interpret the mass of the IMF turnover as a characteristic {\it turbulent} Bonnor-Ebert mass \citep[see][]{Haugbolle+18imf}.
} 
Because this step is dominant for massive stars, and the large-scale converging flow is a local random realization of the MC turbulence and mostly unaffected by 
the gravity of the star or by the self-gravity of the inflow region, as shown above, we refer to this scenario of massive star formation as the \emph{inertial-inflow model}. 

Our IMF model postulates that a prestellar core is assembled as a piece of a postshock gas layer, which results from the compressive component of a large-scale
turbulent eddy. Turbulent eddies of larger scale generate more massive stars, because the gas reservoir for the prestellar core and for the further growth of the 
star is larger. Scaling relations leading to the stellar IMF are derived from the velocity scaling of the turbulent flow, assuming one dimensional MHD shocks and 
self-similarity \citep{Padoan+Nordlund02imf}. The assumption that a prestellar core is a piece of a postshock layer, its size being determined by the thickness 
of the layer (hence by the MHD jump conditions), was inspired by numerical simulations of supersonic turbulence. Long before the filamentary nature of MCs 
was revealed by Herschel's observations \citep{Menshchikov+10}, such simulations had shown that turbulent fragmentation results into a complex filamentary 
morphology \citep{Padoan+Nordlund99MHD}, with dense cores found in knots within filaments \citep{Padoan+2001cores}. 
The knots are the locations of intersection of filaments, and filaments are the location of intersection of 
postshock layers. While the dense postshock gas within layers and filaments is characterized by a relatively strong (possibly supersonic) shear, within an 
intersection of filaments the flow stagnates to subsonic velocities, and a quiescent core emerges, with mass flowing to the core through the filaments. 

Magnetic field lines are mostly aligned along such filaments, because the magnetic field component perpendicular to the compression is amplified\footnote{
This statement applies to very dense, pc-scale filaments whose intersections host the formation of massive stars, and only to the magnetic field {\it inside} the 
filaments. The ambient magnetic field outside of the filaments may very well be mostly perpendicular to the filaments, as that would allow for a higher postshock 
density inside the filaments. Recent sub-mm polarization studies confirm that the ambient field direction outside of the filaments is usually perpendicular to the 
filaments \citep[e.g.][]{Malinen+16,PlanckXXXV16,Liu+18,Soler+Juan19}, while the magnetic field in their interior is less well constrained, and the picture at 
smaller scales less straightforward \citep[e.g.][]{PlanckXXXIII16,Pattle+17}, with some evidence of magnetic field parallel to dense filaments as in our scenario 
\citep[e.g.][]{DallOlio+19}.
}, so the mass 
inflow feeding the core can freely increase the ratio of mass to magnetic flux in the core \citep{Padoan+Nordlund99MHD,Lunttila+08,Lunttila+09,Padoan+10mhd}
and magnetic support cannot affect significantly the core critical mass, even in the absence of ambipolar drift or resistive processes. Thus, when the core reaches 
a mass of the order of a few times the critical Bonnor-Ebert mass, it starts to collapse. The collapse of the prestellar core does not affect the large-scale, filamentary 
mass inflow that originated the core, so the mass supply continues. Instead of accreting on the surface of a prestellar core, the gas now feeds the circumstellar disk 
that drives the accretion onto the surface of the star. As a result, the accretion rate of the star is controlled, or at least constrained, by the rate of the large-scale mass 
inflow. Because the accretion onto the stellar surface is through the circumstellar disk, and the disk is fed by dense filaments, the radiative pressure from the star is 
not a fundamental barrier to the growth of the star \citep[e.g.][]{Krumholz+05,Keto07,Krumholz+09,Kuiper+11,Klassen+16}. Photo-evaporation of the infalling gas may limit or stop 
the stellar growth, but only if the inflow rate is very small \citep[e.g.][]{Tanaka+17}. 

This scenario implies that prestellar cores cannot achieve very large masses, hence their mass function above a few solar masses should be steeper than 
the stellar IMF \citep{Padoan+Nordlund11imf}. In the absence of self-gravity, the process of turbulent fragmentation may indeed produce a core mass function 
that resembles the stellar IMF even at large masses, as demonstrated with simulations of driven supersonic MHD turbulence without self-gravity \citep{Padoan+07imf}. 
We have modeled the stellar IMF assuming that a mass-independent fraction of the core mass is converted into a star, so the core mass function from the turbulence 
determines the stellar IMF, essentially decoupling the effect of the gravity from the effect of the turbulence. When gravity is included, though, the most massive cores 
start to collapse before they are fully assembled, so their predicted mass should be viewed as a mass reservoir for the growth of protostars, not as the mass
of prestellar cores. 

The inertial-inflow scenario described here is qualitatively consistent with the results of our simulation. As shown in \S~\ref{sec_ic}, the prestellar cores
in the simulation are found within dense filaments, and their mass is approximately a few times their critical Bonnor-Ebert mass. The region surrounding
a prestellar core is turbulent and gravitationally unbound, but channels a net mass flow onto the core, as shown in \S~\ref{sec_ic_inflow} (clearly not a consequence of 
gravitational instability or global collapse, as the region is turbulent and gravitationally unbound). The timescale 
of star formation is much larger than a core free-fall time and is an increasing function of the final stellar mass, as found in \S~\ref{sec_time}. The evolution 
of the mass accretion rate is characterized by stochastic time variations and insensitive to the stellar mass (see \S~\ref{sec_history}), consistent with a
large-scale turbulence source. In the following subsection, we demonstrate that the star-formation time and the size of the stellar mass reservoir follow
the relation expected for the MC turbulence, providing further evidence that the star-formation time is proportional to the turbulence turnover time, as
in our scenario. We also show that the final stellar mass is, on average, an increasing function of the size of the mass reservoir, also consistent with the 
general scenario. 

Although inspired by our IMF model \citep{Padoan+Nordlund02imf}, the inertial-inflow scenario presented in this work calls for a fundamental revision of all IMF 
models to date, including our own, with regards to the origin of the power-law tail of the IMF. While our model is based on postshock sheets, the inertial-inflow
scenario stresses the filamentary nature of the mass-accretion process. Furthermore, our model uses a single-valued scaling of the velocity, so it cannot account for 
the large scatter in the star-formation time at any given final stellar mass.

\subsection{Inertial-Inflow Scenario and Velocity Scaling}\label{sec_scaling}

\begin{figure}[t]
\includegraphics[width=\columnwidth]{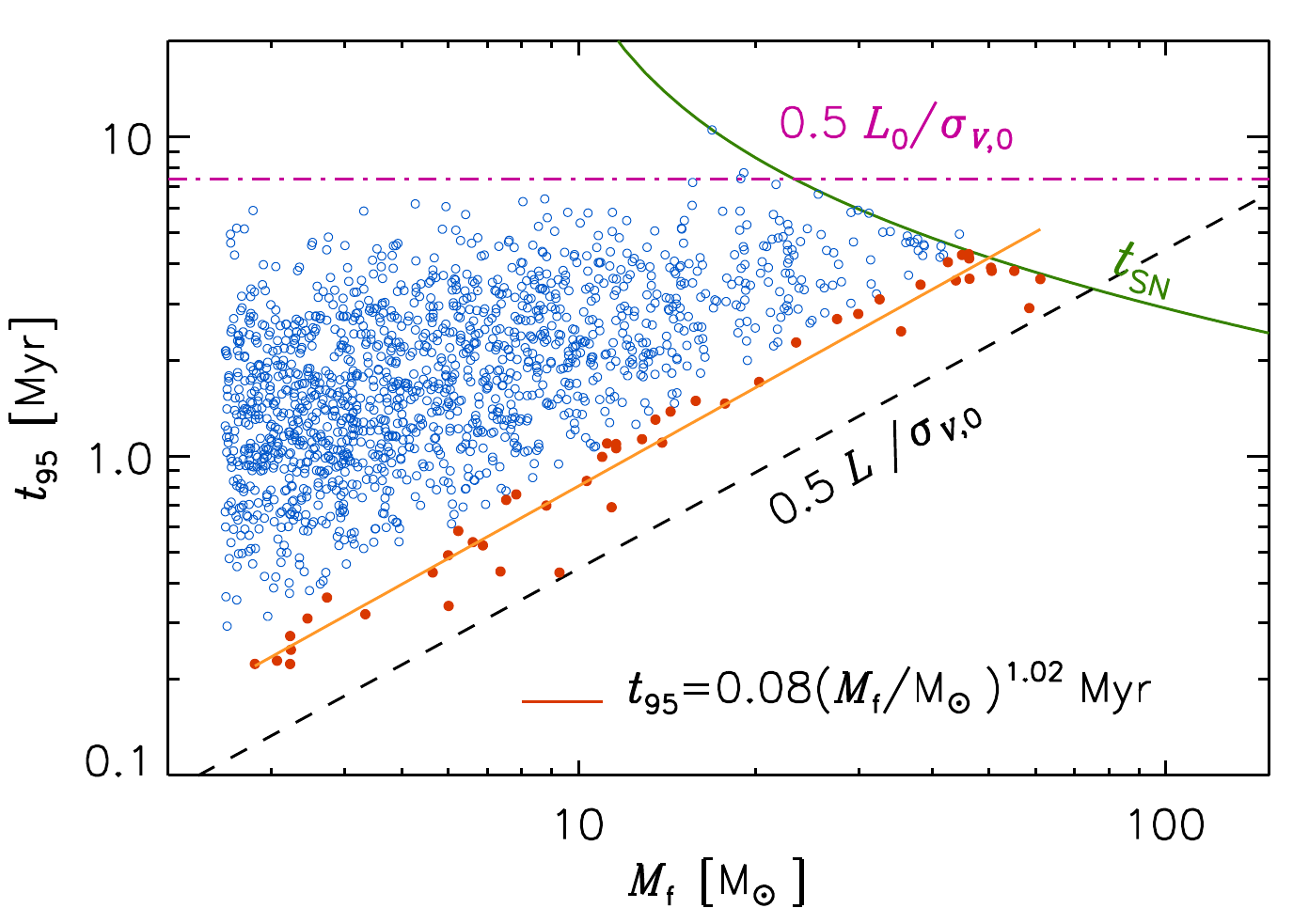}
\caption[]{Star-formation time versus final mass as in Figure~\ref{t95_v1}. Here, we stress the interpretation of the origin of the upper (dashed-dotted line) 
and lower (short-dashed line) envelopes. We also mark as filled, red circles the stars with average infall rate $\dot{M}_{\rm in} \ge 2 \times 10^{-5} M_\odot$ yr$^{-1}$. 
The straight, red, solid line is a least-square fit to the filled, red circles, with slope 1.02.
}
\label{t95_v2}
\end{figure}

A complete understanding of the $t_{95}$--$M_{\rm f}$ relation shown in Figure~\ref{t95_v1} would require a new theoretical model of the stellar IMF, which is 
beyond the scope of the current work. Here, paving the way towards such a new theory, we focus on the origin of the upper and lower envelopes of the 
$t_{95}$--$M_{\rm f}$ plot, which will provide an interpretation of its large scatter. For the purpose of this
discussion, we refer to a new version of the $t_{95}$--$M_{\rm f}$ plot, shown in Figure~\ref{t95_v2}. We interpret the envelopes as the
result of the velocity scaling of supersonic turbulence, which provides further support to our scenario for massive-star formation. 
According to our inertial-inflow scenario, the formation timescale of a star is set by the dynamical time of the turbulent structure out of 
which it forms. In molecular-cloud turbulence, the amplitude of velocity differences, $\sigma_{\rm v, \ell}$, measured at a scale $\ell$, 
goes like $\sigma_{v,\ell}\sim \ell^{\alpha}$, with $\alpha \approx 0.5$ \citep{Larson81,Solomon+87,Padoan+2003scaling,Heyer+Brunt04,Padoan+06perseus}. 
This approximate relation holds also for MCs selected from our simulations \citep{Padoan+16SN_I,Padoan+16SN_III,Padoan+17sfr} and 
for supersonic turbulence in general \citep{Boldyrev+2002scaling,Padoan+04PRL,Federrath13_4096}. 
The dynamical time of turbulent structures of size $\ell$, which is essentially the turnover time of the 
eddies of size $\ell$, scales as $t_{\rm dyn} \simeq \ell/\sigma_{v,\ell} \propto \ell^{1-\alpha} \sim \ell^{0.5}$. If our interpretation 
that the star-formation time, $t_{95}$, is set by the turbulent dynamical time is correct, a similar scaling is expected to hold also for $t_{95}$. 

\begin{figure}[t]
\includegraphics[width=\columnwidth]{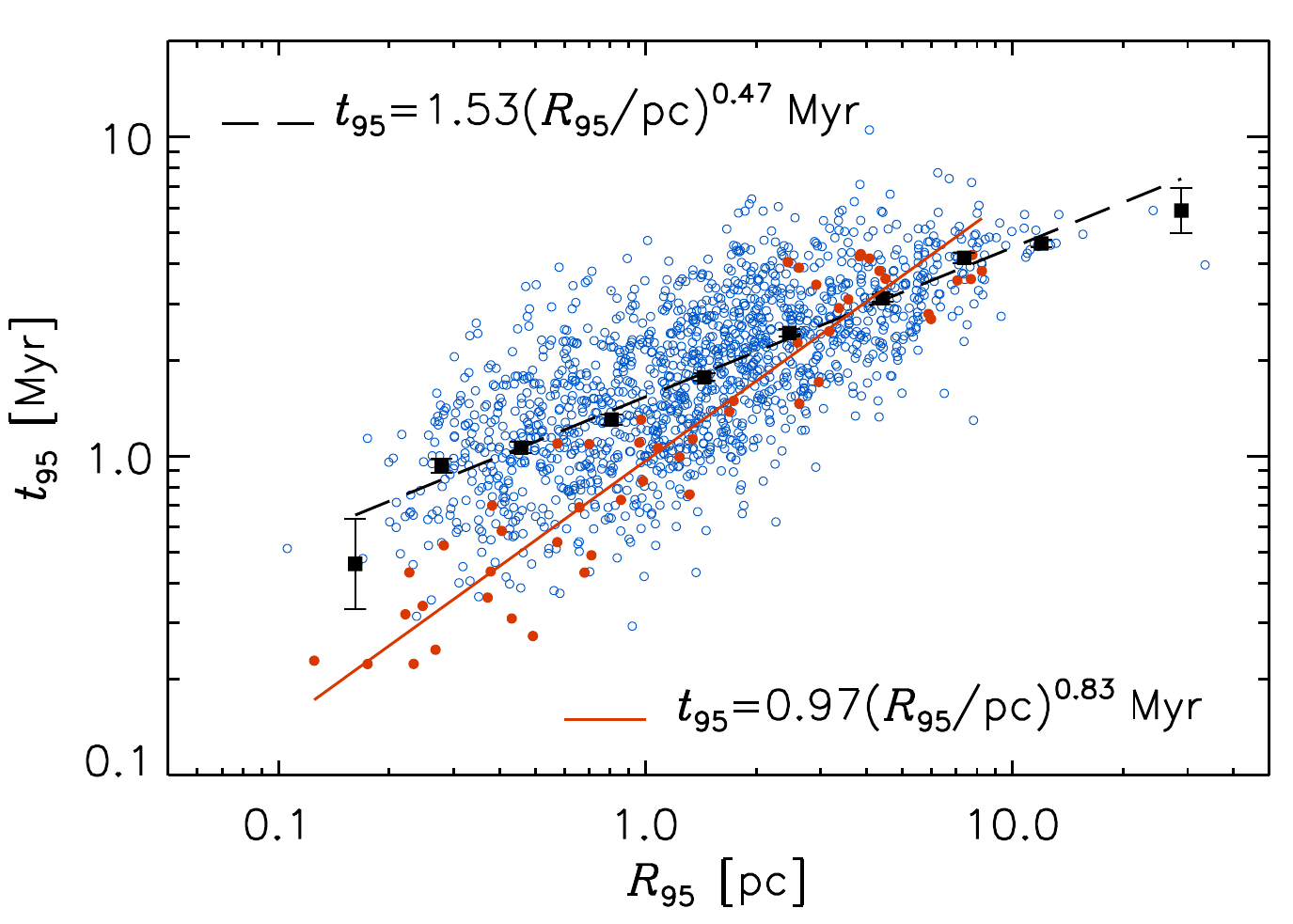}
\caption[]{Star-formation time versus inflow radius. The black squares are the mean values of $t_{95}$ within equal logarithmic intervals of $R_{95}$ (error bars show
the $1-\sigma$ uncertainty of the mean), and the black, long-dashed line is a least-square fit to the black squares, with slope 0.47. The stars with average infall rate  
$\dot{M}_{\rm in} \ge 2 \times 10^{-5} M_\odot$ yr$^{-1}$ are shown as red, filled circles, as in Figure~\ref{t95_v2}, and the red line is a least-square fit to the red,
filled circles, with slope 0.83. 
}
\label{t95_r95}
\end{figure}

To define a proper length scale for the process of star formation in the simulation, we consider the spatial extent of tracer particles that will be
accreted onto a given sink particle, at the moment of the sink creation. As in the computation for the plots in Figure \ref{tracer_cum}, we compute 
the mass in tracer particles found within spheres centered around the sink particle. We then define the characteristic size, $R_{95}$, of the mass 
reservoir for the formation of a star as the radius of the sphere that contains 95\% of the tracer mass. We already referred to 
$R_{95}$ as the {\it inflow radius} in previous sections. In Figure \ref{t95_r95} we plot 
the star-formation time, $t_{95}$, versus the inflow radius, $R_{95}$. The figure indicates a strong correlation between the two quantities. 
The black squares  plot the mean value of $\log(t_{95})$ over the stars that fall in logarithmic intervals of $R_{95}$. The black, long-dashed line, 
\begin{equation}
t_{95}=1.53 ({R_{95}/{\rm pc}})^{0.47} {\rm Myr},
\label{t_R95}
\end{equation}
is a power-law fit to the black squares. This scaling is consistent with the velocity-size relation of MCs selected from our simulation (e.g. Figure~4 of \citet{Padoan+17sfr}) 
and with supersonic turbulence in general, thus providing a strong evidence for our inertial-inflow scenario, i.e., the stellar mass is assembled 
by inertial, compressive motions in the supersonic turbulent flow. 

As seen in Figure~\ref{t95_v2}, the upper envelope of the $t_{95}$--$M_{\rm f}$ plot is almost flat. We interpret this maximum formation time as the
largest turnover time in the turbulent flow, because beyond that time the flow cannot remain coherent, hence the mass-inflow towards the star cannot
continue. Thus, we estimate this maximum time as 
\begin{equation}
t_{\rm 95,max}= \tau_0=L _0/(2\,\sigma_{\rm v,0}),
\label{t_95_max}
\end{equation}
where $L_0$ is the turbulence outer scale and $\sigma_{\rm v,0}$ the velocity dispersion on that scale. 
In our simulation, $L_0 \sim70$~pc, corresponding to the size of the largest GMCs (and the driving scale 
estimated in \citet{Padoan+16SN_I}), and $\sigma_{\rm v,0} \sim4.6$~km\,s$^{-1}$
(the value extrapolated for a cloud radius of 35~pc from Figure~4 of \citet{Padoan+17sfr}), so that $t_{\rm 95,max}  \simeq 7.4$~Myr, which is
shown by the dashed-dotted line in Figure~\ref{t95_v2} and is evidently a reasonable description of the upper envelope of that plot (only a single star
in the plot has a larger value of $t_{95}$). This value of $t_{\rm 95,max}$ derived 
from the velocity-size relation of our MCs is also similar to the value of $t_{95}=8.1$~Myr derived from Equation~(\ref{t_R95}) for $R_{95}=35$~pc.

\begin{figure}[t]
\includegraphics[width=\columnwidth]{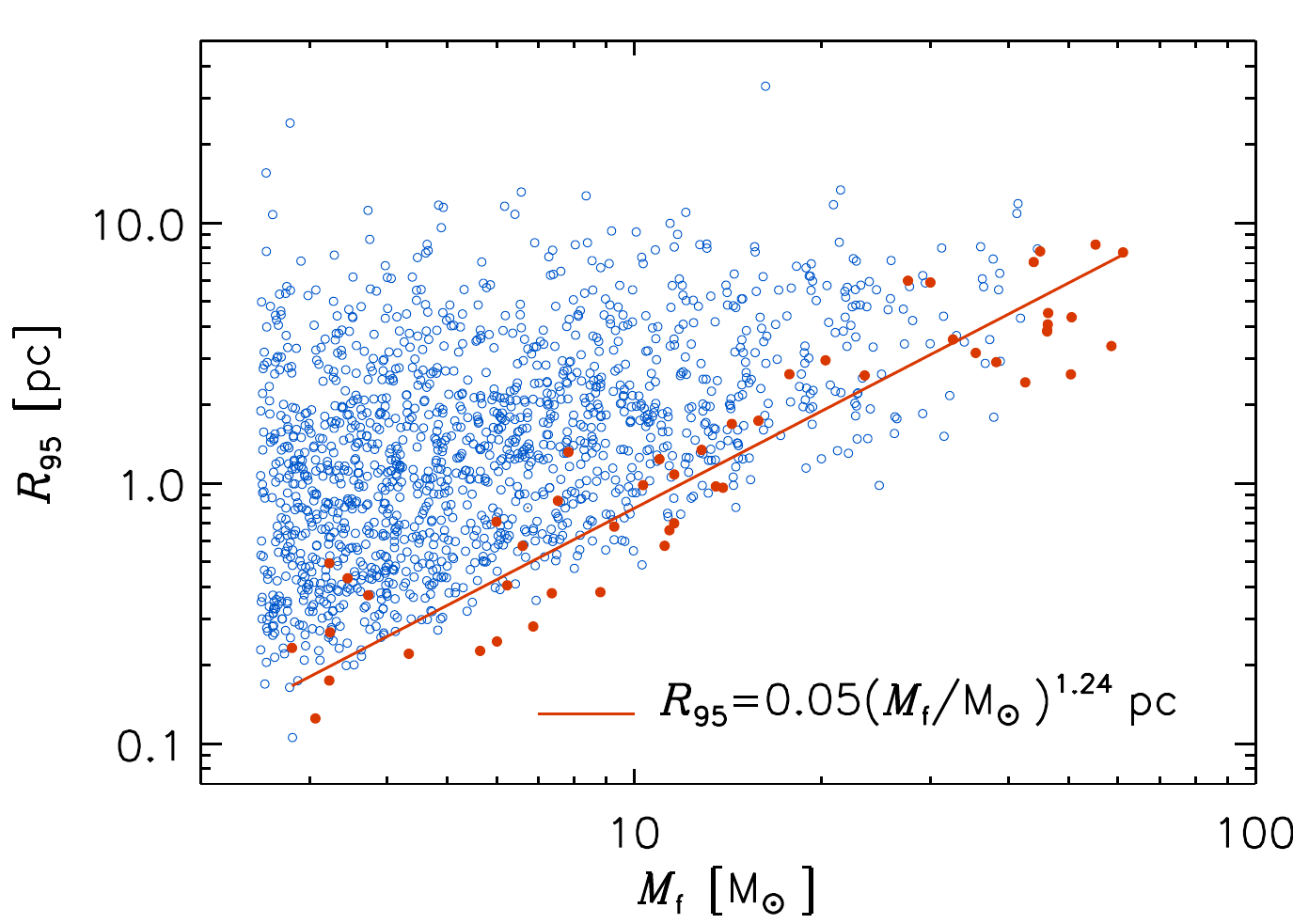}
\caption[]{Inflow radius versus final stellar mass. The stars with average infall rate  $\dot{M}_{\rm in} \ge 2 \times 10^{-5} M_\odot$ yr$^{-1}$
are shown as red, filled circles, as in Figures~\ref{t95_v2} and \ref{t95_r95}, and the red line is a least-square fit to the red, filled circles, with slope 1.24.  
}
\label{mf_r95}
\end{figure}

The lower envelope of the $t_{95}$--$M_{\rm f}$ plot is fit by a linear relation, $t_{\rm 95,min} \propto M_{\rm f}$, shown by  
the dashed line in Figure~\ref{t95_v2}. This line corresponds approximately to the maximum infall rate in the plot, 
$\dot{M} = 4.5 \times 10^{-5} M_{\odot} {\rm yr}^{-1}$.\footnote{
In the simulation, only half of the infalling mass is accreted to the star, that is $\epsilon_{\rm out}=0.5$, so the actual accretion rate 
corresponding to the dashed line is half the value reported above.} 
The origin of the lower envelope is more difficult to understand than that of the upper envelope, and we defer a more detailed study to a future work. 
Here, we make reasonable assumptions to interpret the lower envelope qualitatively. We assume that stars near the lower envelope form from turbulent 
structures with velocity differences of order $\sigma_{\rm v,0}$, even if their size, $L$, can be much smaller than $L_0$. In a turbulent flow, velocity 
differences scale approximately as $L^{0.5}$, but because the lower envelope corresponds to the shortest timescales, it is reasonable to assume that it 
originates from (intermittent) structures with velocity of the order of the overall rms velocity at the outer scale, irrespective of their size $L$. Based on this
assumption, and given that we have already identified the relevant turbulent scale for each star as $L=2\,R_{95}$ (twice the inflow radius), we make the ansatz 
that the lower envelope is given by the smallest scale, $R_{\rm 95,min}$, that can lead to a given final mass, $M_{\rm f}$ (with a velocity $\sigma_{\rm v,0}$):
\begin{equation}
t_{95,\rm min} = R_{\rm 95,min} / \sigma_{\rm v,0}.
\label{t95_min}
\end{equation}
Because $t_{95,\rm min}=t_{95,\rm max}$ for $R_{\rm 95,min}=L_0/2$, the two envelopes intersect at the outer scale, $L_0$.   

Our ansatz implies that the stars in the lower envelope of the $t_{95}$--$M_{\rm f}$ plot should be at the lower envelope of the $R_{95}$--$M_{\rm f}$ plot 
as well, and that their inflow radius should scale linearly with the final mass, $R_{\rm 95,min} \propto M_{\rm f}$. To verify if this is true, we select the stars with
infall rate $\dot{M}_{\rm in} \ge 2 \times 10^{-5} M_\odot$ yr$^{-1}$, and show them as red, filled circles in Figures~\ref{t95_v2} to \ref{mf_r95}. The
$R_{95}$--$M_{\rm f}$ relation is shown in Figure~\ref{mf_r95}, where one can see that the red, filled circles are indeed found at the lower envelope
of the scatter plot, and the lower envelope is very close to linear. Thus, our interpretation of the lower envelope of the $t_{95}$--$M_{\rm f}$ relation is 
qualitatively confirmed. However, the detailed picture must be more complicated, as not all points near the lower envelope are red, filled circles 
(stars with the maximum infall rate), and the red, filled circles alone are a bit steeper than linear, with the least-square fit giving a slope of 1.24.
The lower envelope of the $R_{95}$--$M_{\rm f}$ relation corresponds to the densest turbulent structures (at any given $M_{\rm f}$), and its 
nearly linear nature implies that such structures are nearly filamentary. We may conclude that the stars with the largest infall rates form from the 
densest filamentary structures. This picture is quite different from the idea of gravitational instability of supercritical filaments \citep[e.g.][]{Fiege+Pudritz2000,Andre+14,Andre+16}, which has been proposed to explain the origin of prestellar cores. The initial prestellar core collapse 
only accounts for a small fraction of the final stellar mass, in the case of massive stars. However, the lower envelope of the $R_{95}$--$M_{\rm f}$ relation 
refers to the final stellar mass, which is the result of further inflow of mass, after the initial collapse. The mass reservoir of the stars is distributed over 
a large scale, well beyond the size of the prestellar cores, as shown qualitatively in Figures 18 and 19, and quantitatively in Figures 13 and 20. 
Therefore, the lower envelope of the $R_{95}$--$M_{\rm f}$ relation cannot be explained as a consequence of gravitational instability in filaments.

We can directly verify from the $t_{95}$--$R_{95}$ plot if $R_{\rm 95,min}$ scales linearly with $t_{95}$, as implied by our interpretation of the lower
envelope of Figure~\ref{t95_v2}. The red, filled circles in Figure \ref{t95_r95} provide qualitative support to our interpretation. However, their slope is 
a bit shallower than linear, with the least-square fit giving a slope of 0.83 (the solid red line in Figure \ref{t95_r95}), and they are found somewhat above the 
lower envelope of the $t_{95}$--$R_{95}$ plot, so their corresponding velocity is lower than $\sigma_{\rm v,0}$.  
Thus, the formation process of the stars with the largest infall rate is somewhat more complicated
than in our simple interpretation, and a full understanding of the lower envelope of the $t_{95}$--$M_{\rm f}$ relation 
requires further investigation. The formation process may depend on other factors, such as the preshock density, the flow 
symmetry or compressive ratio, the magnetic fields in the hosting structures.

\subsection{The maximum infall rate} \label{sec_max_infall}

In the absence of a full understanding of the lower envelope of the $t_{95}$--$M_{\rm f}$ relation,\footnote{
Equation~(\ref{t95_min}) gives the relation between $t_{95,\rm min}$ and $R_{\rm 95,min}$ at any given $M_{\rm f}$, 
but does not provide a specific value of the lower envelope as a function of the physical parameters of a star-forming region.}
we propose to interpret the maximum infall rate, $\dot{M}_{\rm in,max}$, as a fraction, $\epsilon_{\rm in}$, of the maximum inflow rate, 
and to express the maximum inflow rate as the ratio of the total mass in the outer scale, $M_0$, and the turnover time of the outer scale, $\tau_0$: 
\begin{equation}
\dot{M}_{\rm in,max} = \epsilon_{\rm in} M_0 / \tau_0 ,
\label{Mdot_max}
\end{equation}
where $\tau_0= L_0/ (2\,\sigma_{\rm v,0})$.
We further assume that $\epsilon_{\rm in}$ is a universal parameter, giving the maximum efficiency with which the largest turbulent motions can be channeled into
a single star\footnote{
The universality of $\epsilon_{\rm in}$ is predicated upon the universality of turbulence. The fundamental assumption of our scenario is that massive stars are assembled by 
converging flows in the turbulent velocity field, so the fraction of the total mass that accumulates at converging points should only depend on the universal statistics of supersonic turbulence.  
However, the independence of $\epsilon_{\rm in}$ on the final stellar mass, $M_{\rm f}$, is not an assumption, but a consequence of the linear nature of the lower envelope of the $t_{95}$--$M_{\rm f}$ scatter plot (see Figure~\ref{t95_v2}).
}. 
Though related to the star-formation efficiency (after one turnover time of the outer scale), $\epsilon_{\rm in}$ should be much smaller than that, 
because the stars that form with the largest infall rate are only a small fraction of all stars formed within a region of size $L_0$. The lower envelope of the 
$t_{95}$--$M_{\rm f}$ plot is then given by 
\begin{equation}
t_{\rm 95,min}=M_{\rm f}/(\epsilon_{\rm out}\dot{M}_{\rm in,max})= \tau_0 (M_{\rm f}/M_0)/(\epsilon_{\rm out} \epsilon_{\rm in}).
\label{t95_min_2}
\end{equation}
In the following, we estimate empirically the value of $\epsilon_{\rm in}$ (based on our simulation), so the lower envelope is fully determined 
by the properties of the outer scale of the turbulence in the star-forming region, namely $M_0$ and $\tau_0$, according to equation~(\ref{t95_min_2}).

Using $M_0 = 2\times 10^5 M_\odot$,  $\sigma_{\rm v,0} \simeq 4.6$~km\,s$^{-1}$, ${L_0} =70$ pc 
for the largest star forming regions in our simulation and $\dot{M}_{\rm in,max}= 4.5 \times 10^{-5} M_{\odot} {\rm yr}^{-1}$, we find that 
$\tau_0 \sim 7.4$~Myr and $\epsilon_{\rm in} = 1.7\times 10^{-3}$.  For the purpose of the subsequent analysis we will assume that this 
coefficient $\epsilon_{\rm in}$ is universal. 

The maximum infall rate can also be expressed in terms of the virial parameter. Equation~(\ref{Mdot_max}) then becomes 
\begin{equation}
\dot{M}_{\rm in,max} = \frac{5 \,\epsilon_{\rm in}}{3\,\alpha_{\rm vir}}\frac{\sigma_{\rm v,0}^3}{G} =2.906 \,\dot{M}_{\rm iso} \left(\frac{{\cal M}_0}{10}\right)^3 \, \alpha_{\rm vir}^{-1}, 
\label{Mdot_max_vir}
\end{equation}
where $\dot{M}_{\rm iso}=0.975\,c_{\rm s}^3/G$ is the accretion rate from the collapse of a critical isothermal sphere \citep{Shu+87}, and 
${\cal M}_0=\sigma_{\rm v,0}/c_{\rm s}$ is the rms Mach number of the turbulence. Thus, the maximum infall rate only depends on  
$\sigma_{\rm v,0}$ and $\alpha_{\rm vir}$. This expression explains the lower value of the maximum infall rate 
in the smaller-scale simulations of \citet{Haugbolle+18imf}, as the rms Mach number was ${\cal M}_0\approx10$, while here we
have ${\cal M}_0\approx 25$. It may be instructive to compare the maximum infall rate with that predicted for the collapse of an isothermal sphere, 
even though the maximum infall rate (as any other lower value of the infall rate) in our scenario is controlled by the {\it inflow} rate, which is the natural 
result of supersonic, gravitationally unbound turbulent flows. Assuming that the large-scale virial parameter is of order 
unity, as it may be the case for the most massive MC complexes, equation (\ref{Mdot_max_vir}) gives $\dot{M}_{\rm in,max} = \dot{M}_{\rm iso}$ 
only for the specific value ${\cal M}_0=7.0$, as a coincidence. For the Mach number of our simulation, ${\cal M}_0\approx 25$, $\dot{M}_{\rm in,max} \gg \dot{M}_{\rm iso}$.
If one adopts the rms velocity instead of the sound speed in the expression for $\dot{M}_{\rm iso}$, $\dot{M}_{\rm iso,turb}=0.975\,\sigma_{\rm v,0}^3/G$, under the assumption 
of turbulent support prior to the collapse, then $\dot{M}_{\rm in,max} \ll \dot{M}_{\rm iso}$ (by three orders of magnitude). More importantly, the value of $\dot{M}_{\rm in,max}$ 
found in our simulation cannot be interpreted as the result of the global collapse of MC complexes because we don't find any evidence of global collapse at large scales. 
In our simulation, the random component of the velocity is always much larger than the inflow velocity (see for example the middle-left panel of Figure~\ref{inflow_profiles}).

\subsection{The maximum stellar mass} \label{sec_mmax}

Based on our star-formation scenario, we can derive the maximum stellar mass set by the turbulence and its dependence on the physical conditions. 
In general, the maximum mass is determined by the intersections of three lines in  Figure~\ref{t95_v2}, namely, the upper envelope, the SN line 
(the curved green line), and the lower envelope. 
The SN line corresponds to the time of SN explosion as a function of stellar mass.  
We denote the stellar mass at the intersection of the lower envelope with the upper envelope 
as $M_{\rm up, low}$ and that at the intersection with the SN line as $M_{\rm SN, low}$. If the lower envelope crosses the upper envelope 
earlier than the SN line, the turbulent structure that hosts the formation of the most massive star becomes decorrelated, 
and the accretion process ends before the life time of the star, so that the maximum stellar mass, $M_{\rm f,max}$, 
is given by $M_{\rm up, low}$. Otherwise, $M_{\rm f,max} = M_{\rm SN, low}$. In other words,  we have $M_{\rm f,max} = \min (M_{\rm up, low}, M_{\rm SN, low})$.

$M_{\rm up, low}$ may be calculated by multiplying the maximum accretion rate with the turnover time at the outer scale of the turbulence 
(the upper envelope of the $t_{95}$--$M_{\rm f}$ plot), which, as mentioned earlier, is the maximum time during which the flow configuration 
remains coherent and correlated. The product is indeed the location where the lower and upper envelopes cross each other.  In our simulation, 
the turnover time of the largest eddies was estimated to be $\tau_0= L_0/ (2\,\sigma_{\rm v,0}) \sim 7.4$~Myr, and 
since the maximum infall rate in our simulation is $\simeq 4.5 \times 10^{-5} M_{\odot} {\rm yr}^{-1}$, 
we  have  $M_{\rm up, low} \simeq 163$~M$_{\odot}$ (accounting for the fact that the maximum accretion rate is half the maximum infall rate,
as $\dot{M}_{\rm max}=\epsilon_{\rm out}\dot{M}_{\rm in,max}$), consistent with the intersection of the upper and lower envelopes in Figure~\ref{t95_v2}. 

$M_{\rm SN, low}$ can be estimated by solving the implicit equation $M_{\rm SN, low} \simeq t_{\rm SN}\, \epsilon_{\rm out} \, \dot{M}_{\rm in,max}$, 
where the SN time,  $t_{\rm SN}$, depends on the mass of the star and $\dot{M}_{\rm in,max}$ is the maximum infall rate,
corresponding to the lower envelope of the $t_{95}$--$M_{\rm f}$ plot. The equation means that the stellar mass is set by how much gas the star may 
accrete during its lifetime. As discussed in \S~4, $M_{\rm SN, low}$ in our simulation is approximately 60~M$_{\odot}$. 
Since $M_{\rm SN, low}$ is significantly smaller than $M_{\rm up, low}$, the largest stellar mass in our simulation is  
$M_{\rm f,max}=M_{\rm SN, low} \simeq 60$~M$_{\odot}$. As shown below, under certain conditions $M_{\rm SN, low}$ may be larger than $M_{\rm up, low}$, 
in which case $M_{\rm f,max} =M_{\rm up, low}$.

\subsection{The maximum stellar mass in different environments} \label{sec_mmax_env}

An immediate implication of assuming the universality of $\epsilon_{\rm in}$ is that the stellar mass $M_{\rm up,low}$ at the intersection of the two envelopes 
is simply proportional to the total mass of gas available for star formation:
\begin{equation}
M_{\rm up,low} \simeq \epsilon_{\rm out} \, \dot{M}_{\rm in,max} \, \tau_0 \simeq 0.9\times 10^{-3} M_0,
\label{m_max_hp}
\end{equation}
where we have used Equation (\ref{Mdot_max}) with $\epsilon_{\rm in} = 1.7\times 10^{-3}$, and have also adopted $\epsilon_{\rm out}=0.5$ as in the simulation. 
Interestingly, this estimate depends only on the total mass $M_0$, and is independent of the turbulent rms velocity. In fact, an increase of the turbulent velocity 
would lower both the upper and lower envelopes of the $t_{95}$--$M_{\rm f}$ plot, decreasing them by the same factor. Thus, their intersection would occur at 
the same value of $M_{\rm f}$. 

The stellar mass, $M_{\rm SN,low}$, at which the lower envelope crosses the SN line,  can be obtained by solving,  
\begin{equation}
M_{\rm SN, low} =t_{\rm SN}  \, \epsilon_{\rm out} \, \dot{M}_{\rm in,max} =  \epsilon_{\rm out} \, \epsilon_{\rm in} \, M_0 \, t_{\rm SN}/\tau_0.
\label{maxmass}
\end{equation}
A comparison of Equations  (\ref{m_max_hp}) and (\ref{maxmass}) shows that the three curves intersect at the same point if  $t_{\rm SN} (M_{\rm up,low}) = \tau_0$. 
If the dynamical time is larger than the lifetime of a star of mass $M_{\rm up,low}$, $\tau_0 > t_{\rm SN} (M_{\rm up,low})$, then  $M_{\rm SN, low} <M_{\rm up,low}$ 
and the maximum stellar mass is set by $M_{\rm SN, low}$  from  Equation (\ref{maxmass}). Since $t_{\rm SN}$ is a decreasing function of the stellar mass, it is 
straightforward to see from Equation  (\ref{maxmass}) that $M_{\rm SN, low}$, hence $M_{\rm f,max}$, increases as $\tau_0$ decreases (with an increasing intensity 
of turbulence). Once $\tau_0$ drops below $\tau_{\rm SN} (M_{\rm up,low})$, $M_{\rm SN, low}$ exceeds $M_{\rm up,low}$
and $M_{\rm f,max} $ is set by $M_{\rm up,low}$ and stops increasing with decreasing $\tau_0$. 

In the case of regions where the turbulence outer scale contains a very large total mass, $M_0$, equation~(\ref{m_max_hp}) gives a very high value for $M_{\rm up,low}$, 
for example $M_{\rm up,low} \gtrsim 850$~M$_{\odot}$ for $M_0  \gtrsim 10^6$~M$_{\odot}$. However, a very large value of $M_0$ would usually imply that 
$M_{\rm SN, low} < M_{\rm up,low}$, so the maximum stellar mass would be determined by $M_{\rm SN, low}$, unless the rms velocity of the turbulence were large
enough to satisfy the condition $\sigma_{\rm v,0} \gtrsim 14\, \alpha_{\rm vir}^{1/3} (t_{\rm SN}/{\rm Myr})^{-1/3} (M_0/10^6 M_{\odot})^{1/3}$~km~s$^{-1}$, where the virial 
parameter is defined as $\alpha_{\rm vir} = 5 \sigma_{\rm v,0}^2 R\, /(3 \, G\,M_0)$, with $R \simeq L_0/2$ the radius of the star-forming region. 

To calculate  $M_{\rm SN, low}$ from Equation (\ref{maxmass}) as a function of the virial parameter we use Equation (\ref{Mdot_max_vir}).
To further simplify the solution to Equation (\ref{maxmass}), we fit the stellar lifetime, $t_{\rm SN}$, as a function of the stellar mass, $M_{\rm f}$, by a power-law, 
$t_{\rm SN} \simeq 30 (M_{\rm f}/M_\odot)^{-0.5}$ Myr, which is a reasonably good approximation for large stellar masses, with errors 
$\lesssim 6\%$, for $ M_{\rm f} \ge 30 M_\odot$. 

Solving Equation (\ref{maxmass}) then gives, 
\begin{equation}
M_{\rm SN, low} \simeq 4.7 \left(\frac{\epsilon_{\rm in}}{1.7\times 10^{-3}} \right)^{0.67} \left(\frac{\sigma_{\rm v,0}}{1 \hspace{1mm} {\rm km}  \hspace{1mm} {\rm s}^{-1}}\right)^2 \alpha_{\rm vir}^{-0.67} M_\odot. 
\label{finalmaxmass}
\end{equation}
In our simulation, we have ${\sigma_{\rm v,0}} = 4.6$ km s$^{-1}$ and $\alpha_{\rm vir} \simeq 1.4$ at the outer scale, $L_0 \simeq 70$~pc, so that  
$M_{\rm SN, low}$ from the above equation is $\simeq 80$~M$_\odot$, consistent with the intersection of the lower envelope and the line of the SN time in Figure~\ref{t95_v2},
and comparable to the actual maximum mass found in our simulation, $M_{\rm SN, low} \simeq 60$~M$_{\odot}$,  mentioned above.\footnote{
Equation (\ref{finalmaxmass}) is valid only for  $M_{\rm SN, low} \ge 30$~M$_\odot$, where the power-law fit for $t_{\rm SN} $ applies. If 
$M_{\rm SN, low}$ turns out to be smaller than 30~M$_\odot$, one needs to solve Equation~(\ref{maxmass}) using a more accurate function for $t_{\rm SN} $.} 

Equation~(\ref{finalmaxmass}) shows that, if the rms turbulent velocity is kept constant, $M_{\rm SN, low}$ decreases with increasing virial parameter, $\alpha_{\rm vir}$. 
Clearly, with $\sigma_{\rm v,0}$ fixed, a larger $\alpha_{\rm vir}$ corresponds to a smaller gas reservoir, i.e., smaller  $M_0$, and/or a larger size, $L_0$, 
of the star-forming region. A smaller $M_0$ would decrease the infall rate due to the lower mass supply, while a larger size, $L_0$, would increase the 
turbulent dynamical time, also leading to a decrease in the infall rate. In either case, the lower envelope of $t_{95}$--$M_{\rm f}$ plot would be lifted upward, 
resulting in a smaller value for  $M_{\rm SN, low}$. 

On the other hand, if the  virial parameter is fixed,  $M_{\rm SN, low}$ increases quite fast, $\propto \sigma_{\rm v,0}^2 $, 
with the turbulent velocity, $\sigma_{\rm v,0}$. There are two contributions to this effect. First, in regions with stronger turbulence, the dynamical time of turbulent motions
is shorter, which increases the infall rate. Second, at a fixed virial parameter, a larger turbulent velocity corresponds to a larger total gas mass, 
$M_0$, or a smaller size, $L_0$, of the region, further increasing the infall rate. Both contributions shift down the lower envelop of the 
$t_{95}$--$M_{\rm f}$ plot, moving its intersection with the SN line to larger stellar masses, increasing $M_{\rm SN, low}$. In other words, a higher accretion 
rate results into a more massive star before the star explodes as a SN. 

Due to the strong dependence of the star-formation rate on $\alpha_{\rm vir}$ \citep[e.g.][]{Padoan+12sfr,Padoan+14ppvi,Padoan+17sfr}, most star-forming regions
may be characterized by values of $\alpha_{\rm vir}$ close to unity, so their maximum stellar mass, $M_{\rm SN, low}$, would be determined primarily
by $\sigma_{\rm v,0}$ (in the regime where $M_{\rm SN, low} < M_{\rm up, low}$). 

\begin{figure}[t]
\includegraphics[width=\columnwidth]{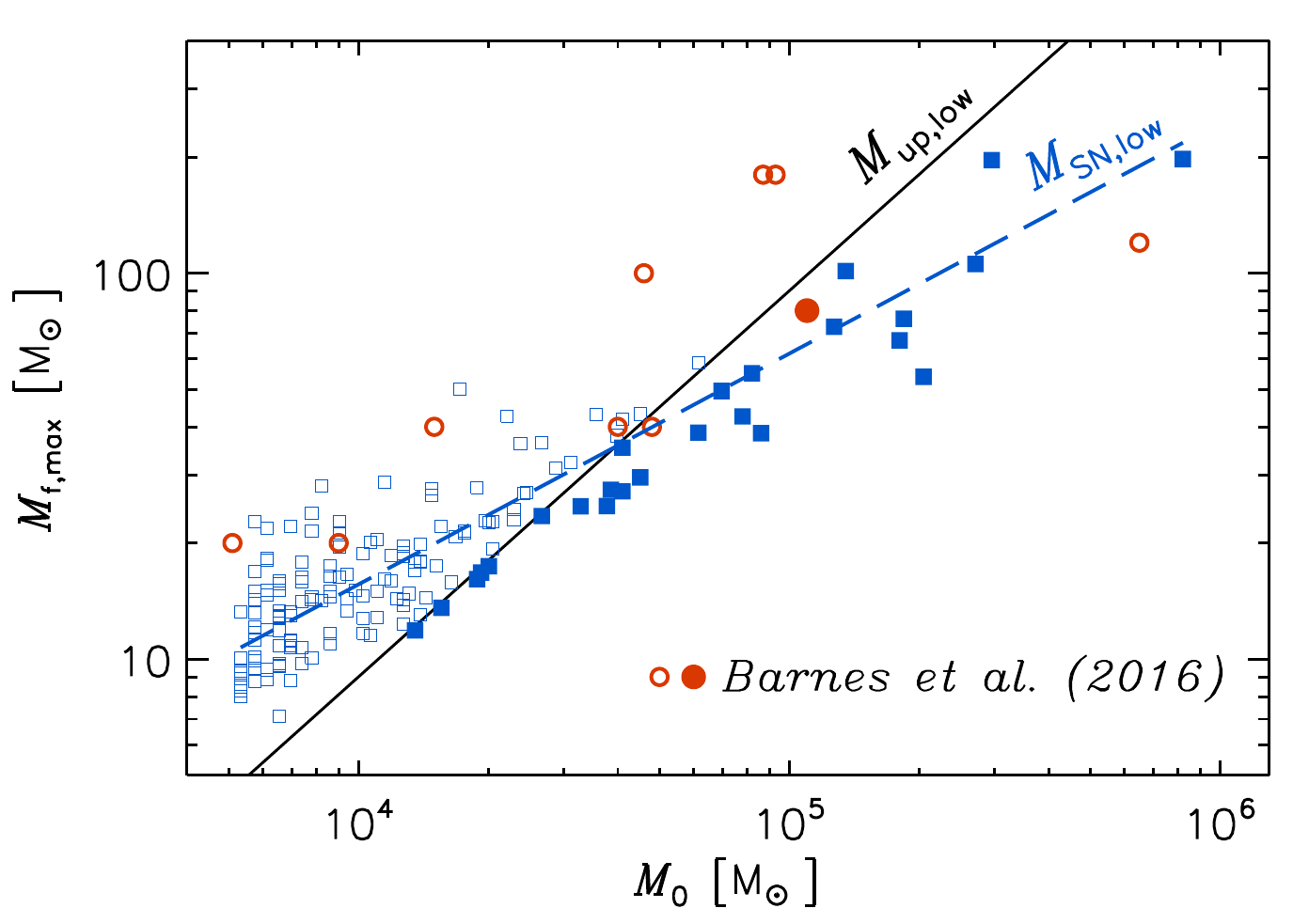}
\caption[]{Solid black line: Predicted maximum stellar mass, $M_{\rm f,max}=M_{\rm up,low}$ (under the condition $M_{\rm up,low} < M_{\rm SN, low}$), as a 
function of the total mass, $M_0$, from Equation~(\ref{m_max_hp}). Square symbols: $M_{\rm SN, low}$ from Equation~(\ref{finalmaxmass}) applied to the 
physical parameters of the MCs in the Outer Galaxy Survey \citep{Heyer+98}. Empty squares satisfy the condition $M_{\rm up,low} < M_{\rm SN, low}$, hence 
$M_{\rm f,max}=M_{\rm up,low}$ (the red solid line) for those clouds. The dashed line is a least-squares fit to all the square symbols, 
$M_{\rm SN, low}=15.6$~M$_{\odot}$~$(M_0/10^4$M$_{\odot})^{0.60}$. Circular symbols: maximum stellar masses in CMZ clouds from Table~4 of  
\citet{Barnes+17}.   
}
\label{barnes17}
\end{figure}

\subsection{The maximum stellar mass in MCs} \label{sec_mmax_mcs}

As an application of the predicted maximum stellar mass, we consider the MC catalog by \citet{Heyer+01}, extracted from a decomposition of the $^{12}$CO FCRAO Outer 
Galaxy Survey \citep{Heyer+98}, one of the largest Galactic MC samples to date. To limit the distance and mass uncertainties, \citet{Heyer+01} considered only MCs with 
circular velocity $<-20$ km s$^{-1}$, which yield a sample of 3901 clouds. Because we are interested in clouds that can host the formation of massive stars, we retain only 
MCs with mass $> 5\times10^3$ M$_{\odot}$, resulting in 157 MCs. Using the MC physical parameters derived by \citet{Heyer+01}, we compute $M_{\rm SN, low}$ for each 
MC based on Equation~(\ref{finalmaxmass}), which we show as blue square symbols in Figure~\ref{barnes17}. The dashed line is a least-squares fit through the data points, 
giving $M_{\rm SN, low}=15.6$~M$_{\odot}$~$(M_0/10^4$M$_{\odot})^{0.60}$. This shallower than linear dependence on $M_0$ is expected as a result of the velocity-size 
and mass-size relations of MCs. The black solid line is the predicted maximum mass, $M_{\rm up,low}$, as a function of the cloud mass, $M_0$, according to 
Equation~(\ref{m_max_hp}). The intersection of the dashed and solid lines shows that $M_{\rm up,low} < M_{\rm SN, low}$, hence $M_{\rm f,max}=M_{\rm up,low}$, for 
$M_0\lesssim 4\times 10^4$~M$_{\odot}$. Thus, the predicted maximum stellar mass in these low-mass clouds has a linear dependence on the cloud mass, as shown by the 
solid line in Figure~\ref{barnes17} (the values of $M_{\rm SN, low}$ for all clouds where $M_{\rm up,low} < M_{\rm SN, low}$ are represented by empty squares to stress
that their predicted maximum stellar mass is not $M_{\rm SN, low}$). 

As a second example, we consider the star-forming MCs in the central molecular zone (CMZ) of the Galaxy that are characterized by a very large velocity dispersion, while 
their virial parameter is not far from unity \citep[e.g.][]{Barnes+17}. With such parameters, the star-formation timescale is very short (e.g. significantly shorter than in our 
simulation) and thus the maximum stellar mass is not limited by $t_{\rm SN}$, but rather by the total cloud mass as in Equation~(\ref{m_max_hp}), that is 
$M_{\rm f,max}=M_{\rm up,low}$ and the maximum stellar mass we predict is that given by the solid line in Figure~\ref{barnes17}. In the specific case of the so-called 
``Brick" cloud, $\sigma_{\rm v,0}=6.8$~km~s$^{-1}$ and $\alpha_{\rm vir}=0.85$, based on the study by \citet{Federrath+16}, giving $M_{\rm SN, low}\simeq 727$~M$_{\odot}$. 
However, given the total estimated mass, $M_0=7.2\times 10^4$~M$_{\odot}$ \citep{Federrath+16}, Equation~(\ref{m_max_hp}) gives $M_{\rm up,low} \simeq 65$~M$_{\odot}$. 
Because  $M_{\rm up,low} \ll  M_{\rm SN, low}$, the predicted maximum stellar mass in the ``Brick" is $M_{\rm f,max}=M_{\rm up,low} \simeq 65$~M$_{\odot}$, not far from the
observational value of 80~M$_{\odot}$ estimated by \citet{Barnes+17}, and shown in Figure~\ref{barnes17} as a filled circle (in Barnes et al. the estimated cloud mass 
is $M_0=11\times 10^4$~M$_{\odot}$, which would increase our predicted maximum mass to 99~M$_{\odot}$). Estimated values of the maximum stellar mass for other CMZ
clouds derived by \citet{Barnes+17} assuming that the infrared luminosity is dominated by a single embedded star are shown in Figure~\ref{barnes17} as empty 
circles. In the case of Sgr B2, the most massive cloud in \citet{Barnes+17}, the predicted mass is much lower than our predicted value, $M_{\rm up,low}$. However, the estimated
value of $M_0$ is highly uncertain, as it depends on the choice of the cloud contour. Most of the star formation in Sgr B2 is concentrated in three dense cores with size of order 1~pc
and masses of order $10^5$~M$_{\odot}$  \citep[e.g.][]{Schmiedeke+16}, which would bring the estimated maximum stellar mass very close to $M_{\rm up,low}$. Furthermore, the 
value of $M_0$ is uncertain also because in our scenario it should be the mass contained in the outer scale of the turbulence, of which the cloud mass is only a very rough 
approximation at best.

The actual maximum stellar mass in a star-forming region is of course affected by stellar radiation and winds, which are not modelled in our prediction 
(except for a constant $\epsilon_{\rm out}$). Nevertheless, it is useful to determine the upper bound to the stellar mass set by the turbulence alone, as discussed above.

\section{Prestellar Cores with Herschel and ALMA} \label{sec_observations}

Prestellar cores are notoriously difficult to observe in regions of massive-star formation, due to the large distances, the low galactic
latitudes, the large column densities, and the high background luminosities. As a result, their observational properties may strongly depend on
spatial resolution and on data-analysis procedures. Although we cannot reproduce the full complexity of the characteristic Galactic 
environment of regions of massive-star formation within our 250 pc volume, we can nevertheless study the basic effects of spatial resolution by positioning 
our simulation at different distances and by simulating Herschel and ALMA observations with very different angular beam sizes. We can also address the 
effect of line-of-sight projection by comparing core masses derived from different directions.

\subsection{Synthetic Observations}

\begin{figure}[t]
\includegraphics[width=\columnwidth]{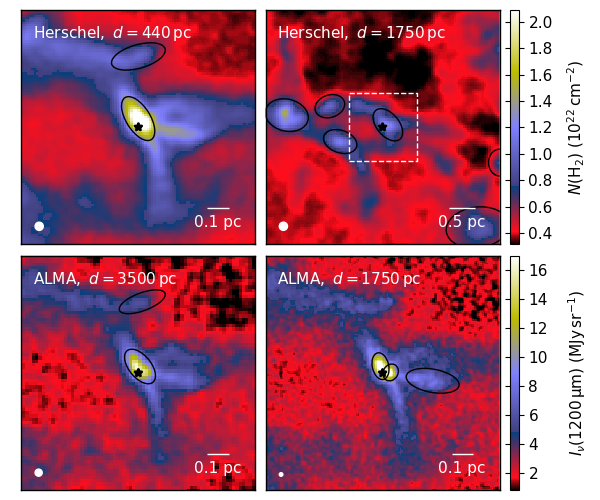}
\includegraphics[width=\columnwidth]{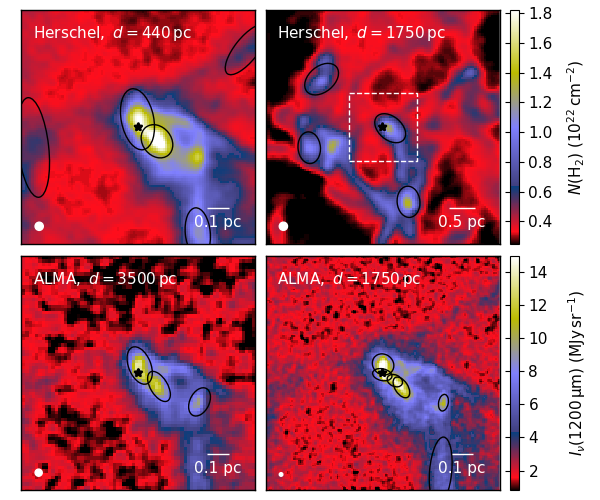}
\caption[]{Herschel and ALMA synthetic observations centered around the progenitors of two stars that will achieve final masses of 14.2~M$_{\odot}$ 
(top four panels) and 32.6~M$_{\odot}$ (bottom four panels). The right column compares Herschel and ALMA at the same distance of 1750~pc. The ALMA 
maps have a size of 1~pc and their perimeter is indicated by the dashed square in the larger Herschel maps. The left column compares Herschel and 
ALMA at different distances, such that the Herschel effective beam size of 18.2'' corresponds to a physical scale only slightly larger than that of the 
ALMA beam size of 2'' (all maps have a size of 1~pc). For Herschel we show the column density map, for ALMA we show the surface birghtness at 
1200~$\mu$m. The ellipses correspond to the {\em getsources} cores that satisfy all the selection criteria. The top four panels are meant to represent 
a case where the increase in angular resolution from Herschel to ALMA results in a mild fragmentation of the progenitor core (from one to two cores), 
while the bottom four panels show a case where a single Herschel progenitor core is broken into four cores at the ALMA resolution.   
}
\label{map_synthetic}
\end{figure}

We are mostly interested in the observational properties of the prestellar cores from our simulation characterized in previous sections,
so we center the observations around the positions of our sink particles at birth. Because of the computational cost, we focus on a small 
subset of 38 sink particles, 17 selected randomly among those with $M_{\rm s,f} > 40$~M$_{\odot}$, and 21 among lower mass ones. 
However, we do not discriminate by final sink mass as it does not seem to play a role in the qualitative results of this section. 

We compute synthetic surface brightness maps with radiative transfer calculations (for details see Appendix~\ref{appendix:synthetic}). The dust model is
adopted from \citet{Compiegne2011} and corresponds to the average dust properties in the Milky Way. However, following observational results on dense cores,
the submillimeter dust opacity is increased to $\tau$(250\,$\mu$m)/$\tau$(K)=$1.6\times 10^{-3}$ by scaling the dust absorption cross sections with a
constant factor for all $\lambda>30\mu$m \citep{Juvela+15}. We compute the dust emission for columns of $10\times10\times250$ pc$^3$ that are centered
around the birth position of each sink particle and are illuminated by the normal interstellar radiation field \citet{Mathis1983}. Three such columns (one
per coordinate direction) are considered for each sink particle, each giving a synthetic map of $10\times10$~pc$^2$ with the line-of-sight direction along
the length of the column.

To simulate Herschel observations, we calculate synthetic surface brightness maps at 160, 250, 350, and 500\,$\mu$m with beam sizes 11.7, 18.2, 24.9,
and 36.3 arcsec and observational noise of magnitude 3.7, 1.2, 0.85, and 0.35\,MJy\,sr$^{-1}$ per beam, for the four bands, respectively. The noise values are typical of
Herschel observations (statistical errors without calibration errors). Simulations are repeated for 440\,pc, 875\,pc, and 1750\,pc cloud distances,
the last one allowing direct comparison with the NGC~6334 region \citep{Tige+17}.

The synthetic ALMA observations are calculated for the 1.2\,mm wavelength, 2$\arcsec$ beam size, and the distances of 1750\,pc, 3500\,pc, and 7000\,pc. The
resolution is partly dictated by the finite resolution of our models (the cell size 0.0076\,pc corresponds to 0.9$\arcsec$ at the distance of 1750\,pc). We
add to the maps noise of magnitude 0.37\,MJy\,sr$^{-1}$, which is achievable in actual ALMA observations \citep[e.g.][]{Beuther+19} and gives similar signal-to-noise
ratios as for example in \citet{Motte+18}, where, unlike in our simulations of the prestellar phase, the signal is boosted by local heating.

We use the {\em getsources} software \citep{Menshchikov+12} to detect cores and compute their properties (see Sect.~\ref{appendix:synthetic}). In the case
of Herschel observations, we follow closely the procedures used in the study of NGC6334 by \citet{Tige+17}, including the requirements of reliable
detections in at least three bands and a good SED fit. The core mass estimates use the temperature obtained from the modified blackbody fit of the SED. In
the case of ALMA cores, the basic selection criteria are only applied to the single 1.2\,mm band and the masses are calculated assuming 10\,K temperature.

\begin{figure}[t]
\includegraphics[width=\columnwidth]{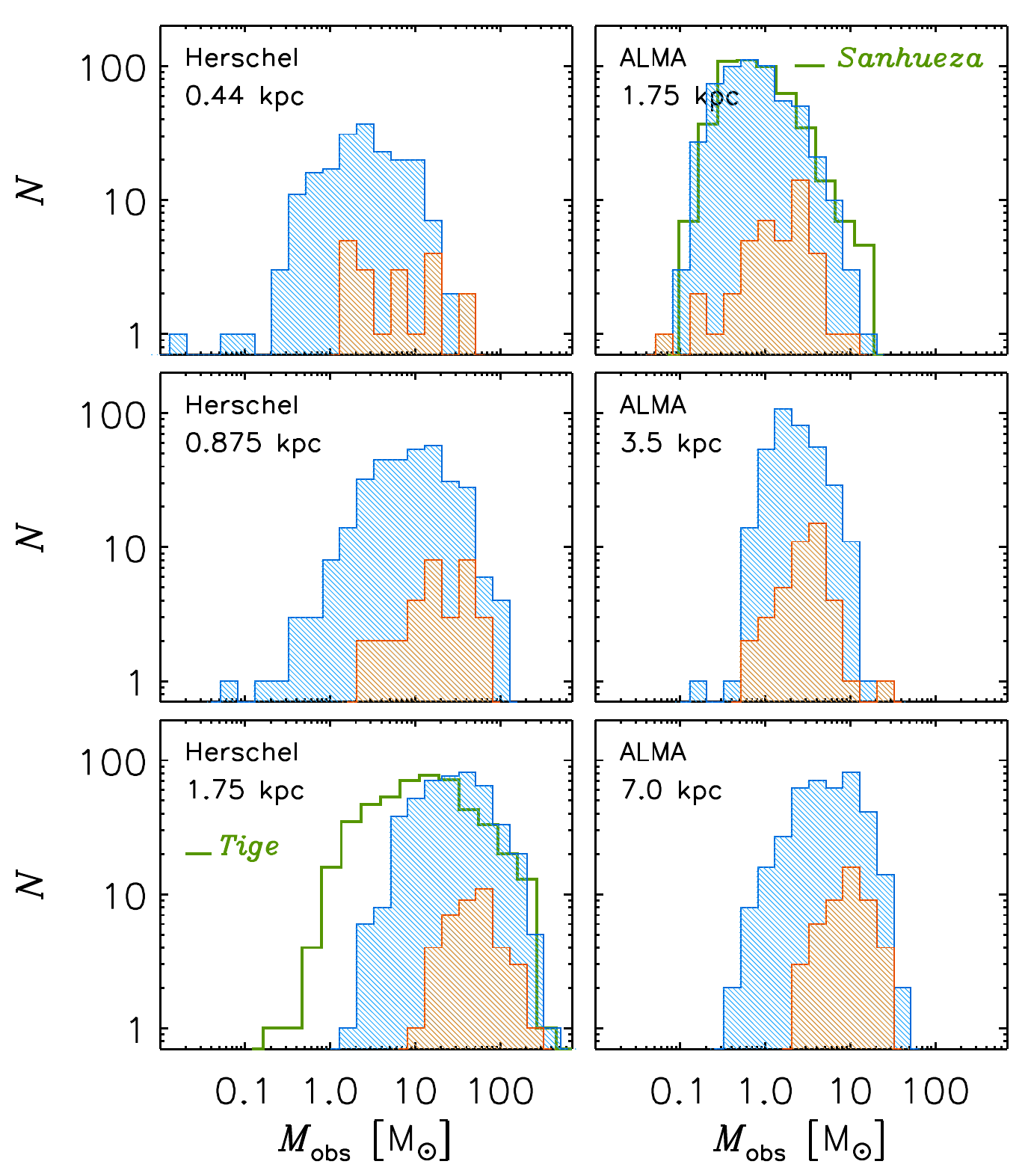}
\caption[]{Mass distribution of {\em getsources} cores selected from the Herschel maps (left column) and ALMA maps (right column), at different distances. Blue
histograms are for the cores found over the whole extension of the map, while red histograms are based only on the cores that contain the sink particle position 
(in the map center). The histograms clearly shift towards larger masses as the distance increases (top to bottom panels). 
The green, unshaded histograms show the prestellar-core mass distributions from \citet{Tige+17} and \citet{Sanhueza+19} in the panels corresponding
approximately to the linear resolution of those interferometric observations.
}
\label{mass_synthetic_histo}
\end{figure}
\begin{figure*}[t]
\centering
\includegraphics[width=\textwidth]{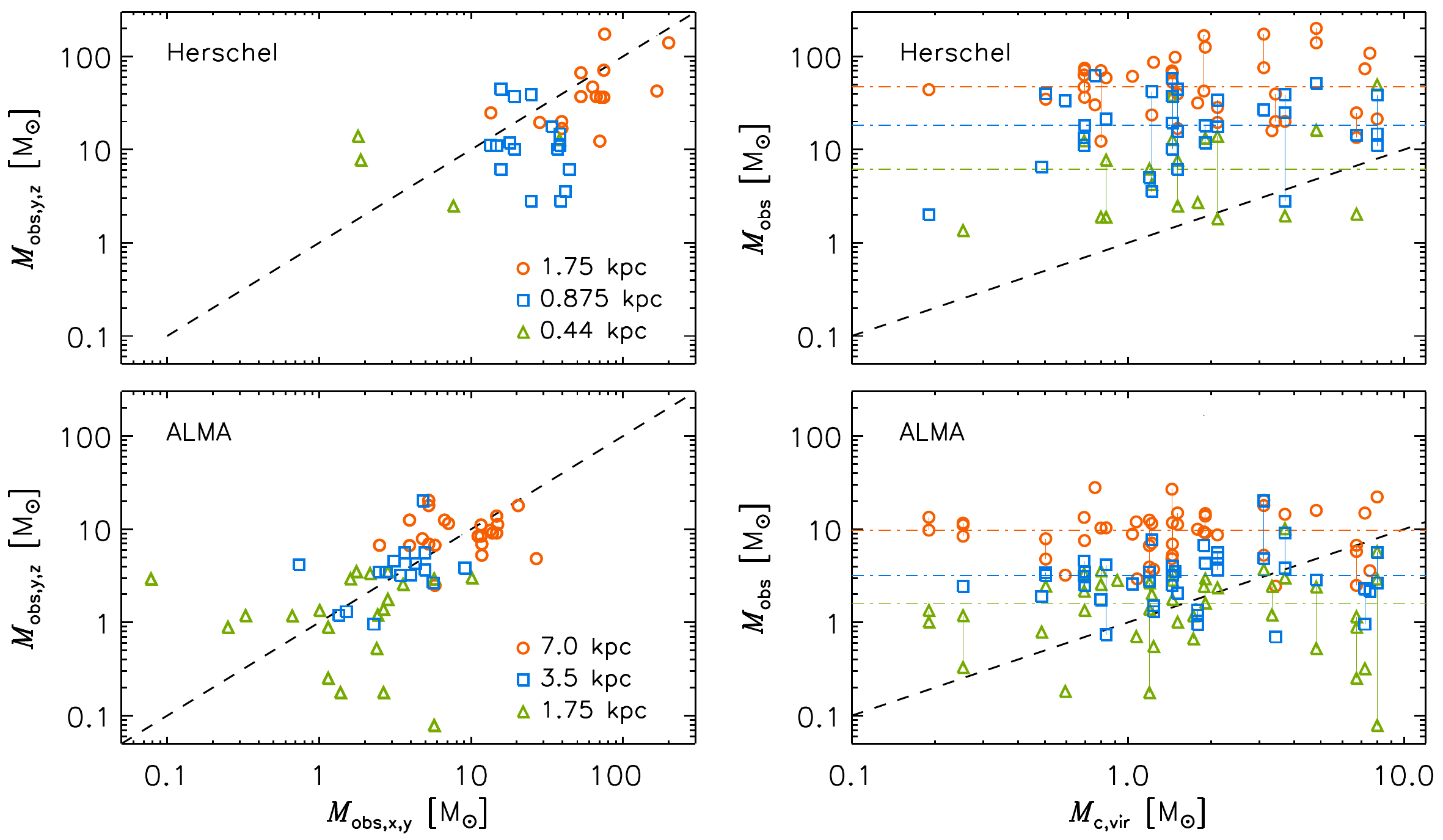}
\caption[]{Effect of line-of-sight direction on the value of the observed core mass (left panels), and comparison of observed versus true core masses
(right panels). Only a fraction of the true cores is detected (see Fig.~\ref{detections}) and thus shown in the right panel, and only a fraction of those 
is detected in more than one direction and thus shown in the left panel. The vertical segments in the right panel join values of the observed mass 
of true cores that are detected in more than one direction. The dashed-dotted horizontal lines in the right panels correspond to the median core mass
at each distance. Observed core masses can differ by more than a factor of 10 for different directions and do not show any correlation with the true
core masses.       
}
\label{mass_synthetic_real}
\end{figure*}

\subsection{Core Masses from Synthetic Observations}

With the birth positions of 38 sink particles, three directions of observation, and three simulated distances, we apply {\em getsources} to 342 
maps (for each instrument), resulting in the selection of hundreds of cores. Among those hundreds of cores, there may be up to 342 prestellar 
cores defined as those containing the sink particles (the central position of each map), though in practice only a fraction of the sink-progenitor cores are detected.

Fig.~\ref{map_synthetic} shows examples of such synthetic observations for the maps around two sink particles with final masses of 
14.2~M$_{\odot}$ (top four panels) and 32.6~M$_{\odot}$ (bottom four panels). In the right column of panels, we compare Herschel 
and ALMA at the same distance of 1750~pc, showing maps with a size of 1~pc for ALMA, and approximately three times larger for
Herschel (the perimeter of the smaller ALMA maps is shown as a dashed white square in the Herschel maps). In the left column of panels, 
we show maps, all with a size of 1~pc, comparing Herschel and ALMA at different distances, such that the spatial resolution of Herschel 
(with an effective beam size of 18.2'') is comparable to that of ALMA (with and angular beam size of 2''). The maps also show the ellipses
corresponding to the {\em getsources} cores that satisfy all the criteria mentioned above.  

These two cases are meant to illustrate different amount of fragmentation with increasing spatial resolution. One can see from the two 
upper panels of the right column that one sink-progenitor core is detected with Herschel at the distance of 1750~pc, and that core is broken into
two pieces in the ALMA observations at the same distance. The two lower panels of the right column show that the sink-progenitor core detected by Herschel
results into four cores when observed with ALMA at the same distance. 

The mass distributions of the selected {\em getsources} cores at different distances are shown in Figure~\ref{mass_synthetic_histo} for both Herschel 
(left column) and ALMA (right column). The blue histograms include cores selected everywhere on the maps, while the red histograms include only
the sink-progenitor cores (those that contain the central position of each map). One can clearly see that the histograms of both the general core sample 
and the sink-progenitor cores shift towards larger masses as the distance increases. We have not tailored the simulation to represent a 
specific region, and our core sample is the result of the superposition of many different maps, separated from each other in time and space across 
the simulation. However, for the Herschel observations we have followed the same core-extraction procedure as in the study of NGC6334 by \citet{Tige+17},
and our reference distance of 1750~pc is the same as the distance to that star-forming region, which justifies a comparison of our core sample with that
from NGC6334. The {\em getsources} cores in NGC6334 have a mass distribution between approximately 0.3 and 300~M$_{\odot}$, with three outliers 
around 1000~M$_{\odot}$, and peaks at approximately 15~M$_{\odot}$ (see the unshaded histogram in the bottom-left panel of Figure~\ref{mass_synthetic_histo}). 
Our Herschel mass distribution at a distance of 1750~pc is not very different, with
a range between approximately 1 and 500~M$_{\odot}$ and a peak around 40~M$_{\odot}$. The slightly larger values are expected because we have not
excluded cores larger than 0.3~pc as in \citet{Tige+17}, and some of our cores are a bit larger than that. 

The ALMA mass distributions shown in the right column of Fig.~\ref{mass_synthetic_histo} are also quite similar to prestellar-core mass distributions derived 
from recent interferometric observations of infrared-dark clumps, for similar angular resolutions and distances \citep[e.g.][]{Sanhueza+19,Li+19,Servajean+19}.
A proper comparison of our synthetic observations with those surveys would require a re-analysis of our synthetic maps using the same methods of
core extractions (dendograms, graphs, etc.) as in those works, like we did for the comparison with \citet{Tige+17} using {\em getsources}. Nevertheless, 
in the top-right panel of Figure~\ref{mass_synthetic_histo}, we show (unshaded histogram) the mass distribution of the prestellar core candidates from the recent ALMA 
survey by \citet{Sanhueza+19}, where the angular resolution is $\sim 1.2''$ and the average clump distance $\sim 4$~kpc. The spatial resolution of that 
survey is nearly the same as in our ALMA synthetic observations at 1.75~kpc, and their core mass distribution is nearly indistinguishable from ours.  
A  further discussion of the most recent interferometric studies is given in \S~\ref{sec_obs}.

\begin{figure}[t]
\includegraphics[width=\columnwidth]{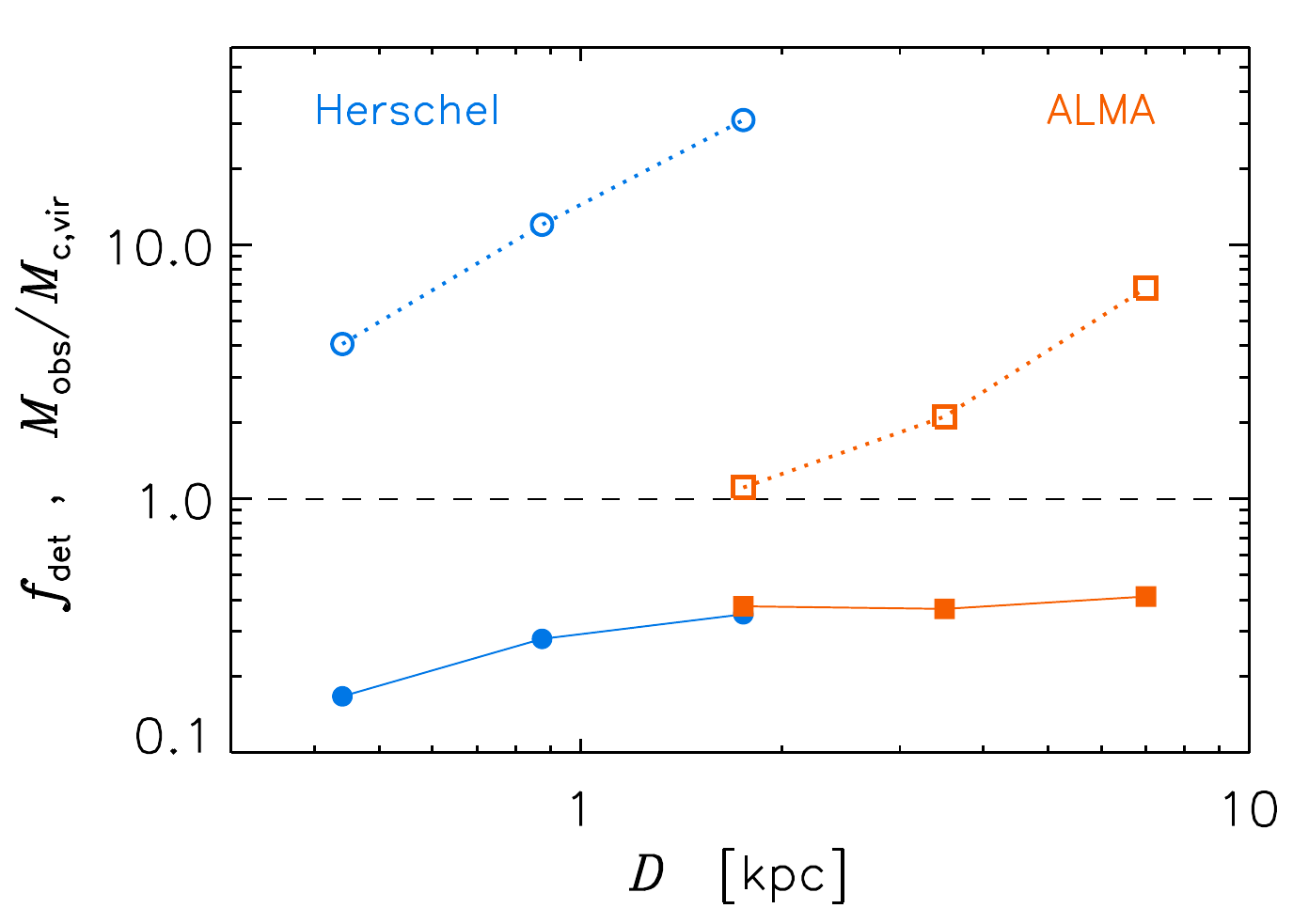}
\caption[]{Observed core masses divided by true core masses (open-symbol plots above the dashed line) and number of detected cores divided
by the total number of true cores, $f_{\rm det}$ (filled-symbol plots below the dashed line) for the progenitor cores, plotted against the distance. 
The masses are the median for each distance, divided by the median mass of the corresponding true cores. The observed masses clearly increase
with increasing distance or increasing beam size. In the ALMA case (right three points at larger distances), approximately 40\% of progenitor cores
are detected, while for Herschel the fraction drops to less than 20\% at the closest distance. The observed core masses are always larger than the
true core masses.  
}
\label{detections}
\end{figure}

The dependence of the core masses on spatial resolution (or distance) implies that many of the {\em getsources} cores are artifacts due to projection
effects or lack of resolution. We can further test this by focusing on the cores that contain the birth positions of the sink particles, because those can be
compared with the corresponding true prestellar cores identified in the simulation. Furthermore, because these sink-progenitor cores should contain the
3D position of the sinks, we can also compare their estimated masses when observed from different directions. The results of these comparisons are 
shown in Fig.~\ref{mass_synthetic_real}, where the left panels plot the observed mass in one direction versus the mass in a different direction (for those
few cores that are detected in at least two orthogonal directions), and the right panels plot the observed mass (from all sink particles and directions where
the cores are detected) versus the true core mass from the simulation. The left panels show that the observed core masses may depend strongly on the 
direction of observations, with differences that can be more than an order of magnitude. It should be stressed that this comparison is possible only when 
cores are selected in more than one direction, so this small subsample is biased towards the most favorable cases where the cores are real 3D entities
(e.g. a filament along the line of sight may appear as a core in that direction, but will not have a counterpart in the other two orthogonal directions), so the
actual uncertainty in the observed mass of the general core population must be even larger than suggested by these plots. Indeed, the right panels of
Fig.~\ref{mass_synthetic_real}, show that there is no clear correlation between the observed mass and the true core mass, with differences between 
the two  of up to two orders of magnitude. 

This artificial nature of the {\em getsources} cores was partly to be expected based on the strong distance dependence of the mass distributions
illustrated in Fig.~\ref{mass_synthetic_histo}. We summarize this mass dependence and the comparison with the true core masses by plotting 
the ratio of the median observed mass to the median true mass for cores at any given distance, as shown by the empty-symbol plots above the 
dashed line in Fig.~\ref{detections}. The median mass grows by approximately a factor of three for every factor of two increase in distance or 
telescope-beam size, and it is always larger than the true mass. In our reference Herschel case at 1750~pc, comparable to NGC6334, the median 
observed mass is more than 30 times larger than the true median mass. Only at the closest of the three ALMA distances the median value approaches
the true one, and in fact the distance dependence at the ALMA resolution seems to be converging to the true value, though this cannot be really
concluded with only three points.\footnote{\citet{Juvela+19} carried out similar synthetic observations of Planck cold clumps. As in this work,
the masses and sizes of extracted sources were found to depend not only on the distance, but also on the amount of line-of-sight confusion.}  

The plots below the dashed line in  Fig.~\ref{detections} give the ratio of the number of detected 
sink-progenitor cores to the total number of true prestellar cores (multiplied by three to account for the three directions of observations), $f_{\rm det}$, 
at each distance. Approximately 40\% of progenitor cores are detected in the ALMA maps, while for Herschel the fraction drops to less than 
20\% at the closest distance.

\section{DIscussion} \label{sec_discussion}

\subsection{Related Works}

Some aspects of our inertial-inflow scenario, like the filamentary nature of the inflow or the large-size of the mass reservoir, have been proposed in previous 
works to interpret numerical results or observational data. However, the fundamental idea of our scenario, the inertial nature of the large-scale inflow as a
result of supersonic turbulence, is very different from those earlier proposals, where the mass reservoir and/or the dense filaments are controlled by gravity, 
essentially through the collapse of pc-size clumps and/or the gravitational instability of dense filaments. In our case, the (spherically-averaged, radial) inflow velocity is only a fraction of the 
random velocity field, while in previous scenarios the infall motion must be dominant. 

\citet{Bonnell+2004} 
studied the formation of massive stars with a smoothed-particle-hydrodynamics (SPH) simulation of a 1-pc region, with a mass of 1,000~M$_{\odot}$, where 
the turbulence is initialized as a random velocity field and left to decay over time. They find that the progenitors of massive stars are cores with mass less
than 1~M$_{\odot}$, so most of the stellar mass is accreted from the larger-scale clump. Later SPH star-formation simulations on larger scales, 
10~pc with $10^4$~M$_{\odot}$, confirmed that most of the mass that accretes onto massive stars originates from the collapse of pc-size clumps 
\citep{Smith+09}. The results from these SPH simulations were interpreted as due to competitive accretion, although no direct comparison with 
competitive-accretion models (for example the predicted time-evolution of the accretion rate) was presented. However, it is indeed possible that competitive 
accretion plays an important role in those simulations due to the numerical setup, as the turbulence is left to decay causing a rapid drop in the virial parameter
of the clumps. 

\citet{Wang+10} followed the formation of a small number of massive stars in a clump of 1,215~M$_{\odot}$ with an MHD simulation in a computational box of 2~pc, 
driving the turbulence with stellar outflows. They find that the growth of the massive stars is fed by the collapse of the pc-scale clump, so it is neither driven by the 
stellar gravity, contrary to the competitive-accretion model, nor due to the collapse of 0.1-pc cores, in contrast with the core-collapse model. However, their conclusions 
that the mass feeding is regulated by stellar outflows and that the building blocks of massive stars are pc-scale clumps are evidently a direct consequence of their 
numerical setup: a pc-scale box driven only by stellar outflows; their work could not assess the role of larger scales or different driving forces. 

The apparent lack of massive prestellar cores in regions of massive star formation, demonstrated with statistical significance in the case of the Herschel survey of 
NGC 6334, led \citet{Tige+17} to propose a scenario where protostars grow into massive stars through infall and accretion from a pc-scale filamentary mass reservoir
\citep[see also][]{Motte+18}, 
similar to the inertial-inflow scenario proposed in this work. They concluded that a high-mass prestellar phase may not be necessary for high-mass star formation, consistent
with the numerical works mentioned above and the evidence from our multi-scale simulation. However, in this observational scenario, the large-scale reservoir is assumed 
to undergo a global free-fall collapse, consistent with recent proposals, based on numerical simulations, that MCs are in a general state of free-fall collapse 
\citep{Ibanez-Mejia+16,Vazquez-Semadeni+17,Vazquez-Semadeni+19}. A global free-fall collapse is not necessary in our inertial-inflow model, as compressive motions 
on all scales are naturally present in supersonic turbulence even in the absence of gravity. 

Furthermore, our simulation demonstrates that SN-driven turbulence results in
MCs that are mostly transient or lightly bound, with all measured properties, including their star formation rate, consistent with the observations. The  SPH simulations 
mentioned above \citep{Bonnell+2004,Smith+09}, where clumps are in global collapse, did not include any turbulence driving, and the hierarchical collapse model for MCs 
simply neglects the existence of SN-feedback in both the simulations \citep{Vazquez-Semadeni+17} and the theoretical modeling \citep{Vazquez-Semadeni+19}. 
Galactic-fountain simulations including SN-driving that fails to reproduce the observed MC turbulence \citep[e.g.][]{Ibanez-Mejia+16} are most likely affected by insufficient 
numerical resolution and unrealistic positioning of SNe. Our simulation has a much higher spatial resolution than Galactic-fountain simulations, yields a realistic star-formation 
rate, both globally and in MCs, and resolves the formation of individual massive stars, hence the timing and position of the SNe, relative to their parent clouds, is correctly 
described for the first time. Under these realistic conditions, we find that MCs are readily dispersed by SNe, so SN feedback should be a fundamental ingredient in MC models.

Although filaments are not essential to our scenario, they naturally arise in supersonic turbulence from the intersection of postshock sheets. Because the intersection of
filaments results in the formation of dense cores, the inflow motion feeding the growth of massive stars is usually channeled through dense filaments. The ubiquity of filaments in 
star-forming clouds has led to the proposal that prestellar cores are the result of the gravitational instability of dense filaments \citep[e.g.][]{Andre+14,Andre+16}. 
This idea is fundamentally different from the scenario we propose here, where the gravitational instability in a section of a dense filament is induced by a dynamical process, 
the convergence of multiple shocks in the turbulent flow, which feeds that section with more mass. The intersections of filaments are regions where the flow dissipates as the 
density is enhanced, so gravity takes over locally, causing the collapse of individual prestellar cores, but gravity is not the trigger for the formation of the cores. 
Furthermore, the stellar mass is not limited by the core mass, as the mass inflow at the filament intersection may continue well after a core has collapsed.

\subsection{Interferometric Studies of IR-Dark Massive Clumps} \label{sec_obs}

The lack of a high-mass prestellar phase in high-mass star formation suggested by \citet{Tige+17} is also consistent with recent interferometric studies of massive, 
IR-dark clumps. These studies usually reveal the presence of dense cores of relatively low mass, often with the most massive ones showing signs of protostellar activities.
The selection of IR-dark clumps already illustrates the scarcity of purely prestellar regions. \citet{Guzman+15} found that only 83 of the 3246 clumps 
from the Millimetre Astronomy Legacy Team 90 GHz Survey \citep[MALT90;][]{Foster+11,Jackson+13,Foster+13} are IR dark from 3.6 to 70~$\mu$m, and thus 
potentially hosting only prestellar cores. In the following, we briefly discuss a few of the most recent interferometric studies where the masses of all detected prestellar 
cores are reported (rather than focusing only on the most massive ones), so the findings have a degree of statistical significance and may be broadly compared with 
our numerical results. 

The ASHES survey \citep{Sanhueza+19} is a systematic ALMA study of IR-dark massive clumps at an angular resolution of $\sim 1.2''$. In their pilot survey of 
12 clumps, \citet{Sanhueza+19} found 210 prestellar core candidates, with masses between 0.1~M$_{\odot}$ and 11~M$_{\odot}$. Given the characteristic 
distance of the clumps (4~kpc) and the angular resolution of the survey, these mass estimates should be compared with our ALMA synthetic observations at the 
smallest distance. This comparison is shown in the top-right panel of Figure~\ref{mass_synthetic_histo}, where the mass distribution of the prestellar core candidates 
from the ALMA survey (unshaded histogram) is nearly identical to the mass distribution from the synthetic observations. In both distributions, the core masses, are 
approximately between 0.1~M$_{\odot}$ and 15~M$_{\odot}$. As shown in Figure~\ref{detections}, for this spatial resolution we expect the average value of
the observed masses to be close to the average value of the real core masses, although with large errors for the mass of each individual core
(see Figure~\ref{mass_synthetic_real}). In \citet{Sanhueza+19}, the core masses are of the order of the clump 
Jeans masses on average, consistent with our results in \S~\ref{core_properties}\footnote{
The ratio of core mass and clump Jeans mass has an average value of 0.6 (using all the masses 
of the prestellar core candidates from the electronic table provided by the authors). However, 
to compare correctly the Jeans mass with the critical BE mass, one should use the gas density just outside of the cores, 
which is usually larger than the mean clump density, hence the Jeans mass would be reduced the a value closer to the mean core mass.
}.
In an earlier Submillimeter-Array (SMA) study of another IR-dark clump, \citet{Sanhueza+17} had 
identified five potential prestellar cores, all less massive than 15~M$_{\odot}$. 

\citet{Li+19} mapped seven IR-dark massive clumps with the SMA and found relatively low mass cores as well, with masses between 1.4~M$_{\odot}$ and 38~M$_{\odot}$ 
(excluding the cores associated with the detected outflows). Given the clump distances and the SMA angular resolution in this work, the spatial resolution is approximately 
three times lower than in \citet{Sanhueza+19}. Thus, based on our results in the previous section, there are indications that the core masses are significantly overestimated  
on average. Our ALMA synthetic observations at the largest distance of 7~kpc have comparable spatial resolution and give masses of prestellar cores in the 
approximate range  2-30~M$_{\odot}$ (see Figure~\ref{mass_synthetic_histo}), consistent with those from the SMA observations. As shown in our Figure~\ref{detections},
the derived masses in that case are on average seven times larger than the masses derived at a distance of 1.75~kpc. \citet{Li+19} found that the core masses are a few 
times larger than the Jeans mass, but they could be of the order of the Jeans mass if the core masses were overestimated as suggested by our synthetic observations.   

\citet{Servajean+19} carried out an ALMA survey of a massive dark clump at an angular resolution of $\sim 2''$, detecting 12 dense cores with masses between 
3 and 50~M$_{\odot}$. Given the angular resolution and the distance of 3.5~kpc of the clump, this survey could be related to the intermediate-distance case of 
our ALMA synthetic observations. As shown in our Figure~\ref{detections}, we expect that the derived masses should be approximately twice larger than the 
real core masses on average, for that distance and angular resolution. The median core mass in the sample of 12 cores is approximately 12~M$_{\odot}$,
and the estimated Jeans mass in the clump is 3.5~M$_{\odot}$. If the masses were indeed overestimated by a factor of two on average, the median core mass 
would be less than a factor of two larger than the mean Jeans mass. The observed core linewidths are highly supersonic, in the range 1.9 to 3.1 km~s$^{-1}$,
also suggesting that the cores are likely to be further fragmented if observed at higher resolution. 

\citet{Kong19cmf} surveyed an infrared dark cloud with ALMA with an angular resolution of $\sim 0.5''$, finding between 197 and 280 cores, using either a dendogram
or a graph method, respectively. The core masses have values between approximately 0.2~M$_{\odot}$ and 30~M$_{\odot}$, assuming a constant temperature 
of 20~K, and somewhat larger using the ammonia-based median temperature of 13.3~K. However, it is not clear what fraction of these cores may be considered as
prestellar candidates, and this fraction may be rather small, considering the large number of molecular outflows detected in that region by \citet{Kong+19outflows}.
Similarly, \citet{Pillai+19} identified nine dense cores in their SMA survey of two massive IR-dark clumps, with masses ranging from 1.3 to 33~M$_{\odot}$, but found 
that they are mostly associated with outflows. Finally, \citet{Svoboda+19} carried out an ALMA survey of 12 massive dark clumps at a resolution of $\sim 0.8''$ and detected 
53 cores without associated outflows (67 in total), but they did not provide the estimated mass of their prestellar core candidates, except in the case of the two most massive 
ones, 14~M$_{\odot}$ and 29~M$_{\odot}$. 

Based on these recent interferometric studies, one might conclude that, if massive prestellar cores are not present, these regions will simply not form massive stars.
However, the estimated core masses are consistent with the ones found in our simulation (or at least with the overestimated masses derived from the corresponding 
synthetic observations), so, from the perspective of our inertial-inflow scenario, some of these cores may be genuine progenitors of massive stars. Furthermore, these 
interferometric cores are embedded in massive clumps that are the IR-dark counterpart of a much larger population of similar massive clumps where massive stars are 
known to be formed. As in their more active counterparts, the total mass reservoir in the dark clumps is more than sufficient to feed the growth of some of the cores to
the point of supporting the formation of massive stars, as also suggested by the mounting evidence of infall motion in the cores and along the filaments connected to 
the cores, briefly summarized in the next subsection.

\begin{figure}[t]
\includegraphics[width=\columnwidth]{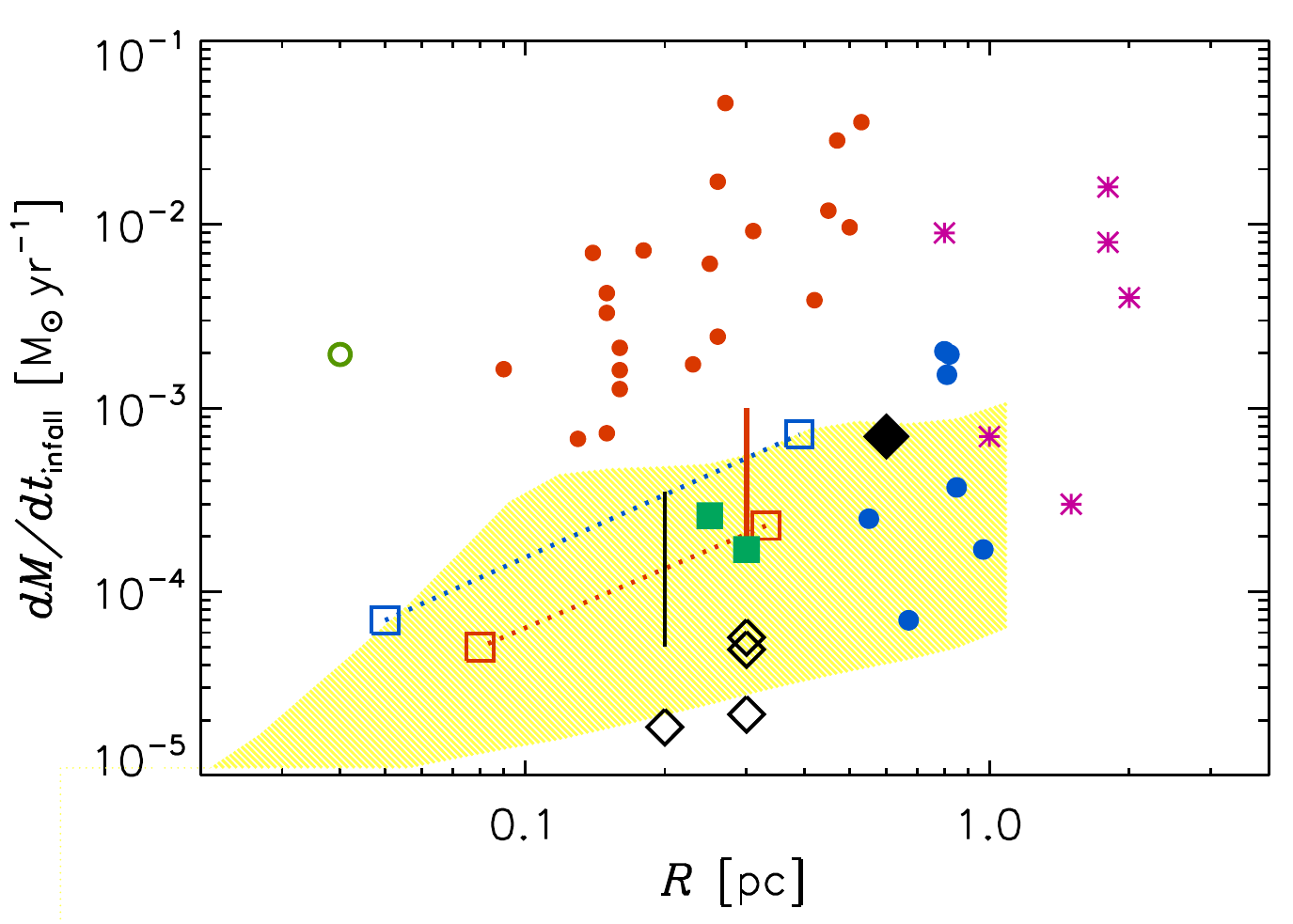}
\caption[]{Observed infall rates in massive star-forming clumps and filaments versus clump radius or filament thickness. The yellow shaded area includes 
the values between the average profile of the cores from the simulation (lower envelope) and the maximum profile (upper envelope), from panel d of 
Figure~\ref{inflow_profiles}. Filled, blue circles: \citet{Traficante+17}, filled, red circles: \citet{Traficante+18}, green asterisks:  \citet{Wyrowski+16}, 
filled, black diamond: \citet{Peretto+13}, empty, black diamonds: \citet{Peretto+14}, filled, green squares: \citet{Chen+19filaments}, empty, green circle:
 \citet{Contreras+18}, empty, blue squares: \citet{Yuan+18},  empty, red squares: \citet{Kirk+13}, black, vertical segment: \citet{Lu+18}, red, vertical 
 segment: \citet{Fuller+05}.         
}
\label{accretion_rates}
\end{figure}

\subsection{Infall-Rate Estimates} \label{sec_rates}

In \S~\ref{sec_ic_inflow}, we found that the mean radial profile of the mass-flow rate around our prestellar cores increases approximately linearly with radius, with an average 
value of $\sim10^{-5}$~M$_{\odot}$\,yr$^{-1}$ at 0.1~pc and $\sim10^{-4}$~M$_{\odot}$\,yr$^{-1}$ at 1~pc (see Figure~\ref{inflow_profiles}). The largest values can 
be a factor of ten above the mean profiles. Similar infall rates, and often much larger ones, have been measured in regions of massive-star formation, although the large 
majority of measurements are for relatively-evolved protostellar sources, while in the simulation we have only considered prestellar cores.

\citet{Fuller+05} found evidence of infall in 22 massive protostellar clumps, out of a total sample of 77 clumps. By analyzing the blue-shifted emission of optically thick lines,
they inferred infall rates between $2\times10^{-4}$~M$_{\odot}$\,yr$^{-1}$ and $10^{-3}$~M$_{\odot}$\,yr$^{-1}$. These values were derived from measured infall velocities
in the range $\sim 0.1-1$~km/s, assuming a density of infalling gas of $5\times10^{4}$~cm$^{-3}$, and a size of $\sim 0.3$~pc for the infalling region. Somewhat larger 
infall rates have been measured in more recent studies of single massive cores, such as  $2.5\times10^{-3}$~M$_{\odot}$\,yr$^{-1}$ on a scale of 0.6~pc by \citet{Peretto+13},
$1.96\times10^{-3}$~M$_{\odot}$\,yr$^{-1}$ on a scale of 8000~au by \citet{Contreras+18}, and $3.5\times10^{-3}$~M$_{\odot}$\,yr$^{-1}$ on a scale of 500~au by 
\citet{Beuther+13}. These estimates have large uncertainties. For example, \citet{Beuther+12} had studied the same massive clump as \citet{Beuther+13}, but had derived 
an approximately 20 times larger accretion rate, primarily because they had assumed a larger size of the infalling region. In \citet{Peretto+13}, the infall rate was actually 
measured along six filaments, giving a total rate of $0.7\times10^{-3}$~M$_{\odot}$\,yr$^{-1}$. The higher value they reported was derived under the assumption of spherical 
symmetry, where the filament infall speed was assigned to the whole volume around the clump.   

We interpreted the radial dependence of the mass-flow rate found in \S~\ref{sec_ic_inflow} as due to the turbulent nature of the inflow region causing the gas to shock 
and accumulate at several secondary converging points. An analogous picture, with very similar infall-rate values as in our mean radial profile, was proposed by 
\citet{Kirk+13} to interpret the kinematics of infalling gas in the Serpens South cluster. They found evidence of accretion onto the main filament of 0.3~pc length, at a rate 
of $2.3\times10^{-4}$~M$_{\odot}$\,yr$^{-1}$, and infall from the filament (with a thickness of 0.08~pc) onto the cluster region at a significantly smaller rate of 
$5.0\times10^{-5}$~M$_{\odot}$\,yr$^{-1}$.\footnote{We have updated these infall-rate values based on the recently determined Gaia distance to Serpens South 
of 436~pc \citep{Ortiz-Leon+18}} Tentative evidence of an increase of the infall rate with scale was also found by \citet{Yuan+18}, with a value of 
$7\times10^{-5}$~M$_{\odot}$\,yr$^{-1}$ on a core scale of 0.05 pc, and $7.2\times10^{-4}$~M$_{\odot}$\,yr$^{-1}$ on a clump scale of 0.39~pc.
Infall along filaments at a scale of 0.2-0.3~pc was also found by \citet{Peretto+14} with rates in the range $1.8-5.6\sim10^{-5}$~M$_{\odot}$\,yr$^{-1}$, by \citet{Lu+18} 
with rates of $0.5-3.5\sim10^{-4}$~M$_{\odot}$\,yr$^{-1}$, and by \citet{Chen+19filaments} with rates in the range $1.7-2.6 \times10^{-4}$~M$_{\odot}$\,yr$^{-1}$. 
 
 \citet{Wyrowski+16} found evidence of infall in six massive clumps at pc scale, using ammonia-line SOFIA observations. They estimated infall rates in the broad range 
 $0.3-16\sim10^{-3}$~M$_{\odot}$\,yr$^{-1}$.
 \citet{Traficante+17} measured infall rates of IR-dark, massive clumps, with radii between approximately 0.54 and 1 pc, and detected infall in 7 out of 18 clumps, with infall
 rates between $4\times10^{-5}$~M$_{\odot}$\,yr$^{-1}$ and $2\times10^{-3}$~M$_{\odot}$\,yr$^{-1}$. In a following survey of 213 more evolved massive clumps,
 \citet{Traficante+18} found evidence of infall in 21 clumps, with infall rates in the range $0.7-45.8\times10^{-3}$~M$_{\odot}$\,yr$^{-1}$. They interpreted tese larger 
 values as evidence that the clump infall rate may increase during the clump evolution.
 
 The infall rates from the works mentioned above are plotted in Figure~\ref{accretion_rates} as a function of the length scale associated with the measurements. In the 
 case where spherical symmetry was assumed, the scale in Figure~\ref{accretion_rates} corresponds to the adopted value of the radius. For infall along filaments we have 
 used the thickness of the filament, or, when one or more filaments are seen to converge onto a clump, the radius of the clump itself. In the case of the infall onto the filament
 studied by \citet{Kirk+13}, the relevant scale is the length of the filament (0.33~pc). The yellow, shaded area in Figure~\ref{accretion_rates} shows the range of infall-rate values
 from our simulation, between the mean and the maximum profiles in panel d of Figure~\ref{inflow_profiles}. Several observed values are found reasonably close to our mean
 profile (the lower envelope of the shaded area), but many observed values are much larger, even exceeding our maximum profile. 
 
\begin{figure}[t]
\includegraphics[width=\columnwidth]{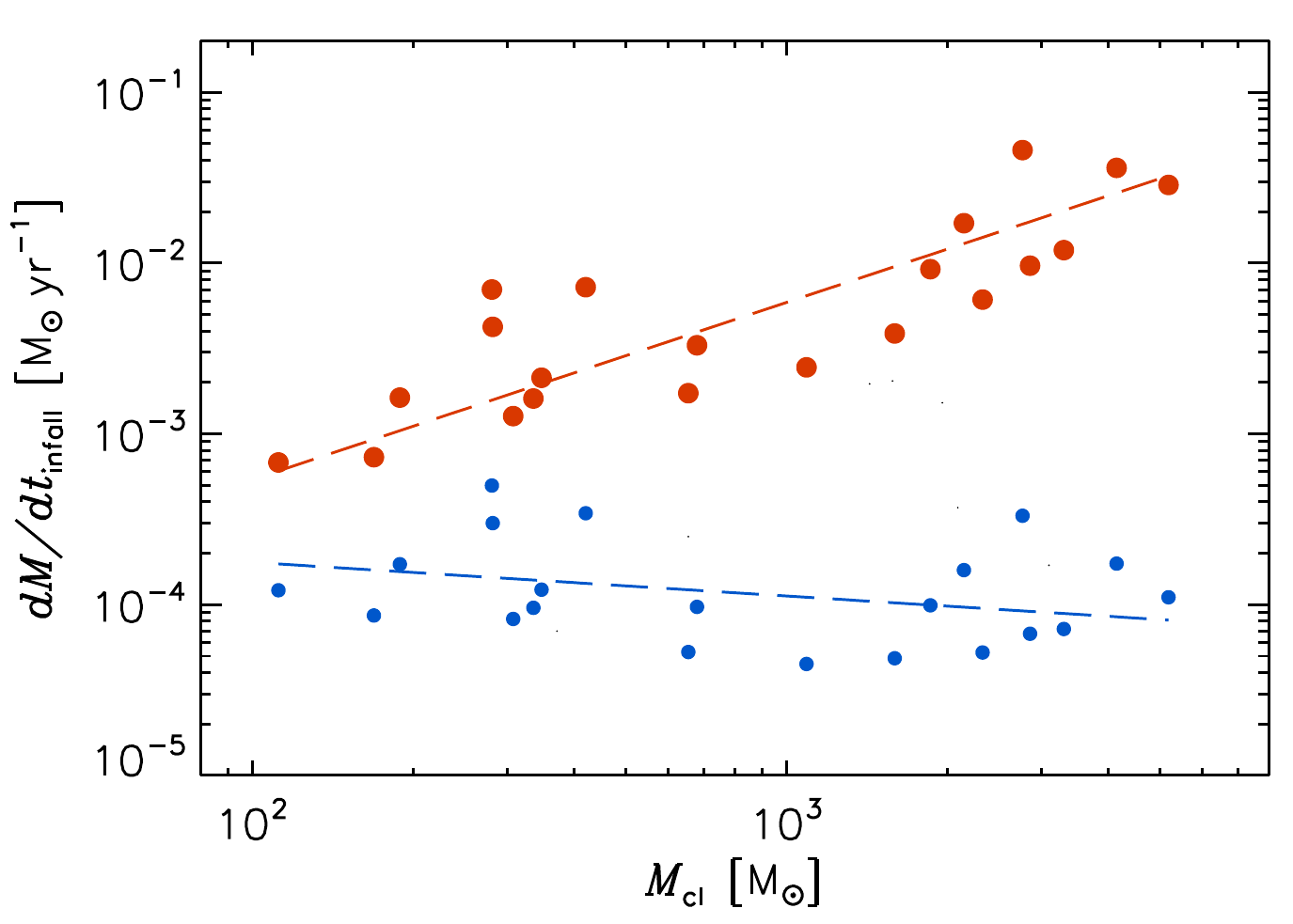}
\caption[]{Red circles: Observed infall rates in massive star-forming clumps versus clump mass from \citet{Traficante+18}. The red dashed line is a least-squares fit 
to the data points, $dM/dt_{\rm infall}\simeq 5.9\times 10^{-3} (M_{\rm cl}/10^3 {\rm M}_{\odot} {\rm yr}^{-1})^{1.04}$~M$_{\odot}$\,yr$^{-1}$. Blue circles: Same infall rates 
as shown by the red circles, but divided by the estimated number of protostellar cores, $N_{\rm p}=M_{\rm cl}/20$\,M$_{\odot}$, as explained in the text. The blue dashed 
line is a least-squares fit to the data points, $dM/dt_{\rm infall,p}\simeq 1.1\times 10^{-4} (M_{\rm cl}/10^3 {\rm M}_{\odot} {\rm yr}^{-1})^{-0.20}$~M$_{\odot}$\,yr$^{-1}$.
}
\label{accretion_mass}
\end{figure}

 Larger values than in 
 our simulation are of course expected in more extreme star-forming regions with higher mean density and/or larger velocity dispersion than in the simulation. However, part of 
 the discrepancy is due to the fact that the observed infall rates of massive clumps apply to a stellar group or cluster, so the actual accretion rate on the individual massive 
 protostars in the clumps must be a small fraction of the global infall rate. In the sample of \citet{Traficante+18}, for example, the clumps with detected infall rates have 
 masses between 100 and 5,000~M$_{\odot}$, as shown in Figure~\ref{accretion_mass}, where the red circles give the infall rates versus clump mass. The red dashed line
 is a least-squares fit to the data, giving $dM/dt_{\rm infall}\simeq 5.9\times 10^{-3} (M_{\rm cl}/10^3 {\rm M}_{\odot} {\rm yr}^{-1})^{1.04}$~M$_{\odot}$\,yr$^{-1}$.
 As discussed in the previous subsection, interferometric observations show that massive clumps are usually fragmented into a number of compact protostellar cores,
 with a characteristic mass of order 10\,M$_{\odot}$. If we assume that the clumps in the sample of \citet{Traficante+18} also contain cores of that characteristic mass,
 accounting for example for half of the clump mass, the other half being outside of the cores, we can estimate an approximate number of protostellar cores per clump as
 $N_{\rm p}=M_{\rm cl}/20$\,M$_{\odot}$. We can then estimate the characteristic infall rate per protostellar core by dividing the clump infall rate by $N_{\rm p}$. The result 
 is shown by the blue circles in Figure~\ref{accretion_mass}, where the blue dashed line is a least-squares fit to the data giving 
 $dM/dt_{\rm infall,p}\simeq 1.1\times 10^{-4} (M_{\rm cl}/10^3 {\rm M}_{\odot} {\rm yr}^{-1})^{-0.20}$~M$_{\odot}$\,yr$^{-1}$. This nearly constant infall rate per protostar is not
 much larger than characteristic values in our simulation.

 A source of uncertainty is related to the procedure usually adopted to derive the infall rate of clumps. 
The infall rate is usually derived by assuming that the measured infall velocity, $v_{\rm in}$, can be interpreted as a mean radial velocity, hence the infall rate is given by 
$4\,\pi\,\rho\,r_{\rm in}^2\,v_{\rm in}$, where $r_{\rm in}$ is the size of the infall region and $\rho$ the mean density. Because of the $r_{\rm in}^2$ dependence, the 
uncertainty on the size of the infall region is a major factor in the uncertainty of the estimated infall rate. The infall radius should be the radial distance where both the infall 
velocity and the gas density are measured. However, in the study of cores or clumps, the infall velocity is usually derived from the blue asymmetry of an optically thick line, 
which may be tracing a surface layer of accreting gas, rather than the bulk of the observed core or clump (depending also on the depletion of the observed molecule), while 
the adopted density value is often the mean density {\it within} the radius $r_{\rm in}$, based on the total mass derived from the dust mass, rather than the density of the 
accreting gas layer whose velocity is measured through the optically thick line. Because the density in the accreting gas may be much smaller than the mean density in the 
core or clump, this procedure may result in a significant overestimate of the infall rate. This may partly explain why the infall rates of massive clumps or cores are typically 
much larger than characteristic infall rates along filaments, where both the choice of the mean density (the filament mean density) and the relevant cross section (the filament width) 
are usually better constrained, and the spherical assumption is not required.

Another source of uncertainty is the fact that unambiguous infall signatures are usually found only on a small fraction of selected clumps, for example
22 ouf of 77 in \citet{Fuller+05}, or 21 out of 213 in \citet{Traficante+18}. It remains to be understood whether the majority of clumps do not show infall because their infall
rates are below the detection limit of the method (in which case the mean infall rate of the clumps could be much lower than the mean of the detected values), or because
even large infall rates are often hard to measure \citep{Smith+13}. However, if large infall rates were hard to measure from most lines of sight, it would imply that
the infall has large deviations from the adopted spherical assumption, hence the infall rate would be significantly overestimated in the cases where it is detected.   
Our mass-flow rate profiles shown in Figure~\ref{inflow_profiles} are shell-averaged first, and then averaged again over many cores. In future studies, we should carry 
out a proper statistical comparison with the observations, by using synthetic observations of individual cores, and measuring their mass-flow rate along different lines of sight.

Outflow rates from massive clumps have also been used to constrain clump infall rates \citep[e.g.][]{Maud+15}. However, deriving infall rates from outflow properties requires 
further assumptions introducing additional uncertainties (for example both the outflow dynamical time and the outflow rate may be hard to interpret if the outflow is
intermittent as a result of episodic accretion events), so we don't consider outflow rates in this work.

\section{Summary of Results and Conclusions} \label{sec_conclusions}

This work presents a new scenario for the origin of massive stars, based on a self-consistent simulation where the conditions controlling star
formation are maintained by the supernovae that result from the star formation.  We propose that massive stars are assembled by large-scale, converging, 
inertial flows that naturally occur in supersonic turbulence. The star-formation timescale and accretion rate are determined by the statistics of supersonic 
turbulence. Because the turbulence turnover time is longer than the postshock free-fall time, the final stellar mass is not set by the mass of the prestellar 
core, which contains only a small fraction of the final stellar mass, in contradiction with the core-collapse models. The stellar accretion rate is controlled 
by the pc-scale inertial inflow that is insensitive to the gravity of the star, in disagreement with the competitive-accretion models. 

The large dynamic range of our simulation in both space and time is well suited to test this multi-scale scenario, and it provides an unprecedented 
statistical sample of massive stars, forming with realistic distributions of initial and boundary conditions. Our analysis is based on the study of the 
physical conditions in the neighborhood of 1,503 stars with mass above 2.5~M$_{\odot}$, formed over a period of 30~Myr and within a volume of 
250~pc. We have focused on the birth time of each star, defined as the beginning of the gravitational collapse of the prestellar core (in practice, the
creation of the sink particle in the simulation). 

The analysis provides clear evidence of the role of inertial inflows in the formation of massive stars, besides refuting basic elements of the core-collapse 
and competitive-accretion models. In what follows, we outline the main results from the simulation and from the synthetic sub-mm observations (derived
quantities apply specifically to our simulation and need to be rescaled in more extreme environments).    

\begin{enumerate}

\item The star-formation timescale, $t_{95}$, increases with the final stellar mass, $M_{\rm f}$, on average, but with a large scatter that increases towards
lower masses. It is mostly in the approximate range 1-6~Myr for stars above 10~M$_{\odot}$. This result contradicts the core-collapse idea, because $t_{95}$ 
is clearly much larger than the free-fall time of prestellar cores, while it is comparable to the turbulence turnover time at the scale of several pc, consistent with 
our scenario.

\item Massive stars achieve their final mass along approximately linear tracks on the $t_{95}-M_{\rm f}$ plane, that is with an approximately constant mean accretion 
rate. Large stochastic or periodic (in binaries) fluctuations of the accretion rate are common, but there is no systematic increase of the accretion rate with time  
as the stellar mass increases. This is in stark contradiction with the prediction of competitive accretion models, and shows that the inertial inflow is not
driven by the gravity of the star, but by large-scale turbulence, as in our scenario.  

\item Prestellar cores that evolve into massive stars have a broad mass distribution, mostly in the range 0.2-40~M$_{\odot}$, with a probability peak just below 
2~M$_{\odot}$. On average, only a small fraction of the final stellar mass is found in the prestellar core. This fraction decreases with increasing stellar mass, 
because there is no correlation between prestellar core mass and final stellar mass, which rules out the fundamental assumption of core-collapse models. 
The lack of correlation with $M_{\rm f}$ applies also to the mass within the gravitationally-bound infall region around the core, so even IMF models where 
stellar masses originate from gravitationally-bound overdensities are ruled out. Prestellar cores start to collapse when they become supercritical at the 
characteristic postshock density, as postulated in our scenario. 

\item The pc-scale region around a prestellar core is turbulent and gravitationally unbound, which makes competitive accretion far too inefficient (the stellar
gravity is too weak). However, it also exhibits a net mass inflow that feeds the infall region of the growing star. For the most massive stars, this inflow region
has an extension, $R_{95}$, of order 10~pc on average, showing again that most of the stellar mass initially is far away from the prestellar core. 

\item In the inflow region, the net inflow velocity (the shell-averaged radial velocity) is generally much smaller than the turbulent velocity, contrary to the prediction 
of other models where the inflow region is dominated by gravity and the infall velocity must be comparable to or larger than the turbulent velocity. The comparison
between inflow and turbulent velocity components is thus an important observational test of the inertial-inflow model.  

\item The average radial profile of the mass-flow rate around a prestellar core is a growing function of radius. While the infall rate onto the core is, on 
average, of order $10^{-5}$~M$_{\odot}$yr$^{-1}$, the inflow rate at a radius of 1~pc is of order $10^{-4}$~M$_{\odot}$yr$^{-1}$. The largest rates are nearly 
ten times higher than these mean values. Because the accretion rate of a star does not grow systematically with time, this radial dependence of the mass-flow 
rate at the end of the prestellar phase indicates that only a fraction of the inflow is destined to that given star. The turbulence is highly supersonic at 
parsec scale, so the inflow region must be strongly fragmented with dense filaments feeding different stars. We show that the feeding of multiple stars
in a typical massive clump explains the apparent discrepancy between the infall rates measured in the simulation and some of the highest infall rates 
inferred from observations. 

\item The star-formation time, $t_{95}$, and the stellar mass-reservoir size, or inflow radius, $R_{95}$, are found to be well correlated, with 
$t_{95}\propto R_{95}^{0.47}$. Both the normalization and the slope of the $t_{95}-R_{95}$ relation are virtually indistinguishable from the 
relation between turbulence turnover time and eddy size inferred from the velocity-size relation of the MCs in the simulation. This is a most 
direct and definite proof that massive stars are assembled by random compressive motions naturally occurring in supersonic turbulence,
the fundamental assumption of our scenario. 

\item The $R_{95}-M_{\rm f}$ relation, as the analogous $t_{95}-M_{\rm f}$ relation, is characterized by a very large scatter that increases with decreasing 
$M_{\rm f}$, due to an approximately linear lower envelope. The lower envelopes of these two relations correspond to the stars with the maximum accretion 
rate, which we have interpreted as a universal fraction, $\epsilon_{\rm in}$, of the ratio of total mass to turnover time of the outer scale, $M_0/\tau_0$. 
This interpretation also predicts the correct value of the maximum accretion rate found in our smaller-scale simulations. 

\item From our inertial-inflow scenario, as well as from the results of the analysis of the simulation, we conclude that observational surveys should fail to find
true prestellar cores of very large mass. However, our synthetic observations in the Herschel and ALMA bands show that cores selected with the same 
procedure as in the observational studies may appear to be much more massive than the true prestellar cores. The masses derived from the observations 
are highly uncertain due to both the line-of-sight projection and the lack of resolution. 

\item The median value of the observationally-derived core masses grows by approximately a factor of three for every factor of two increase in distance or 
telescope-beam size, and only our highest ALMA resolution seems to be converging to the median mass of the actual prestellar cores. However, even in
this best case, only 40\% of our prestellar cores are successfully selected with ALMA, and their observationally-derived masses show no correlation at all 
with the true core masses (they can be a factor of ten larger or smaller than the true masses).  

\item Recent interferometric surveys of massive, IR-dark clumps yield candidate prestellar cores with mass ranges consistent with the results of our 
synthetic observations, suggesting that core masses from surveys with angular resolution worse than $\sim 1''$ (for typical distances of 3-5~kpc) may 
be significantly overestimated on average. On the other hand, in ALMA surveys with angular resolution of $\sim 1''$ or better, 
the average core mass may be spatially resolved (independent of resolution), although individual core masses may still have errors of up to a factor 
of ten due to projection effects.

\end{enumerate}

The most striking result of this study is the surprising similarity of the prestellar phases of high-mass and intermediate-mass stars, suggesting that the 
observational quest for the elusive prestellar cores of high-mass stars may be a lost cause.  Our analysis of the initial conditions for high-mass stars 
did not yield any distinctive traits of such cores, nor of their surrounding inflow regions. {\it Essentially, the final stellar mass is unpredictable based on
prestellar properties.} On the other hand, this also implies that a fraction of prestellar cores already
identified in dark massive clumps are true progenitors of high-mass stars, we just cannot set them apart from the rest. 
Besides focusing on individual cores, future observational programs should aim at characterizing the pc-scale structure and kinematics around the cores,
to constrain the role of inertial inflows in the process of star formation.

\acknowledgements
PP acknowledges support by the Spanish MINECO under project AYA2017-88754-P. 
The work of {\AA}N was supported by grant 1323-00199B from the Danish Council for Independent 
Research (DFF), and by the Centre for Star and Planet Formation, which is funded by the Danish National Research Foundation (DNRF97).
LP acknowledges financial support from NSFC under grant No. 11973098.
MJ acknowledges the support of the Academy of Finland Grant No. 285769.
Computing resources for this work were provided by the NASA High-End Computing (HEC) Program through the NASA Advanced 
Supercomputing (NAS) Division at Ames Research Center. 
We thankfully acknowledge the computer resources at MareNostrum and the technical support provided by 
Barcelona Supercomputing Center (AECT-2018-3-0019). 
We acknowledge PRACE for awarding us access to Curie at GENCI@CEA, France.
Storage and computing resources at the University of Copenhagen HPC centre, funded in part by Villum Fonden (VKR023406), were used 
to carry out part of the data analysis. 

We thank the anonymous referee for a very useful report and the following colleagues for reading the first draft and providing comments: 
Philippe Andr\'{e}, Javier Ballesteros-Paredes, Henrik Beuther, Gilles Chabrier, Lee Hartmann, Patrick Hennebelle, Alessio Traficante, Enrique Vazquez-Semadeni. 
Special thanks to Patricio Sanhueza and Delphine Russeil for providing the electronic tables to compute the observational core mass functions 
shown in Figure~\ref{mass_synthetic_histo}, and to Alessio Traficante for many instructive discussions on the observations of infall rates in 
massive clumps, and for providing the electronic tables to produce the plots in Figures~\ref{accretion_rates} and \ref{accretion_mass}.

\appendix

\section{Details of synthetic observations} \label{appendix:synthetic}

The radiative transfer calculations are made with the SOC code \citep{Juvela19}, which has been tested against other codes in the
TRUST\footnote{http://ipag.osug.fr/RT13/RTTRUST/} benchmark project \citep{Baes2016, Gordon2017}. SOC is able to treat also stochastically heated grains
\citep{Camps2015}, but, to speed up the calculations, all grains are here assumed to be in thermal equilibrium with the radiation field. This is a good
approximation for large dust grains and thus for emission observed at submillimeter wavelengths.

The model clouds are illuminated only by an isotropic external radiation field, similar to the local interstellar radiation field \citet{Mathis1983}. The
area-averaged extinction through the full 250\,pc model is A$_{\rm V}\sim 1.9$\,mag and, because of the inhomogeneity of the density field, the effective
optical depth $\tau_{\rm eff}$ (defined by $\exp(-\tau)=\langle \exp(-\tau(\theta,\phi)) \rangle$ where the averaging is over the full solid angle) is below
one at optical wavelengths. Therefore, there are no significant large-scale gradients in the radiation field energy density between the boundary and the
center of the models and the main dust temperature variations take place inside individual dense regions. The simulations describe the prestellar phase and
thus no embedded radiation sources are included. As noted in Sect.\ref{sec_observations}, the dust model was adopted from \citet{Compiegne2011} but modified
to have a higher submillimeter opacity, for a better agreement with observations of prestellar cores \citep{Juvela+15}.

The radiative transfer simulations use $\sim 10^9$ photon packages on each of 52 selected discrete frequencies between $10^{10}$\,Hz and $3\times
10^{15}$\,Hz. The volume discretisation is the same as in the original MHD runs and the smallest cell size in high-density areas is thus $\sim$0.0076\,pc.
The overall rms error of the computed dust temperature values of individual cells is well below 0.1\,K, but increases with the refinement level. At low
temperatures this could result in relative errors of several percent in the surface brightness values. In practice, the effect is smaller because the final
synthetic surface brightness maps correspond to the total emission along a line of sight and are further averaged by the convolution with the instrument
beam.

Cores are extracted and analyzed using the {\em getsources} program \citep{Menshchikov+12}, as described in the study of NGC6334 by \citet{Tige+17}.  The
analysis of the Herschel data uses the four surface brightness maps and a column density map. In the detection phase, the 160\,$\mu$m and 250\,$\mu$m
maps are scaled with factors
\begin{equation}
k_{\nu} = I_{\nu}(17\,{\rm K})/I_{\nu}(T_{\rm d}).
\end{equation}
This scaling converts the surface brightness maps to an approximation of the column density. The dust color temperatures, $T_{\rm d}$, are obtained by
fitting the SED of each pixel with a modified blackbody with a fixed dust spectral index value of $\beta=1.8$. A separate column density map is also used in
the source detection. It is calculated with the method described by \citet{Palmeirim+13}, where the analysis of different band combinations results in a
column density map at the resolution of the Herschel 250\,$\mu$m data, that is $18.2\arcsec$.  Once the detections have been made, {\em getsources}
measures the source properties using the original surface brightness maps. The results include the flux density, the physical core size (major and minor
axes), the position angle, and estimates of the detection signal-to-noise ratio and other quality flags.

The source fluxes are fitted with a modified blackbody with a fixed dust opacity spectral index of $\beta=1.8$. \citet{Tige+17} used a value of $\beta=2.0$,
but $\beta=1.8$ is closer to the actual value of the dust model used in our simulations. The SED fits use flux measurements of all the bands with reliable
object detections. To account for the different effective aperture sizes, the flux densities are corrected by the linear scaling
\citep{Motte+10,NguyenLuong+11}
\begin{equation}
F_{\nu}^{\prime} = F_{\nu} \times \frac{FWHM_{\nu}}{FWHM_{\rm ref}},
\end{equation}
\\
where the FWHM values are the deconvolved sources sizes in the band and in a reference band. The reference band is 160\,$\mu$m or, if there is no reliable
160\,$\mu$m detection, the 250\,$\mu$m band.

Following \citet{Tige+17} and \cite{Russeil+19}, a number of criteria are applied to the {\em getsources} extraction to select the most reliable
detections. In each band, a reliable detection requires that the signal-to-noise ratio is above 2 for both the peak and integrated fluxes and that the
monochromatic significance (reported by {\em getsources}) is above 5. To increase the core sample, we do not enforce the criteria of \citet{Tige+17} that
would exclude objects with sizes above 0.3\,pc or aspect ratios above 2. In our final sample, over 50\% of the sources have aspect ratios above 2 but only
some 10\% are larger than 0.3\,pc. Finally, we only accept sources with reliable flux measurements (as defined above) in three or four bands (including at
least one of the 160\,$\mu$m and 250\,$\mu$m bands) and where the reduced $\chi^2$ value of the SED fit is below 10. Sources with unrealistically low
temperatures, $T_{\rm d}<8$\,K, are also rejected \citep[cf.][]{Tige+17}.

The selection of ALMA cores follows the same procedure, except that there is no SED fit and the basic selection criteria are only applied to the single
1.2\,mm band. Thus, the main criteria are the signal-to-noise ratio of the measured fluxes and the monochromatic significance provided by {\em getsources}.

\bibliographystyle{apj}
\bibliography{nsf08,MC,padoan,paper_SN}

\end{document}